\documentclass[reqno,12pt,a4paper]{amsart}
\usepackage{mathrsfs}
\usepackage{amssymb}
\usepackage{amsfonts}
\usepackage{latexsym}
\usepackage{amsthm}
\usepackage{graphicx}
\usepackage{pgfplots}
\usepackage{pgf}
\usepackage{url}
\usepgfplotslibrary{groupplots}
\pgfplotsset{compat=newest}
\def\lb{\label}

\newcommand{\er}[1]{\textrm{(\ref{#1})}}

\begin{document}


\renewcommand{\theequation}{\arabic{section}.\arabic{equation}}
\theoremstyle{plain}
\newtheorem{theorem}{\bf Theorem}[section]
\newtheorem{lemma}[theorem]{\bf Lemma}
\newtheorem{corollary}[theorem]{\bf Corollary}
\newtheorem{proposition}[theorem]{\bf Proposition}
\newtheorem{definition}[theorem]{\bf Definition}
\newtheorem{condition}[theorem]{\bf Condition}
\newtheorem{remark}[theorem]{\it Remark}

\def\a{\alpha}  \def\cA{{\mathcal A}}     \def\bA{{\bf A}}  \def\mA{{\mathscr A}}
\def\b{\beta}   \def\cB{{\mathcal B}}     \def\bB{{\bf B}}  \def\mB{{\mathscr B}}
\def\g{\gamma}  \def\cC{{\mathcal C}}     \def\bC{{\bf C}}  \def\mC{{\mathscr C}}
\def\G{\Gamma}  \def\cD{{\mathcal D}}     \def\bD{{\bf D}}  \def\mD{{\mathscr D}}
\def\d{\delta}  \def\cE{{\mathcal E}}     \def\bE{{\bf E}}  \def\mE{{\mathscr E}}
\def\D{\Delta}  \def\cF{{\mathcal F}}     \def\bF{{\bf F}}  \def\mF{{\mathscr F}}
\def\c{\chi}    \def\cG{{\mathcal G}}     \def\bG{{\bf G}}  \def\mG{{\mathscr G}}
\def\z{\zeta}   \def\cH{{\mathcal H}}     \def\bH{{\bf H}}  \def\mH{{\mathscr H}}
\def\e{\eta}    \def\cI{{\mathcal I}}     \def\bI{{\bf I}}  \def\mI{{\mathscr I}}
\def\p{\psi}    \def\cJ{{\mathcal J}}     \def\bJ{{\bf J}}  \def\mJ{{\mathscr J}}
\def\vT{\Theta} \def\cK{{\mathcal K}}     \def\bK{{\bf K}}  \def\mK{{\mathscr K}}
\def\k{\kappa}  \def\cL{{\mathcal L}}     \def\bL{{\bf L}}  \def\mL{{\mathscr L}}
\def\l{\lambda} \def\cM{{\mathcal M}}     \def\bM{{\bf M}}  \def\mM{{\mathscr M}}
\def\L{\Lambda} \def\cN{{\mathcal N}}     \def\bN{{\bf N}}  \def\mN{{\mathscr N}}
\def\m{\mu}     \def\cO{{\mathcal O}}     \def\bO{{\bf O}}  \def\mO{{\mathscr O}}
\def\n{\nu}     \def\cP{{\mathcal P}}     \def\bP{{\bf P}}  \def\mP{{\mathscr P}}
\def\r{\rho}    \def\cQ{{\mathcal Q}}     \def\bQ{{\bf Q}}  \def\mQ{{\mathscr Q}}
\def\s{\sigma}  \def\cR{{\mathcal R}}     \def\bR{{\bf R}}  \def\mR{{\mathscr R}}
\def\S{\Sigma}  \def\cS{{\mathcal S}}     \def\bS{{\bf S}}  \def\mS{{\mathscr S}}
\def\t{\tau}    \def\cT{{\mathcal T}}     \def\bT{{\bf T}}  \def\mT{{\mathscr T}}
\def\f{\phi}    \def\cU{{\mathcal U}}     \def\bU{{\bf U}}  \def\mU{{\mathscr U}}
\def\F{\Phi}    \def\cV{{\mathcal V}}     \def\bV{{\bf V}}  \def\mV{{\mathscr V}}
\def\P{\Psi}    \def\cW{{\mathcal W}}     \def\bW{{\bf W}}  \def\mW{{\mathscr W}}
\def\o{\omega}  \def\cX{{\mathcal X}}     \def\bX{{\bf X}}  \def\mX{{\mathscr X}}
\def\x{\xi}     \def\cY{{\mathcal Y}}     \def\bY{{\bf Y}}  \def\mY{{\mathscr Y}}
\def\X{\Xi}     \def\cZ{{\mathcal Z}}     \def\bZ{{\bf Z}}  \def\mZ{{\mathscr Z}}
\def\O{\Omega}
\def\th{\theta}

\newcommand{\gA}{\mathfrak{A}}
\newcommand{\gB}{\mathfrak{B}}
\newcommand{\gC}{\mathfrak{C}}
\newcommand{\gD}{\mathfrak{D}}
\newcommand{\gE}{\mathfrak{E}}
\newcommand{\gF}{\mathfrak{F}}
\newcommand{\gG}{\mathfrak{G}}
\newcommand{\gH}{\mathfrak{H}}
\newcommand{\gI}{\mathfrak{I}}
\newcommand{\gJ}{\mathfrak{J}}
\newcommand{\gK}{\mathfrak{K}}
\newcommand{\gL}{\mathfrak{L}}
\newcommand{\gM}{\mathfrak{M}}
\newcommand{\gN}{\mathfrak{N}}
\newcommand{\gO}{\mathfrak{O}}
\newcommand{\gP}{\mathfrak{P}}
\newcommand{\gQ}{\mathfrak{Q}}
\newcommand{\gR}{\mathfrak{R}}
\newcommand{\gS}{\mathfrak{S}}
\newcommand{\gT}{\mathfrak{T}}
\newcommand{\gU}{\mathfrak{U}}
\newcommand{\gV}{\mathfrak{V}}
\newcommand{\gW}{\mathfrak{W}}
\newcommand{\gX}{\mathfrak{X}}
\newcommand{\gY}{\mathfrak{Y}}
\newcommand{\gZ}{\mathfrak{Z}}

\newcommand{\gm}{\mathfrak{m}}
\newcommand{\gn}{\mathfrak{n}}
\newcommand{\gf}{\mathfrak{f}}
\newcommand{\gh}{\mathfrak{h}}
\newcommand{\mg}{\mathfrak{g}}
\newcommand{\gb}{\mathfrak{b}}
\newcommand{\gp}{\mathfrak{p}}
\newcommand{\gu}{\mathfrak{u}}
\newcommand{\ga}{\mathfrak{a}}
\newcommand{\gt}{\mathfrak{t}}
\newcommand{\gs}{\mathfrak{s}}

\def\ve{\varepsilon}   \def\vt{\vartheta}    \def\vp{\varphi}    \def\vk{\varkappa}

\def\Z{{\mathbb Z}}    \def\R{{\mathbb R}}   \def\C{{\mathbb C}}    \def\K{{\mathbb K}}
\def\T{{\mathbb T}}    \def\N{{\mathbb N}}   \def\dD{{\mathbb D}}


\def\la{\leftarrow}              \def\ra{\rightarrow}            \def\Ra{\Rightarrow}
\def\ua{\uparrow}                \def\da{\downarrow}
\def\lra{\leftrightarrow}        \def\Lra{\Leftrightarrow}


\def\lt{\biggl}                  \def\rt{\biggr}
\def\ol{\overline}               \def\wt{\widetilde}
\def\no{\noindent}


\let\ge\geqslant                 \let\le\leqslant
\def\lan{\langle}                \def\ran{\rangle}
\def\/{\over}                    \def\iy{\infty}
\def\sm{\setminus}               \def\es{\emptyset}
\def\ss{\subset}                 \def\ts{\times}
\def\pa{\partial}                \def\os{\oplus}
\def\om{\ominus}                 \def\ev{\equiv}
\def\iint{\int\!\!\!\int}        \def\iintt{\mathop{\int\!\!\int\!\!\dots\!\!\int}\limits}
\def\el2{\ell^{\,2}}             \def\1{1\!\!1}
\def\sh{\sharp}
\def\wh{\widehat}
\def\bs{\backslash}

\def\sh{\mathop{\mathrm{sh}}\nolimits}
\def\Area{\mathop{\mathrm{Area}}\nolimits}
\def\arg{\mathop{\mathrm{arg}}\nolimits}
\def\const{\mathop{\mathrm{const}}\nolimits}
\def\det{\mathop{\mathrm{det}}\nolimits}
\def\diag{\mathop{\mathrm{diag}}\nolimits}
\def\diam{\mathop{\mathrm{diam}}\nolimits}
\def\dim{\mathop{\mathrm{dim}}\nolimits}
\def\dist{\mathop{\mathrm{dist}}\nolimits}
\def\Im{\mathop{\mathrm{Im}}\nolimits}
\def\Iso{\mathop{\mathrm{Iso}}\nolimits}
\def\Ker{\mathop{\mathrm{Ker}}\nolimits}
\def\Lip{\mathop{\mathrm{Lip}}\nolimits}
\def\rank{\mathop{\mathrm{rank}}\limits}
\def\Ran{\mathop{\mathrm{Ran}}\nolimits}
\def\Re{\mathop{\mathrm{Re}}\nolimits}
\def\Res{\mathop{\mathrm{Res}}\nolimits}
\def\res{\mathop{\mathrm{res}}\limits}
\def\sign{\mathop{\mathrm{sign}}\nolimits}
\def\span{\mathop{\mathrm{span}}\nolimits}
\def\supp{\mathop{\mathrm{supp}}\nolimits}
\def\Tr{\mathop{\mathrm{Tr}}\nolimits}
\def\BBox{\hspace{1mm}\vrule height6pt width5.5pt depth0pt \hspace{6pt}}
\def\as{\text{as}}
\def\all{\text{all}}
\def\where{\text{where}}
\def\Dom{\mathop{\mathrm{Dom}}\nolimits}
\def\Var{\mathop{\mathrm{Var}}\nolimits}
\def\Argmax{\mathop{\mathrm{Argmax}}\nolimits}
\def\Argmin{\mathop{\mathrm{Argmin}}\nolimits}
\def\Lip{\mathop{\mathrm{Lip}}\nolimits}
\def\osc{\mathop{\mathrm{osc}}\nolimits}


\newcommand\nh[2]{\widehat{#1}\vphantom{#1}^{(#2)}}
\def\dia{\diamond}

\def\Oplus{\bigoplus\nolimits}



\def\qqq{\qquad}
\def\qq{\quad}
\let\ge\geqslant
\let\le\leqslant
\let\geq\geqslant
\let\leq\leqslant
\newcommand{\ca}{\begin{cases}}
\newcommand{\ac}{\end{cases}}
\newcommand{\ma}{\begin{pmatrix}}
\newcommand{\am}{\end{pmatrix}}
\renewcommand{\[}{\begin{equation}}
\renewcommand{\]}{\end{equation}}
\def\eq{\begin{equation}}
\def\qe{\end{equation}}
\def\[{\begin{equation}}
\def\bu{\bullet}

\newcommand{\fr}{\frac}
\newcommand{\tf}{\tfrac}

\title[Metadynamics and Replicator Equations]
{Metadynamics and Replicator Equations}

\date{\today}
\author[Andrey Badanin]{Andrey Badanin}
\author[Olga Rogacheva]{Olga Rogacheva}
\address{Saint-Petersburg
State University, Universitetskaya nab. 7/9, St. Petersburg,
199034 Russia,
an.badanin@gmail.com,\  a.badanin@spbu.ru,\
o.rogacheva@spbu.ru}


\begin{abstract}
We study the long-time behavior of metadynamics with the standard
Gaussian hills in the adiabatic
instantaneous-equilibrium limit. The evolution equation for the bias
potential is transformed into a measure-valued replicator equation
with a Gaussian interaction kernel. We investigate three settings: a
periodic collective variable, a collective variable on a finite
interval, and the INTERVAL scheme.
 We establish global existence and
uniqueness of solutions, including weak solutions in spaces of
probability measures. In the periodic case, we prove convergence of
the sampling density to the uniform distribution and recovery of the
free-energy profile, with stronger modes of convergence under
additional regularity assumptions. On a finite interval, the weak
solutions converge to the unique minimizer of an interaction-energy
functional, which is a finitely atomic measure; we also quantify the
influence of the boundaries. For the INTERVAL scheme, we establish a
lower bound on the lifetime of a quasi-stationary state and
characterize the structure of stationary measures and possible
limits. These results show that the convergence of the bias
potential is an exceptional phenomenon. In most cases, we observe a
weak concentration of the sampling distribution in the vicinity of a
few points.
\end{abstract}

\maketitle

\section{Introduction and main results}
\setcounter{equation}{0}

\subsection{Introduction}
Metadynamics is an enhanced-sampling method introduced by
Laio and Parrinello \cite{LP2} to investigate metastable systems
and reconstruct free-energy landscapes. In molecular dynamics,
the microscopic state $q$ typically belongs to a high-dimensional
configuration space and, at equilibrium, follows
a Gibbs distribution proportional to $e^{-\beta U(q)}$,
where $U$ is the potential energy and $\beta$ is the inverse temperature.
Rather than describing the full microscopic state, one selects a
low-dimensional collective variable
$x(q)\in S,$
chosen to distinguish the relevant metastable states.
If $\rho(s)$ denotes the equilibrium marginal density
of the collective variable, its free energy is defined,
up to an additive constant, by
$F(s)=-\beta^{-1}\log \rho(s).$
The minima of $F$ correspond to frequently visited metastable states,
whereas high free-energy barriers can make transitions between
them exceedingly rare on time scales accessible to direct simulation.

The basic idea of metadynamics is to overcome this sampling problem
by adding a history-dependent bias potential $V(s,t)$.
During the simulation, localized positive functions, usually
Gaussian ``hills'' $G_\s$ of height $h_k$,
are deposited at successive values of the collective variable
$x(q_{t_k})$ at times $t_k$,
$$
V(s,t)\simeq
\sum_{t_k<t}h_k
G_\sigma\big(s-x(q_{t_k})\big),\qq h_k>0,\qq t_1<t_2<...,
$$
see Fig.~\ref{Figmd}.
Consequently, regions that have already been visited become
progressively less favorable
and the dynamics is driven toward less-explored regions of the
collective-variable space $S$. In the idealized picture underlying
standard metadynamics, this progressive filling tends to compensate
the underlying free-energy landscape, and the accumulated bias can
therefore be used to reconstruct $F$. Thus, metadynamics serves two purposes: it accelerates transitions between metastable
states and provides information about the free energy as a function
of the chosen collective variable
 \cite{LG08}, \cite{LP2}, \cite{BL20}.

A widely used modification is well-tempered metadynamics \cite{BBP8},
in which the height of newly deposited hills decreases as the bias grows.
This prevents indefinite filling and leads, under appropriate conditions,
to an asymptotic bias proportional to the negative free energy rather than
to its full compensation. The convergence properties of well-tempered
metadynamics have subsequently been studied in detail, including an
asymptotic convergence result by Dama, Parrinello, and Voth \cite{DPV14}.
From a mathematical viewpoint, metadynamics is therefore a self-interacting
stochastic or deterministic dynamical system: the instantaneous dynamics
depends on a potential generated by its own past trajectory. In the adiabatic
regime considered in the present work, this feedback can be reduced to an
evolution equation for the probability distribution of the collective variable,
leading to the measure-valued replicator equation studied below.

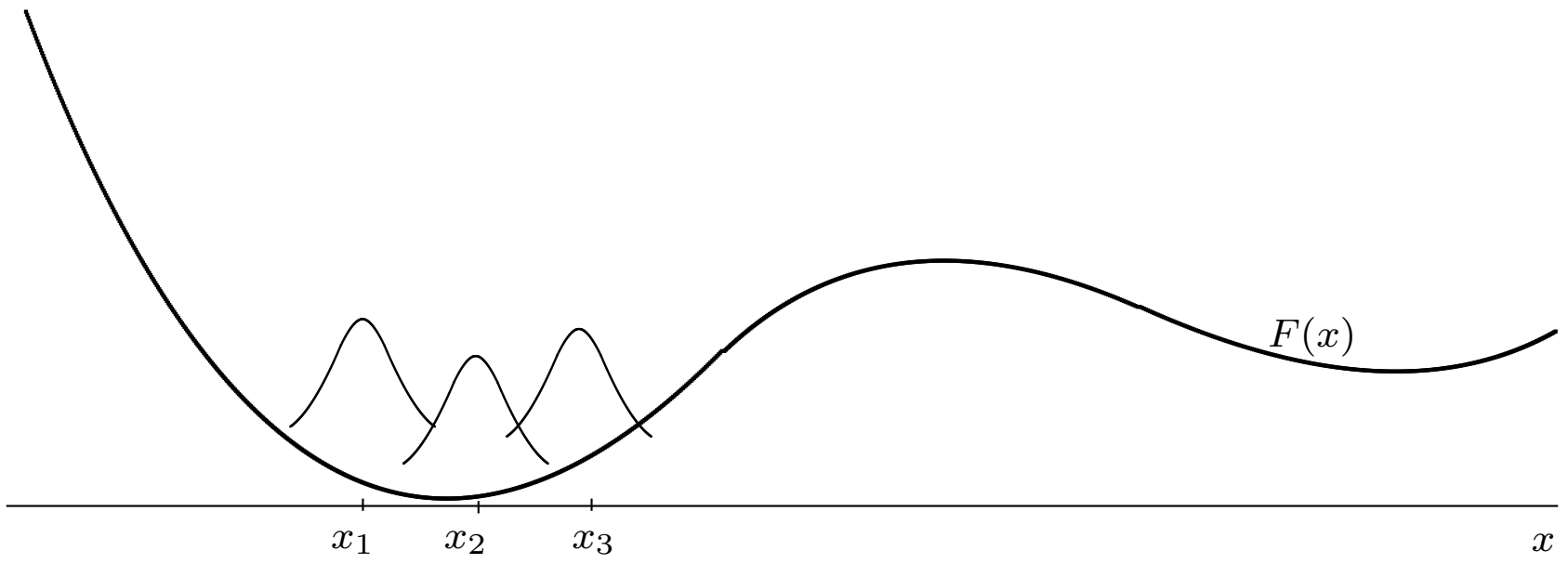
\begin{figure}
    \tiny
    \unitlength 0.70mm
    \begin{picture}(167.25,69.75)(0,0)
        \put(9.5,19.5){\line(1,0){157.75}}
        \thicklines
        \qbezier(11.5,69.75)(40.625,-7)(82.25,35.25)
        \qbezier(82.5,35.25)(99,51)(124.5,39.75)
        \qbezier(124.75,39.75)(150.625,28)(167,37.25)
        \thinlines
        \qbezier(38.375,27.563)(40.688,29.188)(43.5,35.813)
        \qbezier(49.875,23.813)(52.188,25.438)(55,32.063)
        \qbezier(60.375,26.563)(62.688,28.188)(65.5,34.813)
        \qbezier(53.063,27.563)(50.75,29.188)(47.938,35.813)
        \qbezier(64.563,23.813)(62.25,25.438)(59.438,32.063)
        \qbezier(75.063,26.563)(72.75,28.188)(69.938,34.813)
        \qbezier(43.5,35.813)(44.781,38.469)(45.688,38.5)
        \qbezier(55,32.063)(56.281,34.719)(57.188,34.75)
        \qbezier(65.5,34.813)(66.781,37.469)(67.688,37.5)
        \qbezier(47.938,35.813)(46.656,38.469)(45.75,38.5)
        \qbezier(59.438,32.063)(58.156,34.719)(57.25,34.75)
        \qbezier(69.938,34.813)(68.656,37.469)(67.75,37.5)
        \put(165.75,15.75){\makebox(0,0)[cc]{$x$}}
        \put(142.25,36.75){\makebox(0,0)[cc]{$F(x)$}}
        \put(45.75,20.25){\line(0,-1){1.25}}
        \put(57.5,20){\line(0,-1){1.25}}
        \put(69,20.25){\line(0,-1){1.25}}
        \put(44.75,15.75){\makebox(0,0)[cc]{$x_1$}}
        \put(56.25,15.75){\makebox(0,0)[cc]{$x_2$}}
        \put(69.25,15.75){\makebox(0,0)[cc]{$x_3$}}
    \end{picture}
    \linethickness{0.4pt}
    \caption{Illustration of the metadynamics framework,
    $x=x(q)$ is the 1-dimensional collective variable,
    $F(x)$ is the unknown free energy profile.
    Three successive hills are deposited, each centered at a different point:
    $x_1=x(q_{t_1})$, $x_2=x(q_{t_2})$, and $x_3=x(q_{t_3})$.}
    \lb{Figmd}
\end{figure}

Replicator equations originated in mathematical biology
and evolutionary game theory as dynamical models for the evolution
of frequencies of competing types or strategies.
For a finite strategy set, the classical equation has the form
$$
\begin{aligned}
&\dot x_i=x_i\bigl(f_i(x)-\bar f(x)\bigr),\qq i=1,2,...,N,\ \ N\ge 1,
\\
&x=(x_1,x_2,...,x_N),\ \ x_i\ge 0,\ \ \sum_{j=1}^Nx_j=1,
\ \
f_i(x)\in\R,\qq
\bar f(x)=\sum_{j=1}^N x_j f_j(x),
\end{aligned}
$$
so that strategies $i$ whose fitness $f_i(x)$ exceeds the population average
$\bar f(x)$
increase their relative frequency $x_i$. The relation of this dynamics
to evolutionary stability was developed in the seminal works of Smith
and Price \cite{MP73}, Taylor and Jonker \cite{TJ78}, and
Hofbauer, Schuster, and Sigmund \cite{HSS79},
while Bomze \cite{Bo83}, \cite{Bo90}
established several of the basic qualitative properties
of finite-dimensional replicator systems,
see also the book by Hofbauer and Sigmund \cite{HS98}.
 The theory is closely related
to the notion of Nash equilibrium.
For a symmetric game with payoff matrix $A$, $f_i(x)=(Ax)_i$,
$\bar f(x)=x^\top Ax$, and
a state $x$ in the probability simplex is a Nash equilibrium if
$(Ax)_i\le x^\top Ax$ for all $i$,
with equality for every strategy $i$ such that $x_i>0$.
Thus no pure strategy
has a payoff against the equilibrium population larger
than the mean equilibrium payoff. Every Nash equilibrium
is therefore a stationary state of the replicator dynamics
$$
\dot x_i=x_i\left((Ax)_i-x^{T}Ax\right),
$$
although the converse is generally false,
since stationarity only imposes the equality condition
on strategies belonging to the support of $x$.

For continuous strategy spaces, the population state is naturally
described by a probability measure. Oechssler and Riedel \cite{OR01},
\cite{OR02} formulated the replicator equation
$$
\dot\mu=\bigl(A_\mu-\bar A_\mu\bigr)\mu
$$
on spaces of probability measures $\mu$ and proved
well-posedness under boundedness assumptions on the payoff function
$f(x,\mu)=A_\mu(x)$,
including existence, uniqueness, and invariance of the probability
simplex in the measure-valued setting. Stability and weak-topology
questions for continuous trait spaces were further studied by
Cressman, Hofbauer, and Riedel \cite{CHR06},
Shaiju and Bernhard \cite{SB09}, while more general selection
and selection-mutation equations in measure spaces were considered
by Ackleh, Cleveland, and Thieme \cite{ACT16}.
Kravvaritis, Papanicolaou, Xepapadeas, and
Yannacopoulos \cite{KPXY10} considered stationary solutions
to the replicator dynamics equations.
This measure-valued framework
is especially relevant in the present paper, since the long-time behavior may
involve weak concentration of initially regular distributions
toward singular, and in particular atomic, limiting measures.
Note that, in this interpretation, the Nash condition becomes
$A_\mu(x)\le \bar A_\mu$,
with equality on the support of $\mu$.

We show that metadynamics provides a new physical realization
of replicator dynamics: instead of describing competition
between biological types or game-theoretic strategies,
the equation governs redistribution of probability mass over
the collective-variable space through feedback from the accumulated bias.
In both settings, the common mechanism is relative multiplicative growth: the local rate
is determined by the deviation of an effective fitness from its
population average, while the subtraction of this average preserves
the total mass.
Applying the theory of replicator equations to metadynamics
reveals new features of its convergence behavior.

In this paper, we rigorously investigate the convergence of metadynamics,
a question raised in \cite{DPV14}.
Numerical experiments on simple models illustrating these results
will be presented in a companion paper \cite{BR26}.

We consider three settings that arise in molecular modeling:

1. The collective variable is periodic.

2. The collective variable takes values in a finite interval.

3. The collective variable is treated using the ``INTERVAL'' scheme \cite{BL12}.

More precisely, the ``INTERVAL'' scheme of \cite{BL12} works as follows.
The collective variable $x$ ranges over the entire real line.
Gaussian hills are deposited at all values of the collective variable
visited by the trajectory. Outside the prescribed finite interval, however,
the trajectory is governed solely by the initial potential $F$;
the added bias does not affect the dynamics there. The initial potential is also assumed to grow sufficiently rapidly
at infinity, thereby confining the trajectory and ensuring adequate
decay of the equilibrium distribution.

Our approach begins with a differential equation for the time evolution
of the bias potential. We transform this equation into a
{\it replicator equation} and study its solutions. Our main questions concern the existence
of stationary and quasi-stationary states and their long-time behavior. When suitable stationary or quasi-stationary states exist,
they can provide information about the free-energy profile.
The long-time behavior of solutions determines the form,
and possible failure, of convergence in metadynamics.

Throughout the paper, the reference interval for the
collective variable is $[0,1]$,
and we normalize the free energy $F$ according to the condition
$$
\int_0^1F(x)dx=0.
$$

\subsection{Basic equations}
To treat the periodic case, the finite-interval case,
and the ``INTERVAL'' scheme in a common framework,
we introduce the following notation.

i) Define the domains
\[
\lb{I_j}
I_1=\T:=\R/\Z,\qq I_2=[0,1],\qq I_3=\R.
\]

ii) Define the hill kernels $ G_j$, $j=1,2,3$,
with center $s\in I_j$, by
\[
\lb{kernels}
 G_1(x,s)=\sum_{n\in\Z}g(x-s-n),\qq  G_2(x,s)=g(x-s),\qq G_3(x,s)
=\ca g(-s),x<0\\g(x-s),0\le x\le 1\\ g(1-s),x>1\ac,
\]
for all $(x,s)\in I_j^2$,
where the Gaussian $g$ is given by \er{gaussp}
\[
\lb{gaussp}
g(x)={e^{-{x^2\/2\s^2}}\/\sqrt{2\pi}\s},\qq
\s>0.
\]

iii) Define the spaces $\cQ_j$, $j=1,2,3$,
for the zero-mean free energy $F$ by
\[
\lb{defcQjcXj}
\begin{aligned}
&\cQ_1=C_0(\T),\qq \cQ_2=C_0([0,1]),
\\
&\cQ_3=\{f\in C(\R):\int_0^1f(x)dx=0,\int_\R e^{-{f(s)\/T}}ds<\iy,\inf_{x\in\R}f(x)>-\iy\},
\end{aligned}
\]
Here $C_0(\T)$ denotes the space of continuous,
zero-mean functions on the circle $\T$:
\[
\lb{defCoT}
C_0(\T)=\{f\in C(\T):\int_0^1f(x)dx=0\},
\]
Similarly, $C_0([0,1])$ denotes the space of continuous,
zero-mean functions on $[0,1]$:
\[
\lb{defC001}
C_0([0,1])=\{f\in C([0,1]):\int_0^1f(x)dx=0\}.
\]

iv) Define the spaces $\cX_j$, $j=1,2,3$,
for the zero-mean bias potential $ V$ by
\[
\lb{defcXj}
\cX_1=C_0(\T),\ \ \cX_2=C_0([0,1]),\ \
\cX_3=\left\{f\in C(\R):\int_0^1f(x)dx=0;f(x)=\ca f(0),x\le 0\\
f(1),x\ge 1\ac\right\},
\]
Thus, $\cX_3$ consists of continuous functions on $\R$
that are constant on each component of $\R\setminus[0,1]$
and have zero mean on $[0,1]$.

Throughout the paper,

{\it
\no $j=1$ corresponds to the periodic case,

\no $j=2$ corresponds to  the case of finite interval,

\no $j=3$ corresponds to  the method ``INTERVAL''.
}

Following Dama, Parrinello, and Voth \cite{DPV14},
we write the basic equation for the zero-mean bias potential $ V$ as
\[
\lb{ideq2}
{d V(x,\theta )\/d\theta }=\int_{I_j}K_j(x,s) p(s,\theta; V)ds,\qq
(x,\theta)\in I_j\ts\R_+,\qq j\in\{1,2,3\},
\]
\[
\lb{defKjxs}
K_j(x,s)= G_j(x,s)-\int_0^1 G_j(u,s)du,\qq (x,s)\in I_j^2,
\]
\[
\lb{defKrho}
 p(x,\theta; V)
=\frac{e^{-\a V(x,\theta )-\b F(x)}}{\int_{I_j}e^{-\a V(s,\theta )-\b F(s)}ds},
\qq x\in I_j,
\]
where $\a$ and $\b$ are positive parameters,
$\a=\frac{1}{T}+\frac{1}{\D T}$,
$\b=\frac{1}{T}$, $T$ is the temperature, $\D T$ is an energy-valued positive
scalar parameter of the well-tempered metadynamics.
The space $\cX_j$ is invariant under the evolution
generated by Eq.~\er{ideq2}.

Let $j\in\{1,2,3\}$ and let $\cQ_j$ and $\cX_j$ be given by \er{defcQjcXj}
and \er{defcXj}.
For any fixed $F\in\cQ_j$, the identity
\[
\lb{pcV}
p={e^{-\a V-\b F}\/\int_{I_j} e^{-\a V(s)-\b F(s)}ds},
\]
see \er{defKrho},
establishes a one-to-one correspondence between the set
$\cX_j$ of biased potentials $ V$ and the set $\mP_j$
of continuous probability densities $p$, defined by
\[
\lb{spmP}
\begin{aligned}
\mP_1=\{p\in C(\T):p>0,\int_\T p(x)dx=1\},
\qq \mP_2=\{p\in C([0,1]):p>0,\int_0^1p(x)dx=1\},
\\
\mP_3=\Bigg\{p\in C(\R):p>0,\int_\R p(x)dx=1;
p(x)=\ca p(0)e^{\b(F(0)-F(x))},x\le 0\\
p(1)e^{\b(F(1)-F(x))},x\ge 1\ac\Bigg\}.
\end{aligned}
\]
In Lemma~\ref{Lmglclsol}, we show that the function $ p$,
given by \er{defKrho}, satisfies the equation
\[
\lb{eqgpwti}
{d p(x,\theta)\/d\theta}
=-\a p(x,\theta)
\big(U_j(x,\theta)
-\langle U_j\rangle_{ p,j}(\theta)\big),\qq
(x,\theta)\in I_j\ts\R_+,
\]
where
\[
\lb{djavg}
U_j(x,\theta)=\int_{I_j} G_j(x,s) p(s,\theta)ds,\qq
\langle f\rangle_{p,j}=\int_{I_j} f(x)p(x)dx,
\]
$ G_j$ are given by \er{kernels}.
This is precisely the {\it replicator equation} with the payoff function
$-\a U_j$; see \cite{OR01}.
Note that $F$ enters Eq.~\er{eqgpwti} only through
the initial condition,
$$
 p(x,0)={e^{-\a V(s,0)-\b F(s)}\/\int_0^1e^{-\a V(s,0)-\b F(s)}ds}.
$$
We prove global existence and uniqueness for Eqs.~\er{ideq2}
and \er{eqgpwti} in the corresponding spaces $\cX_j$ and $\mP_j$.

\begin{theorem}
\lb{ThGS}
 Let $j\in\{1,2,3\}$ and let  $F\in\cQ_j$.

i) Let, in addition, $ V_0\in\cX_j$.
Then Eq.~\er{ideq2} has a unique solution
$ V\in C^1([0,+\iy),\cX_j)$
satisfying the condition $ V(x,0)= V_0(x)$, for all $x\in I_j$.

ii) Let, in addition, $ p_{0}\in\mP_j$.
Then Eq.~\er{eqgpwti} has a unique solution
$ p\in C^1(\R_+,\mP_j)$,
satisfying the condition
$ p(x,0)= p_{0}(x)$, for all $x\in I_j$.
\end{theorem}

\no {\bf Remark.} 1) Global solvability of Eq.~\er{eqgpwti} in spaces of probability measures
also follows from the general theory of replicator equations; see
\cite[Th~1]{OR01}.
For completeness, and to obtain the estimates needed below,
we provide direct proofs for the three settings considered here.

2) The argument uses standard Lipschitz estimates for the kernel $K_j$.

\subsection{Properties of the solutions}

In the periodic case $j=1$, we obtain the following convergence results.

\begin{theorem}
\lb{ThPC}
Let  $F\in C_0(\T)$, $ V_0\in C_0(\T)$,
let $ V\in C^1(\R_+,C_0(\T))$ be
 the solution to Eq.~\er{ideq2}, for $j=1$,
satisfying the condition
$ V(x,0)= V_0(x),x\in\T$, and let $ p$ be given by \er{defKrho}.
As $\theta\to\iy$,
Then
\[
\lb{Rto00}
\|\a V(\cdot,\theta)+\b F\|_{L^1(\T)}\to 0,\qq
\| p(\cdot,\theta)-1\|_{L^1(\T)}\to 0.
\]
If  $F\in C_0^1(\T)$, $ V_0\in C_0^1(\T)$, then
\[
\lb{Rto02}
\|\a V(\cdot,\theta)+\b F\|_{C(\T)}\to 0,\qq
\| p(\cdot,\theta)-1\|_{C(\T)}\to 0.
\]
If $F\in C_0^2(\T)$, $ V_0\in C_0^2(\T)$, then
\[
\lb{Rto0s2}
\|\a V(\cdot,\theta)+\b F\|_{C^1(\T)}\to 0,
\qq
\| p(\cdot,\theta)-1\|_{C^1(\T)}\to 0.
\]
\end{theorem}

\no {\bf Remark.}
1) Dama, Parrinello, and Voth \cite{DPV14} analyze the differential equation
\er{ideq2} for the periodic case $j=1$ using the Lyapunov method.
They use the Kullback-Leibler (KL) divergence between
the current tempering-reweighted distribution $p_w$
and the target distribution $p=1$ as the Lyapunov function.
The Lyapunov argument in \cite{DPV14} does not by itself
justify the required convergence in the infinite-dimensional setting.
In particular, compactness of the orbit in the relevant Banach space
is not established, so a direct appeal to finite-dimensional
Lyapunov theory is insufficient. As the proof of Theorem~\ref{ThPC} shows, control of the
high-frequency components of the bias potential is essential.
Our argument combines Lyapunov estimates with Fourier analysis,
which is particularly effective for the periodic problem.

2) The convergence involves a loss of one degree of regularity.
For continuous initial data, convergence is proved in $L^1(\T)$;
for $C^1(\T)$ initial data, in $C(\T)$; and for $C^2(\T)$
initial data, in $C^1(\T)$. The treatment of high-frequency
components is responsible for this regularity requirement;
see the proof of Theorem~\ref{ThPC}.

3) A standard molecular example involving periodic collective variables
is alanine dipeptide.
In this case, the free energy $F$ is a smooth function on the torus.
In some applications, however, the assumption $F\in C^1(\T)$
may be restrictive. For example, periodic extension of a function
originally defined on $[0,1]$ need not preserve smoothness
at the identified endpoints.

\medskip

The following theorem shows that the oscillation of the bias potential
is bounded below by a linearly growing function of time,
in both the finite-interval case and the ``INTERVAL'' scheme.
In particular, the bias cannot converge uniformly to a finite profile in these settings.

\begin{theorem}
\lb{Thosc}
 Let $j\in\{2,3\}$.
Let the function $ V(\cdot,\theta)\in\cX_j$ satisfy Eq.~\er{ideq2}.
Then
\[
\lb{oscVinc}
\osc_{[0,1]} V(x,\theta)\ge\theta \d-\osc_{[0,1]} V(x,0),
\qq\text{some}\ \ \d>0,
\]
where
\[
\lb{defosc}
\osc_{[0,1]}f(x)=\max_{x\in [0,1]}f(x)-\min_{x\in [0,1]}f(x).
\]
\end{theorem}

\no {\bf Remark.}
Theorem~\ref{Thfs} yields a stronger result for the finite interval:
the measure-valued solution converges weakly to a finite sum
of Dirac measures. For the ``INTERVAL'' scheme,
we characterize the stationary measures and give conditional
conclusions for trajectories with a weak limit.

\medskip

Consider a collective variable taking values in a finite interval.
The following proposition describes the so-called ``edge effects''.
When $\sigma$ is small, the state $ p=1$ is quasi-stationary
in the following sense:
the distribution changes at an order-one rate near the
boundaries, whereas sufficiently far into the interior
the initial rate is small. The boundary layers therefore
produce a rapid departure from the quasi-stationary state.

\begin{proposition}
\lb{Prop1qs}
 Let $ p\in C^1(\R_+,\mP_2)$
be the solution to Eq.~\er{eqgpwti} for $j=2$, satisfying
the condition $ p(x,0)=1$ for all $x\in[0,1]$.
Then for any fixed $M>0$ and for all $\s>0$ small enough
the following estimates hold:
\[
\lb{asdergpp=1edge}
{d p(x,\theta)\/d\theta}\big|_{\theta=0}
=-2\a\F\Big({x\/\s}\Big)+\ve_1,\qq x\in[0,M\s],
\]
\[
\lb{asdergpp=1edge+}
{d p(x,\theta)\/d\theta}\big|_{\theta=0}
=2\a\F\Big({1-x\/\s}\Big)+\ve_2,\qq x\in[1-M\s,1],
\]
\[
\lb{asdergpp=1centre}
\Big|{d p(x,\theta)\/d\theta}\big|_{\theta=0}\Big|
\le\a\big(2\F(M)+\s\ve_0\big),\qq
x\in[M\s,1-M\s],
\]
where
$$
\F(x)={1\/\sqrt{2\pi}}\int_x^\iy e^{-{s^2\/2}}ds,
\ \
\ve_0=\sqrt{2\/\pi}+2\int_{1\/\s}^\iy\F(x)dx,
\ \
\sup\{|\ve_1|,|\ve_2|\}\le\a\Big(2\F\Big({1\/\s}-M\Big)
+\s\ve_0\Big).
$$
\end{proposition}

\medskip

\no {\bf Remark.}
1) By choosing $M$ sufficiently large in \er{asdergpp=1centre},
the derivative ${d p\/d\theta}|_{\theta=0}$
away from the boundary layers is small, of order $\s$
when the Gaussian tail term is negligible at that scale.
By contrast, estimates \er{asdergpp=1edge} and
\er{asdergpp=1edge+} show an order-one initial rate
of change near the endpoints.

2) We have $g(x)=\d(x)$ in the limit $\s=0$. In this limit the solution
$ p=1$ is a stationary solution and
$\lim_{\theta\to+\iy} p(x,\theta)=1$ for all solutions
to Eq.~\er{eqgpwti}, $j=2$.

\medskip

Now consider the ``INTERVAL'' scheme.
Let $j=3$, let $F\in\cQ_3$, $\cQ_3$ is defined in \er{defcQjcXj}.
Assume also that $F$ is constant on the intervals $[-L,0]$ and $[1,L+1]$
for some sufficiently large $L>0$:
\[
\lb{Fint}
F(x)=F(0),\ \ \forall \ \ x\in[-L,0],\qq
F(x)=F(1),\ \  \forall\ \ x\in[1,L+1].
\]
Let $ V_L$ be defined by
\[
\lb{defcVL}
 V_L(x)=-{\b\/\a}\ca
F(0), x<-L,\\
F(x),-L\le x\le L+1,\\
F(1), x>L+1\ac,\qq
x\in\R.
\]
In the set $\cX_3$, consider a ball centered at $ V_L$:
\[
\lb{bincX}
\cB_r=\{ V\in\cX_3:\sup_{x\in[0,1]}| V(x)- V_L(x)|<r\},\qq r>0.
\]
The following theorem shows that, for large $L$,
the solution of Eq.~\er{ideq2} with $j=3$ and
initial condition $ V(\cdot,0)= V_L$ remains
near $ V_L$ for a long time.

\begin{theorem}
\lb{ThQSSg}
Let $j=3$.
Let $L>0$ be large enough, let $F\in\cQ_3$ satisfy \er{Fint},
let $ V$ be the solution to Eq.~\er{ideq2}
satisfying the condition $ V(\cdot,0)= V_L$, and let $\cT>0$
be such that $ V(\cdot,\theta)\in\cB_R$ for some $R>0$ and for all
$\theta\in[0,\cT]$, where $\cB_R$ is defined by \er{bincX}.
Then
\[
\lb{estcV-cV0}
\sup_{x\in[0,1]}| V(x,\theta )- V_L(x)|
\le{e^{K_R\theta}-1\/\sqrt{2\pi}\s LK_R},
\qq\forall\ \ \theta\in[0,\cT],
\]
where
\[
\lb{defkRLc}
K_R={2\a e^{4\a R-5\b\inf_{x\in\R}F(x)}\/\int_{\R}e^{-\b F(s)}ds},
\]
 with $R>0$.
Moreover, for $\theta$, satisfying
\[
\lb{tstgtVinBR}
\theta\le{\log(\sqrt{2\pi}\s L K_R R+1)\/K_R},
\]
the solution  $ V$ remains in the ball $\cB_R$.
\end{theorem}

{\no \bf Remark.}
The estimate \er{tstgtVinBR} shows that the lifetime $\theta_*$
of the quasistationary state is at least
$\theta_*\ge{1\/K_R}\log(\sqrt{2\pi}\s L K_R R+1)$.
When $L\s$ is sufficiently large, the solution
remains near the quasi-stationary state for a long time.

\subsection{Weak solutions}
Equation~\er{eqgpwti} is a replicator equation.
For $j=1,2$, the interaction energy defined in \er{defcE}
is a strict Lyapunov functional; see Lemma~\ref{LmLF}.
Consequently, nontrivial periodic orbits are excluded;
the limit behavior is governed by stationary states.
For $j=3$, the long-time behavior is less fully determined.

The natural setting for Eq.~\er{eqgpwti} is the space
of probability measures \cite{SB09}.
In the finite-interval case, stationary solutions do not exist
among strictly positive continuous densities, but do exist
as probability measures; see Lemmas~\ref{Lmstsolfi}
and \ref{Lmfsmdi}. The atomic structure of these measures
makes a weak measure-valued formulation essential.
The ``INTERVAL scheme also permits stationary measures
with prescribed absolutely continuous tails.

Define the spaces of probability measures
$$
\cP_j=\cP(I_j),\qq j=1,2,
$$
where $\cP_1=\cP(\T)$ is the set of probability measures on the torus $\T$,
$\cP_2=\cP[0,1]$ is the set of probability measures on the interval  $[0,1]$.

For the ``INTERVAL'' scheme, consider the following measures.
define the set $\cP_{\le 1}[0,1]$
of non-negative Borel measures $\eta$ on $[0,1]$, such that
$\eta([0,1])\le 1$.
Define the functions
\[
\lb{wpm}
w_-(x)=e^{-\b(F(x)-F(0))},\ \ x\le 0,\qq
w_+(x)=e^{-\b(F(x)-F(1))},\ \ x\ge 1,
\]
and the constants
\[
\lb{Apm}
A_-=\int_{-\iy}^0w_-(x)dx,\qq A_+=\int_1^{+\iy}w_+(x)dx.
\]
Assume that $F\in\cQ_3$, where the space $\cQ_3$ is defined by \er{defcQjcXj}.
Then $0<A_\pm<\iy$.
Define $\cP_3$ to be the space of probability measures of the form
\[
\lb{mcMF}
\m(dx)=\eta(dx)+c_-w_-(x)\1_{(-\iy,0)}dx+
c_+w_+(x)\1_{(1,+\iy)}dx,
\]
where $\eta\in\cP_{\le 1}[0,1]$, $c_\pm\ge 0$,
satisfying the condition
\[
\lb{nmcMF}
\m(\R)=\eta([0,1])+c_-A_-+c_+A_+=1.
\]

Each measure $\m\in\cP_j$, $j\in\{1,2,3\}$,
defines a linear functional $\vp \mapsto\int_{I_j}\vp (x)\m(dx)$
on the space $\mC_j$
of test functions, given by
\[
\lb{sptf}
\mC_1=C(\T),\qq \mC_2=C([0,1]),\qq \mC_3=C_b(\R).
\]
Here $C(\T)$ is the space of continuous functions on the torus $\T$,
 $C([0,1])$ is the space of continuous functions on the interval $[0,1]$,
and $C_b(\R)$ is the space of bounded continuous functions on $\R$.
We equip $\cP_j$ with the topology of weak convergence, defined by
\[
\lb{wcoonvt}
\m_n\rightharpoonup\m\qq\Leftrightarrow\qq
\int_{I_j}\vp (x)\m_n(dx)\to\int_{I_j}\vp (x)\m(dx),\qq\forall\ \
\vp \in \mC_j.
\]

Let $j\in\{1,2,3\}$.
For $\m_j\in\cP_j$ we define the Gaussian potentials
\[
\lb{gp}
\cU_{\m_j,j}(x)=\int_{I_j} G_j(x,s)\m_j(ds),\qq x\in I_j,
\]
where $ G_j$ and $I_j$ are given by \er{kernels} and \er{I_j}, respectively.
In particular, the kernel definitions \er{kernels} imply
\[
\lb{Uconstout}
\cU_{\m_3,3}(x)=
\ca \cA_{\m_3}(0),x<0\\\cA_{\m_3}(x),0\le x\le 1\\ \cA_{\m_3}(1),x>1\ac,\qq
\cA_{\m_3}(x)=\int_{\R}g(x-s)\m_3(ds).
\]
The potential $\cU_{\m,j}$ extends the potential $U_j$ defined in \er{djavg}
from absolutely continuous to general probability measures.

A weak solution of Eq.~\er{eqgpwti} on $[0,\cT]$
with values in $\cP_j$ is a family $\m_{\theta,j}\in\cP_j$
such that, for every test function $\vp\in\mC_j$
(as defined in \er{sptf}), the function
$\theta\mapsto\int_{I_j}\vp(x)\m_{\theta,j}(dx)$ is absolutely continuous
 on  $[0,\cT]$ and the equation
\[
\lb{eqgpwtfj}
{d\/d\theta}\int_{I_j}\vp(x)\m_{\theta,j}(dx)
=-\a\int_{I_j}\vp(x)
\big(\cU_{\m_{\theta,j},j}(x)
-\langle \cU_{\m_{\theta,j},j}\rangle_{\m_{\theta,j},j}\big)
\m_{\theta,j}(dx),\qq\forall\ \ \vp\in \mC_j,
\]
holds for almost every $\theta\in[0,\cT]$.

\begin{theorem}
\lb{ThwS}
Let $j\in\{1,2,3\}$ and let $\m_0\in\cP_j$.
Then there exists a unique solution
$\m_\theta\in AC_w([0,+\iy),\cP_j)$,
to Eq.~\er{eqgpwtfj},
satisfying the initial condition $\m_\theta|_{\theta=0}=\m_0$.
The solution $\m_\theta$ satisfies
\[
\lb{idmutheta}
\mu_\theta (dx)=
e^{-\a\int_0^\theta( \cU_{\mu_s}(x)-\langle \cU_{\mu_s}\rangle_{\mu_s})ds}\mu_0(dx),
\qq\theta>0,
\]
\[
\lb{suppmu=0}
\supp\mu_\theta=\supp\mu_0,\qq \forall\ \ \theta\ge 0.
\]
Moreover, if the initial measure is absolutely continuous with respect to
the Lebesgue measure:
$\mu_0(dx)= p_0(x)dx$,
where
$ p_0\in L^1(I_j)$, $ p_0\ge0$, $\int_{I_j} p_0(x)dx=1$,
then

a) for each $\theta\ge0$ the measure $\mu_\theta$
is absolutely continuous:
$\mu_\theta(dx)= p(x,\theta)dx$,

b) $ p\in C^1\bigl([0,\infty);L^1(I_j)\bigr)$,
$ p$ satisfies Eq.~\er{eqgpwti} in $L^1(I_j)$ for all $\theta\ge0$
and the initial condition
$ p(\cdot,0)= p_0 $ in $L^1(I_j)$.

c) The function $ p$ satisfies
\[
\lb{2estp}
e^{- 2\a {\alpha_j\/\s}\theta} p_0(x)
\le  p(x,\theta)\le
e^{ 2\a {\alpha_j\/\s}\theta} p_0(x),\qq \alpha_j={1\/\sqrt{2\pi}}
\ca {1+e^{-{1\/8\s^2}}\/1-e^{-{1\/8\s^2}}},&j=1\\ 1,&j=2,3\ac,
\]
for a.e. $x\in I_j$ and for all finite $\theta$.

\end{theorem}

\medskip

\no {\bf Remark.}
Identity~\er{idmutheta} shows that $\m_\theta$ and
$\mu_0$ are equivalent measures at each finite time.
This identity is a particular case
of the general identity from \cite[Th~3]{OR01}
for the replicator dynamics equation.

\medskip

The next theorem describes the long-time behavior of weak solutions.
For the periodic and finite-interval settings, the interaction energy
is dissipated, and each trajectory converges weakly to its uniquely
determined energy minimizer. The result for the ``INTERVAL scheme
is stated separately, because convergence of every trajectory
has not been established in that case.

\begin{theorem}
\lb{Thfs}
 Let $j\in\{1,2,3\}$, let $\m_0\in\cP_j$, $S=\supp\m_0$, and let
$\m_\theta\in AC_w([0,+\iy),\cP_j)$
be the solution to Eq.~\er{eqgpwtfj},
satisfying the initial condition $\m_\theta|_{\theta=0}=\m_0$.

i) For $j\in\{1,2\}$, the function $\theta\mapsto\cE_j[\m_{\theta}]$ is non-increasing
and
\[
\lb{trgtmin}
\m_\theta\rightharpoonup\m_\iy,
\]
where $\m_\iy$ is the unique minimizer of the functional $\cE_j$
on the set $\cP_j(S)$ defined by \er{defcPjS}.
The measure $\m_\iy$ satisfies the relation
\[
\lb{smiysssm02}
\supp\m_\iy\ss\supp\mu_0,
\]
and the Nash condition
\[
\lb{NC12}
\cU_{\m_\iy,j}(x)\ge \langle \cU_{\m_\iy,j}\rangle_{\m_\iy}\ \ \forall\  x\in S,\qq
\cU_{\m_\iy,j}= \langle \cU_{\m_\iy,j}\rangle_{\m_\iy}=\const\ \ \text{on}\  \supp\m_\iy.
\]
Moreover,

a) if $j=2$ or if $j=1$ and $\supp\m_0\ne\T$, then $\m_\iy$
is a finitely atomic measure,

b) if $j=1$ and $\supp\m_0=\T$, then $\m_\iy$ is the uniform measure.

ii) For $j=3$,
the $\o$-limit set $\o(\m_0)$ of the trajectory
$\m_{\theta}$,  given by
\[
\lb{omlj}
\o(\m_0)=\cap_{t>0}\ol{\{\m_\theta:\theta>t\}}^w,
\]
is non-empty, compact, and connected.
If, in addition, $\m_\theta\rightharpoonup\m_\iy$, as $\theta\to\iy$, for some $\m_\iy\in\cP_3$,
then $\m_\iy$ is a weak stationary measure satisfying \er{stm3}.

\end{theorem}

\no {\bf Remark.}
1) Theorem~\ref{ThwS} shows that an initially absolutely continuous
measure cannot develop atoms at finite time $\theta$. Moreover, the theorem implies that
the set of zeros of the density $ p$ is the same for all finite $\theta\ge 0$.

2) For full-support initial data in the case $j=2$,
the limiting measure is independent of $\m_0$, and hence of $F$.
For narrow kernels, its support contains the endpoints and
additional interior points; their number is expected to increase
as $\s$ decreases. This observation will be explored numerically
in \cite{BR26}.

3) Formally, in the limit $\s\to 0$, the solution $ p=1$ is a stationary solution
to Eq.~\er{eqgpwti} for $j=2$.
Moreover, in this case $\o(\m_0)$ consists of the unique element $\m_\iy=1$
for all $\m_0\in\cP[0,1]$, see also \cite{TDP15}.

4) The proof is based on LaSalle's invariance principle.
The existence of limit sets follows from the compactness
of the space of measures, and convergence to a weakly stationary solution
is obtained from an analysis of the Lyapunov function.

5) Shaiju and Bernhard proved in \cite[Th~9]{SB09} for the large class
of replicator equations that if the Kullback-Leibler
divergence between $\m_0$ and $\m_\iy$ is finite,
then $\o(\m_0)=\m_\iy$.
For Eq.~\er{eqgpwtfj}, we prove a stronger result:
even if the initial measure $\m_0$ is absolutely continuous
(and even smooth),
in the limit we obtain the atomic measure $\m_\iy$.
Note that the Kullback--Leibler
divergence between an absolutely continuous measure and an atomic measure
is infinite, so our result does not follow from \cite[Th~9]{SB09}.

\section{Local and global solvability}
\setcounter{equation}{0}

\subsection{Preliminaries}
Equip the spaces $\cX_j$ defined in \er{defcXj} with the norms
$$
\|f\|_{\cX_j}=\sup_{x\in I_j}|f(x)|,\qq j=1,2,3;
$$
The kernels $ G_j$, $j=1,2,3$, are defined in \er{kernels},
and $K_j$ is defined in \er{defKjxs}.

\begin{lemma}
\lb{Lmgp}
i) Each function $ G_j$ satisfies
\[
\lb{estgaus}
\sup_{x,s\in I_j} G_j(x,s)\le {\alpha_j\/\s},
\]
\[
\lb{estgaus'}
\sup_{s\in I_j}\Lip_{I_j} G_j(\cdot,s)
\le{\beta_j\/\s^2},\qq \beta_j={1\/\sqrt{2\pi e}}\ca
1+{2\sqrt{e}\/\s}\sum_{n=1}^\iy(n+{1\/2})e^{-{(n-{1\/2})^2\/2\s^2}},
&j=1\\ 1,&j=2,3\ac,
\]
where $\Lip_{(a,b)}f$ is
the Lipschitz constant of a function $f:\R\to\R$
on the interval $(a,b)$, $-\iy\le a<b\le+\iy$,
$\alpha_j$ is given by \er{2estp}.

ii) The function $K_j,j=1,2,3$, given by \er{defKjxs}, satisfies
\[
\lb{estKjIj}
\sup_{x\in I_j}\int_{I_j} |K_j(x,s)|ds\le 2.
\]

\end{lemma}

\no {\bf Proof.}
i) The definitions \er{kernels}, \er{kernels}, and \er{gaussp} imply
$$
\sup_{x,s\in I_1} G_1(x,s)
=\sup_{x\in\T}\sum_{n\in\Z}g(x-n)
\le {1\/\sqrt{2\pi}\s}\Big(1+2\sum_{n=1}^\iy e^{-{(n-{1\/2})^2\/2\s^2}}\Big)
={1\/\sqrt{2\pi}\s}\Big(1+{2e^{-{1\/8\s^2}}\/1-e^{-{1\/8\s^2}}}\Big),
$$
$$
\sup_{s\in I_j}\Lip_{I_j} G_j(\cdot,s)
=\sup_{x\in\T}\sum_{n\in\Z}g'(x-n)
\le {1\/\sqrt{2\pi}\s^2}\Big({1\/\sqrt{e}}
+{2\/\s}\sum_{n=1}^\iy\Big(n+{1\/2}\Big) e^{-{(n-{1\/2})^2\/2\s^2}}\Big),
$$
$$
\sup_{x,s\in I_j} G_j(x,s)\le \sup_{x\in\R}g(x)= {1\/\sqrt{2\pi}\s},\qq
\sup_{s\in I_j}\Lip_{I_j} G_j(\cdot,s)
\le \sup_{x\in\R}|g'(x)|= {1\/\sqrt{2\pi e}\s^2},
$$
for $j=2,3$.
These estimates yield \er{estgaus} and \er{estgaus'}.

ii) The definition \er{defKjxs} gives
$$
\int_{I_j} |K_j(x,s)|ds\le
\int_{I_j} G_j(x,s)ds+\int_{I_j}\Big(\int_0^1 G_j(u,s)du\Big)ds.
$$
The definitions \er{kernels} imply $\int_\T G_1(x,s)ds=1$,
$\int_0^1 G_2(x,s)ds\le 1$, for all $x\in\R$, which proves \er{estKjIj}
for $j=1,2$. Moreover, the definition \er{kernels} gives
$$
\int_\R G_3(x,s)ds
=\ca \int_\R g(-s)ds,&x<0\\\int_\R g(x-s)ds,&0\le x\le 1\\
\int_\R g(1-s)ds,&x>1\ac.
$$
Then $\int_\R G_3(x,s)ds=1$ for all $x\in\R$,
which proves \er{estKjIj} for $j=3$.~\BBox

\medskip

Define closed balls $\mB_r( V_0)\ss\cX_j$
centered at $ V_0\in\cX_j$, by
\[
\lb{defmBcV0}
\mB_r( V_0)=\{ V\in \cX_j:\| V- V_0\|_{\cX_j}\le r\},\qq r>0.
\]
We use the following estimates.

\begin{lemma} Let $j\in\{1,2,3\}$, let $F\in\cQ_j$, and let
\[
\lb{defpYZ}
p(x, V)=\frac{Y(x, V)}{  Z( V)},\qq
\qq (x, V)\in I_j\ts\cX_j,
\]
where
\[
\lb{defYcZw}
Y(x, V)=e^{-\a V(x)-\b F(x)},
\qq   Z( V)=\int_{I_j}e^{-\a V(s)-\b F(s)}ds.
\]
Then
\[
\lb{estgp_j}
\sup_{x\in I_j}p(x, V)\le A_1,
\qq
\sup_{x\in I_j}\big|p(x, V_1)-p(x, V_2)\big|\le  A_2\| V_1- V_2\|_{\cX_j},
\]
for all $ V, V_1, V_2\in\mB_M$, with $M>0$, where
\[
\lb{defmA}
 A_1={B_2\/B_1}e^{-\b\inf_{x\in I_j}F(x)},\qq
 A_2={\a B_2\/ B_1}(  B_2+1)e^{-\b\inf_{x\in I_j}F(x)},
\]
\[
\lb{defmA1}
 B_1=\int_{I_j}e^{-\b F(s)}ds,\qq
 B_2=e^{2\a(\| V_0\|_{\cX_j}+M)}.
\]
\end{lemma}

\no {\bf Proof.}
Let  $ V\in\mB_M$.
Then the definition \er{defmBcV0} gives $\sup_{x\in I_j}| V(x)|
=\sup_{x\in[0,1]}| V(x)|\le \| V_0\|_{\cX_j}+M$.
Let $s\in I_j$.
The definitions \er{defpYZ} and \er{defYcZw} imply
\[
\lb{estZtau}
{ B_1\/ B_2^{1\/2}}
\le   Z( V)
\le
 B_1 B_2^{1\/2},\qq
Y(s, V)\le m B_2^{1\/2},\qq
m=e^{-\b\inf_{x\in I_j}F(x)},
\]
where $ B_1, B_2$ are given by \er{defmA1}.
Moreover, for $ V_1, V_2\in\mB_M$, we obtain
\[
\lb{Y-Y}
|Y(s, V_1)-Y(s, V_2)|\le
\Big|e^{-\b F(s)}\big(e^{-\a V_1(s)}-e^{-\a V_2(s)}\big)\Big|
\le \a mB_2^{1\/2}| V_1(s)- V_2(s)|,
\]
where we have used the estimate $|e^x-e^y|\le e^{\sup\{|x|,|y|\}}|x-y|$.
Then the estimate \er{estZtau} gives
\[
\lb{Z-Z}
\begin{aligned}
\Big|{1\/  Z( V_1)}-{1\/  Z( V_2)}\Big|
=\Big|{  Z( V_2)
-  Z( V_1)\/  Z( V_2)  Z( V_1)}\Big|
\le { B_2\/ B_1^2}\Big|\int_{I_j}e^{-\b F(s)}\big(e^{-\a V_1(s)}
-e^{-\a V_2(s)}\big)ds\Big|
\\
\le { B_2\/ B_1}\sup_{s\in I_j}\big|e^{-\a V_1(s)}
-e^{-\a V_2(s)}\big|
\le{\a B_2^{3\/2}\/ B_1}
\big\| V_1- V_2\big\|_{\cX_j}.
\end{aligned}
\]
The definition \er{defKrho} gives
$$
p(s, V)\le\frac{|Y(s, V)|}{|  Z( V)|},
$$
$$
\begin{aligned}
\big|p(s, V_1)-p(s, V_2)\big|
\le
Y(s, V_1)\Big|{1\/  Z( V_1)}-{1\/  Z( V_2)}\Big|
+{|Y(s, V_1)-Y(s, V_2)|\/  Z( V_2)},
\end{aligned}
$$
Then the estimates \er{estZtau}, \er{Y-Y}, and \er{Z-Z} yield
\er{estgp_j}.~\BBox

\subsection{Local solvability}
We first establish local existence and uniqueness for the initial-value
problem associated with Eq.~\er{ideq2}.

\begin{lemma}
\lb{Lmuthvpc}

 Let $j\in\{1,2,3\}$.
Let $F\in\cQ_j$, $ V_0\in\cX_j$,
let $M>0$, and let $\cT\in(0,\min\{{M\/2 A_1},{1\/2 A_2}\})$, where
 $ A_1, A_2$ are given by \er{defmA}.
Then Eq.~\er{ideq2} has a unique solution
$ V\in C^1([0,\cT],\cX_j)$
satisfying the conditions:

1) $ V(x,0)= V_0(x),x\in[0,1]$,

2) $\sup_{\theta\in[0,\cT]}\| V(\cdot,\theta)- V_0\|_{\cX_j}\le M$.

\end{lemma}

\no {\bf Proof.} Let $j\in\{1,2,3\}$.
Consider the space $\mM_j= C([0,\cT],\cX_j)$, equipped
with the norm
$$
\| V\|_{\mM_j}=\sup_{\theta\in[0,\cT]}\| V(\cdot,\theta)\|_{\cX_j}.
$$
Suppose that $ V\in\mM_j$ satisfies
the integral equation
\[
\lb{ieqvpc}
 V(x,\theta )=(\Phi_j  V)(x,\theta),\qq
(x,\theta)\in I_j\ts[0,\cT],
\]
where the operator $\Phi_j$, acting on $\mM_j$, is given by the identity
$$
(\Phi_j  V)(x,\theta)
= V_0(x)+\int_0^\theta\int_{I_j}K_j(x,s) p(s,t; V)dsdt.
$$
For any $ V_1,V_2\in\mM_j$
and for all $(x,\theta)\in I_j\ts[0,\cT]$ we have
$$
(\Phi_j  V_1)(x,\theta)-(\Phi_j  V_2)(x,\theta)
=\int_0^\theta\int_{I_j} K_j(x,s)
\big( p(s,t; V_1)- p(s,t; V_2)\big)dsdt.
$$
This identity and the estimate \er{estKjIj} give
\[
\lb{estwtVn-V0}
\|(\Phi_j  V)(\cdot,\theta )- V_0(\cdot,\theta )\|_{\cX_j}
=2\int_0^\theta\sup_{s\in I_j}\big| p(s,t; V)\big|dt,
\]
\[
\lb{estwtVn-Vn-1}
\|(\Phi_j  V_1)(\cdot,\theta )-(\Phi_j  V_2)(\cdot,\theta )\|_{\cX_j}
=2\int_0^\theta\sup_{s\in I_j}
\big| p(s,t; V_1)- p(s,t; V_2)\big|dt,
\]
for all $\theta\in[0,\cT]$.

Assume that $ V\in\cB_M$, where $\cB_M$ is the closed set in the space $\mM_j$,
given by
$$
\cB_M=\{ V\in \mM_j:\| V- V_0\|_{\mM_j}\le M\}.
$$
Substituting the estimates \er{estgp_j} into \er{estwtVn-V0},
we obtain
$$
\|\Phi_j V- V_0\|_{\mM}
\le 2\cT A_1.
$$
Thus, if $\cT<{M\/2 A_1}$, then $\Phi_j$
maps the set $\cB_M$ into itself. Moreover,  the identity
 \er{estwtVn-Vn-1} gives
\[
\lb{Phi-Phi}
\|(\Phi_j V_1)(\cdot,\theta )-(\Phi_j V_2)(\cdot,\theta )\|_{\cX_j}
\le 2 A_2\int_0^\theta\| V_1(\cdot,t )- V_2(\cdot,t )\|_{\cX_j} dt,
\]
which yields
$$
\|\Phi_j V_1-\Phi_j V_2\|_{\mM}
\le 2 A_2 \cT\| V_1- V_2\|_{\mM},
$$
If $\cT<\min\{{M\/2 A_1},{1\/2 A_2}\}$, then the mapping $\Phi_j$ is contractive
on the set $\cB_M$. Therefore, the integral equation
\er{ieqvpc} has the unique solution in $\cB_M$.
This solution is continuously differentiable with respect to $\theta$
and hence solves Eq.~\er{ideq2}
on $[0,\cT]$, satisfying the conditions~1)-2).~\BBox

\subsection{Global solvability}
We now extend the local solution of Eq.~\er{ideq2} globally in time.

\medskip
\no {\bf Proof of Theorem~\ref{ThGS}~i).}
Let $j\in\{1,2,3\}$ and let $ V(x,\theta)$
be the solution to Eq.~\er{ideq2},
satisfying the condition $ V(x,0)= V_0(x),x\in I_j$.
Suppose, for a contradiction, that $T_{max}<\iy$ is such that
the solution $ V(\cdot,\theta )$ exists on the interval
$\theta\in[0,T_{max}]$, and this solution cannot be extended
onto the larger interval $\theta\in[0,T_{max}+\d]$, for any $\d>0$.

Eq.~\er{ideq2} and the estimate \er{estKjIj} give
$$
\sup_{x\in I_j}\Big|{d V(x,\theta )\/d\theta }\Big|
\le \sup_{(x,s)\in I_j^2}|K_j(x,s)|\le 2,\qq
\forall\ \ \theta\ge 0.
$$
This yields
\[
\lb{estcVt1-cVt2}
\sup_{x\in I_j}\big| V(x,\theta_1)- V(x,\theta_2)\big|\le
2|\theta_1-\theta_2|,\qq \forall\ \ \theta_1,\theta_2\ge 0.
\]
Since the space $\cX_j$ is complete, the estimate \er{estcVt1-cVt2}
yields that the limit
$ V_*=\lim_{\theta\to T_{max}} V(\cdot,\theta)\in\cX_j$
exists. We can apply Theorem~\ref{ThGS}~i) to Eq.~\er{ideq2} with the initial
conditions $ V|_{\theta=T_{max}}= V_*$.
This implies that there exists the unique solution $ V(\cdot,\theta)$,
$\theta\in[0,T_{max}+\d]$, for some $\d>0$, satisfying the condition
$ V(\cdot,T_{max})= V_*$.
Therefore, the solution $ V(x,\theta)$ can be extended
uniquely from $[0,T_{max}]$  onto a larger interval.
This contradicts the maximality of $T_{max}$ and proves the assertion.~\BBox

\subsection{Replicator form of the equations}

The following lemma derives the replicator equation
for the density $ p$ defined in \er{defKrho}.

\begin{lemma}
\lb{Lmglclsol}
Let $j\in\{1,2,3\}$. Then

i) The function $ p$, given by \er{defKrho},
satisfies the equation \er{eqgpwti}.

ii) Let, in addition, $ p_{0}\in\mP_j$.
Then Eq.~\er{eqgpwti} has a unique solution
$ p\in C^1(\R_+,\mP_j)$,
satisfying the condition
$ p(x,0)= p_{0}(x)$, for all $x\in I_j$. Moreover,
$ p$ satisfies the equation
\[
\lb{gpgp0}
 p(x,\theta)=
e^{-\a\int_0^\theta(U_j(x,s)-\langle U_j\rangle_{ p,j}(s))ds} p_{0}(x),
\qq\theta>0,
\]
where the average $\langle\cdot\rangle_{ p,j}$ is defined by \er{djavg}.

\end{lemma}

\no {\bf Proof.}
i) Let $j\in\{1,2,3\}$.
Let $(x,\theta)\in I_j\ts\R_+$. The definition \er{defKrho} gives
\[
\lb{difgpwti}
{d p(x,\theta)\/d\theta}
=-\a p(x,\theta){d V(x,\theta )\/d\theta }
-{ p(x,\theta)\/  Z(\theta)}{d  Z(\theta)\/d\theta }
=-\a p(x,\theta)\Big({d V(x,\theta )\/d\theta }
-\Big\langle{d V\/d\theta }\Big\rangle_{ p,j}(\theta )\Big),
\]
where $ Z=\int_{I_j}e^{-\a V(s,\theta )-\b F(s)}ds$, cf.~\er{defYcZw}.
Eq.~\er{ideq2} and the definition \er{defKrho}  give
$$
\Big\langle{d V\/d\theta }\Big\rangle_{ p,j}(\theta )
=\int_{I_j}\Big( U_j(x,\theta)
-\int_0^1 U_j(s,\theta)ds\Big) p(x,\theta )dx
=\langle U_j\rangle_{ p,j}(\theta)-\int_0^1 U_j(s,\theta)ds.
$$
Then Eq.~\er{ideq2} implies
$$
{d V(x,\theta )\/d\theta }-\Big\langle{d V\/d\theta }\Big\rangle_{ p,j}(\theta )
=U_j(x,\theta)-\langle U_j\rangle_{ p,j}(\theta).
$$
Substitution into \er{difgpwti} yields Eq.~\er{eqgpwti}.

ii) Let $ p_{0,j}\in\mP_j$, for some $j=1,2,3$,
and let $ V_0$ be related to $ p_{0,j}$
by the identity \er{defKrho}. Then $ V_0\in\cX_j$.
Theorem~\ref{ThGS}~i) shows that Eq.~\er{ideq2} has the unique solution
$ V\in C^1(\R_+,\cX_j)$, satisfying the condition $ V(\cdot,0)= V_0$.
Then the function $ p$, given by \er{defKrho}, belongs to
$C^1(\R_+,\mP_j)$ and satisfies Eq.~\er{eqgpwti}
and the condition $ p(\cdot,0)= p_{0}$.
Moreover, the solution to
Eq.~\er{gpgp0} satisfies Eq.~\er{eqgpwti} and the condition
$ p(\cdot,0)= p_{0}$. ~\BBox

\medskip

\no {\bf Proof of Theorem~\ref{ThGS}~ii).}
Lemma~\ref{Lmglclsol}~ii) completes the proof.~\BBox

\subsection{Energy functional}

Let $j\in\{1,2,3\}$.
Define the interaction-energy functional $E_j$
on the set $\mP_j$, given by \er{spmP},
by the identity
\[
\lb{defcE}
E_j[ p ]={1\/2}\int_{I_j}\int_{I_j} G_j(x,y) p (x) p (y)dxdy,
\]
where the average $\langle f\rangle_{p,j}$ with respect to $p$
on the interval $I_j$
is defined by \er{djavg}.
We will show that $E_j$ is a Lyapunov functional
for Eq.~\er{eqgpwti}.
Introduce the functional $D_j$ on $\mP_j$ by the identity
\[
\lb{defcD}
D_j[p]=\int_{I_j}p(x)\big(U_j(x)-\langle U_j\rangle_p\big)^2dx
=\langle  U_j^2(x)\rangle_{p,j}-\langle U_j(x)\rangle_{p,j}^2,
\]
where $ U_j(x)=\int_{I_j} G_j(x,s)p(s)ds$, cf.~\er{djavg}.
The functional $D_j$ is the variance of $ U_j(x)$
with respect to the measure $p$.
By definition, $D_j[p]\ge 0$ for all $p\in\mP_j$.

For the densities $p,q\in\mP_j$,
we introduce the Kullback-Leibler divergence
\[
\lb{defKLqp}
D_{j,KL}(q\|p)
=\int_{I_j}q(x)\log\frac{q(x)}{p(x)}dx.
\]
Let $ p\in C^1(\R_+,\mP_j)$ be a solution of Eq.~\er{eqgpwti},
satisfying the condition
$ p(x,0)= p_{0}(x)$, for all $x\in I_j$.
Assume, in addition, that $D_{j,KL}(q\| p_{0})<\iy$.
Substituting \er{gpgp0} into \er{defKLqp}
and using the identity
$$
\int_{I_j}\big( U_j(x,s)-\langle  U_j\rangle_{ p,j}(s)\big)q(x)dx
=\int_{I_j} U_j(x,s)q(x)dx-\int_{I_j}  U_j(x,s) p(x,s)dx,
$$
we obtain
\[
\lb{estKLdivpr}
D_{j,KL}(q\| p(\cdot,\theta))
=D_{j,KL}(q\| p_{0})
+\a\int_0^\theta\Big(\int_{I_j} U_j(x,s)q(x)dx
-\int_{I_j}  U_j(x,s) p(x,s)dx\Big)ds,\ \
\theta\ge 0.
\]

\begin{lemma}
\lb{LmLF}
Let $j\in\{1,2\}$ and let
$ p\in C^1(\R_+,\mP_j)$ be the solution to Eq.~\er{eqgpwti}. Then

i) The energy functional $E_j$ satisfies:
\[
\lb{derLF}
{d E_j[ p(\cdot,\theta)]\/d\theta}
=-\a D_j[ p(\cdot,\theta)]\le 0,\qq\forall\ \ \theta>0.
\]
In particular, the function $E_j[ p(\cdot,\theta)]$
is non-increasing
and has non-negative limit as $\theta\to\iy$.

ii) The functional $D_j$ satisfies
\[
\lb{varcD}
\int_0^\iy D_j[ p(\cdot,\theta)]d\theta<\iy.
\]

iii) Let, in addition, $q\in\mP_j$ and $D_{j,KL}(q\| p_{0,j})<\iy$, where
$ p_{0}= p(\cdot,0)$. Then
\[
\lb{A-AE-E}
\int_{I_j} U_j(y,\theta) p(y,\theta)dy-\int_{I_j} U_j(y,\theta)q(y)dy\ge
E_j[ p(\cdot,\theta)]-E_j[q].
\]
\[
\lb{KLineq}
E_j[ p(\cdot,\theta)]\le E_j[q]+\frac{ D_{j,KL}(q\| p_{0})}{\a\theta },\qq
\theta\ge 0,
\]
where $ U$ is defined by \er{djavg}.
\end{lemma}

\no {\bf Proof.} i) Let $j\in\{1,2\}$. The definition \er{defcE}
and the symmetry $ G_j(x,y)= G_j(y,x)$ give
$$
\begin{aligned}
{E_j[ p(\cdot,\theta)]\/d\theta}
={1\/2}\int_{I_j}\int_{I_j} G_j(x,y)\Big({\pa p(x,\theta)\/\pa\theta} p(y,\theta)
+ p(x,\theta){\pa p(y,\theta)\/\pa\theta}\Big)dxdy
\\
=\int_{I_j}dx{\pa p(x,\theta)\/\pa\theta}\int_{I_j} G_j(x,y) p(y,\theta)dy
=\int_{I_j}{\pa p(x,\theta)\/\pa\theta} U_j(x,\theta)dx,
\end{aligned}
$$
where $ U_j$ are given by \er{djavg}.
Substituting \er{eqgpwti}, we obtain
\[
\lb{derLFpr}
\begin{aligned}
{E_j[ p(\cdot,\theta) ]\/d\theta}
=-\a\int_{I_j} p(x,\theta)\big( U_j(x,\theta)
-\langle  U_j \rangle_{ p,j}(\theta)\big) U_j(x,\theta)dx
\\
=-\a\int_{I_j} p(x,\theta)\big( U_j^2(x,\theta)
-2 U_j(x,\theta)\langle  U_j \rangle_{ p,j}(\theta)
+\langle  U_j \rangle_{ p,j}^2(\theta)\big)dx,
\end{aligned}
\]
where we used the identity
$$
\int_{I_j} p(x,\theta)
\big( U_j(x,\theta)-\langle  U_j \rangle_{ p,j}(\theta)\big)
\langle  U_j \rangle_{ p,j}(\theta)dx=0.
$$
The identity \er{derLFpr} gives \er{derLF}.

ii) Let $j\in\{1,2\}$. Integrating \er{derLF}, we obtain
$$
\int_0^\iy D_j[ p(\cdot,\theta)]d\theta
={1\/\a}\big(E_j[ p(\cdot,0)] -E_j[ p(\cdot,\iy)] \big),
$$
which yields \er{varcD}.

iii) Let $j\in\{1,2\}$. The definitions \er{defcE} and \er{djavg}
and the symmetry $ G_j(x,y)= G_j(y,x)$ imply
$$
\begin{aligned}
0\le{1\/2}\int_{I_j}\int_{I_j} G_j(x,y)
\big( p(x,\theta)-q(x)\big)\big( p(y,\theta)-q(y)\big)dxdy
=E_j[ p(\cdot,\theta)]+E_j[q]
\\
-\int_{I_j} U_j(y,\theta)q(y)dy
=\int_{I_j} U_j(y,\theta) p(y,\theta)dy-E_j[ p(\cdot,\theta)]
+E_j[q]-\int_{I_j} U_j(y,\theta)q(y)dy,
\end{aligned}
$$
where we used the identity
$\int_{I_j} U_j(y,\theta) p(y,\theta)dy=2E_j[ p(\cdot,\theta)]$.
This yields \er{A-AE-E}.
Substituting the estimate \er{A-AE-E} into \er{estKLdivpr}, we obtain
$$
 D_{j,KL}(q\| p_0)\ge
D_{j,KL}(q\| p(\cdot,\theta))+\a\int_0^\theta (E_j[ p(\cdot,s)]-E_j[q])ds
\ge\a\int_0^\theta (E_j[ p(\cdot,s)]-E_j[q])ds,\qq
\theta\ge 0,
$$
where we used $D_{j,KL}(q\| p(\cdot,\theta))\ge 0$.
Since $E_j[ p(\cdot,s)]$ is not increasing, we obtain
$$
 D_{j,KL}(q\| p_0)\ge
\a\theta E_j[ p(\cdot,\theta)]-\a\theta E_j[q],\qq
\theta\ge 0,
$$
which yields \er{KLineq}.~\BBox

\section{Periodic case}
\setcounter{equation}{0}

\subsection{Stationary solution}
We now consider the periodic replicator equation
\er{eqgpwti} with $j=1$.
The corresponding stationary equation has the form
\[
\lb{depcst}
\int_0^1\big( G_1(x,s)-1\big) p(s)ds=0,\qq x\in\T,
\]
where $ G_1$ is given by \er{kernels}, $ p$ and $ V$ are related by
\er{pcV}.

For $f\in C(\T)$, define the Fourier coefficients $\wh f_m,m\in\Z$:
\[
\lb{FC} f(x)=\sum_{m\in\Z}e^{i2\pi mx}\wh f_m,\qq \wh f_m
=\int_\T f(x)e^{-i2\pi mx}dx.
\]
In particular,
\[
\lb{FCG}
 G_1(x,s)=\sum_{m\in\Z}e^{i2\pi m(x-s)}\wh G_m,\qq \wh G_0=1,\qq
\wh G_m={1\/\s\sqrt{2\pi}}\int_\T e^{-{x^2\/2\s^2}}e^{-i2\pi mx}dx
=e^{-2\s^2\pi^2m^2}>0,
\]
for all $m\in\Z$.

\begin{lemma}
Eq.~\er{depcst} has the unique solution $ p=1$
in the class $\mP_1$, defined by \er{spmP}.
The corresponding potential $ V$, defined by \er{pcV},
satisfies: $ V\in C_0(\T)$ and
\[
\lb{solsteq}
 V(x)={\ol F-F(x)\/{T\/\D T}+1},\qq \ol F=\int_0^1F(x)dx.
\]
\end{lemma}

\no {\bf Proof.}
Rewrite Eq.~\er{depcst} in the form
\[
\lb{steq}
\int_0^1  p(s)ds=\int_0^1 G_1(x,s) p(s)ds,\qq x\in\T.
\]
We seek a solution in the class $ p\in \mP_1$.
Since $\int_0^1 G_1(x,s)ds=1$, the function $ p(x)=\const$
satisfies Eq.~\er{steq}. Conversely,
let $ p\in \mP_1$ be a solution to Eq.~\er{steq}.
In terms of the Fourier coefficients $\wh  p_m$,
Eq.~\er{steq} has the form
$$
\wh  p_0=\sum_{m\in\Z}e^{i2\pi mx}\wh G_m\wh  p_m.
$$
This gives $\wh  p_m=0,m\ne 0$, therefore, $ p(x)=\wh  p_0$.
Thus, $ p$ solves Eq.~\er{steq} if and only if
$ p(x)=\const=1$.
The identity \er{pcV} implies \er{solsteq}.~\BBox

\subsection{Kullback-Leibler divergence}
Let $ p\in\mP_1$. Define the Kullback--Leibler divergence
$H=D_{1,KL}(1\| p)$ by
\[
\lb{defKL}
H(\theta)=-\int_0^1 \log p(x,\theta) dx,\qq \theta\ge 0.
\]
By Jensen's inequality, $\int_0^1\log p dx\le\log\int_0^1 p dx=0$,
therefore, $H(\theta)\ge 0$.
The following lemma shows that $H$ is a Lyapunov function
for Eq.~\er{eqgpwti}, see also \cite{DPV14}.

\begin{lemma}
\lb{LmKL}
Let  $ p\in\mP_1$ satisfy Eq.~\er{eqgpwti}.
Then the function $H$, given by \er{defKL}, satisfies
\[
\lb{H'Fourier}
H'(\theta)=-\a\sum_{n\in\Z\sm\{0\}}\wh G_n|\wh p_n(\theta)|^2\le 0.
\]
\end{lemma}

\no {\bf Proof.}
Eq.~\er{eqgpwti} implies
\[
\lb{H'1}
H'(\theta)=-\int_0^1 {\pa p(x,\theta)\/\pa\theta}{dx\/ p(x,\theta)}
=\a\int_0^1
\big( U_1(x,\theta)
-\langle  U_1(\cdot,\theta)\rangle_{ p}\big)dx.
\]
where $ U_1$ is given by \er{djavg}.
Substituting the identities
$$
\int_0^1 U_1(x,\theta)dx
=\int_0^1\int_0^1 G_1(x,s) p(s,\theta)dsdx=1,
$$
and
$$
\int_0^1\langle U_1(\cdot,\theta)\rangle_{ p}dx
=\langle  U_1(\cdot,\theta)\rangle_{ p}
=\int_0^1\int_0^1 G_1(x,s) p(x,\theta) p(s,\theta)dsdx,
$$
into \er{H'1}, we obtain
$$
H'(\theta)=
\a\Big(1-\int_0^1\int_0^1 G_1(x,s) p(x,\theta) p(s,\theta)dsdx\Big).
$$
This yields \er{H'Fourier}.~\BBox

\medskip

By Lemma~\ref{LmKL}, the function $H$ is non-increasing
and non-negative.
However, these properties alone do not imply that $H'$ tends to zero as
$\theta\to +\iy$,
which would give us $\wh p_n\to 0$ for all $n\ne 0$.
We therefore need a more detailed analysis of $H$.

Define the functions
$$
u=\log p,\qq v=u-\int_0^1 udx.
$$
Eq.~\er{eqgpwti} implies
\[
\lb{eqforu}
\frac{\pa u}{\pa\theta}={1\/ p}\frac{\pa p}{\pa\theta}
=\a(\langle K p\rangle_{ p}-K p).
\]
Introduce the functional
\[
\lb{deffuncFi}
\F[v]=\log\int_0^1 e^vdx
\]
on the space of functions
$v\in C_0(\T)$. Due to Jensen's inequality
$\F[v]\ge\int_0^1 vdx=0$ and  $\F[v]=0$ iff $v=0$.

We have $u=v-H$ and
$1=\int_0^1 p dx=\int_0^1 e^udx=e^{-H}\int_0^1 e^vdx$,
then
$$
H=\log\int_0^1 e^vdx=\F[v].
$$
We begin with the following auxiliary result.

\begin{lemma}
Let $ p_0\in\mP_1$, let $ p\in C^1(\R_+,\mP_1)$ satisfy Eq.~\er{eqgpwti}
and the initial condition $ p(\cdot,0)= p_0$.
Let
\[
\lb{defv0}
v_0=\log p_0-\int_0^1 \log p_0 dx.
\]
Then $v_0\in C_0(\T)$ and

i) The function
\[
\lb{vissol}
v(x,\theta)=\log p(x,\theta)-\int_0^1 \log p(x,\theta) dx
\]
satisfies: $v\in C^1(\R_+,C_0(\T))$ and
\[
\lb{derv}
{\pa v(x,\theta)\/\pa\theta}=-\a K\nabla\F[v](x,\theta),\qq
v(x,0)=v_0(x),\qq (x,\theta)\in\T\ts\R_+,
\]
where
\[
\lb{defK}
K f(x)= G_1\star f (x)=\int_0^1 G_1(x,s) f(s)ds,\qq f\in C_0(\T).
\]

ii) Let, in addition, $w\in C_0(\T)$ satisfies:
$\|v_0-w\|_{K^{-1}}<\iy$, where
$$
\|z\|_{K^{-1}}^2=\sum_{n\ne 0}{|\wh z_n|^2\/\wh G_n},\qq\forall\ \
z\in C_0(\T).
$$
Then
\[
\lb{defDw}
D_w(\theta):={1\/2}\|v(\cdot,\theta)-w\|_{K^{-1}}^2<\iy,
\]
and
\[
\lb{estHtheta}
H(\theta)\le\F[w]+{D_w(0)\/\alpha\theta},
\]
for all $\theta>0$, where
$v$ and $H$ are given by \er{vissol} and \er{defKL}, respectively.

\end{lemma}

\no {\bf Proof.}
i) The identity \er{eqforu} gives
$$
{d\/d\theta}\int_0^1u dx=\a(\langle K p\rangle_{ p}-1),
$$
and then
\[
\lb{dervpr}
\frac{\pa v}{\pa\theta}=\frac{\pa u}{\pa\theta}-{d\/d\theta}\int_0^1udx
=\a(1-K p)=-\a K( p-1).
\]
Let $v,\d v\in C_0(\T)$ and let $\|\d v\|_{\cX_1}\to 0$.
Then the definition \er{deffuncFi} implies
\[
\lb{Fv+dv}
\begin{aligned}
\F[v+\d v]=\log\int_0^1 e^{v+\d v}dx
=\log\big(\int_0^1 e^{v}dx+\int_0^1 e^{v}\d vdx+o(\|\d v\|_{\cX_1})\big)
\\
=\F[v]+\int_0^1 \frac{e^{v}\d v}{\int_0^1 e^{v}dx}dx+o(\|\d v\|_{\cX_1})
=\F[v]+\int_0^1 ( p-1)\d vdx+o(\|\d v\|_{\cX_1}),
\end{aligned}
\]
where we used $\int_0^1\d vdx=0$ and
$$
\frac{e^{v}}{\int_0^1 e^{v}dx}=\frac{e^{u}}{\int_0^1 e^{u}dx}=e^u= p.
$$
The identity \er{Fv+dv} gives $\nabla\F[v]= p-1$.
Substituting this identity into \er{dervpr}, we obtain \er{derv}.

ii) Let $0\le\theta<\iy$.  Let $w\in C_0(\T)$ and $\|v_0-w\|_{K^{-1}}<\iy$.
Eq.~\er{dervpr} implies
\[
\lb{idv-w}
v(\cdot,\theta)-w=v_0-w-Kh,
\qq h=\a \int_0^\theta( p(\cdot,s)-1)ds,\qq h\in C_0(\T).
\]
Using $\|f\|_{K^{-1}}^2=(K^{-1}f,f)$,
where $(f,g)=\int_0^1fgdx$ is the inner product in $C_0(\T)$,
we obtain
$$
\|Kh\|_{K^{-1}}^2
=(h,Kh)\le \|K\|_{L^2(\T)\to L^2(\T)}\|h\|_{L^2(\T)}^2<\iy.
$$
Then the identity \er{idv-w} gives $\|v(\cdot,\theta)-w\|_{K^{-1}}<\iy$.

The definition \er{defDw} gives
$$
D_w(\theta)
={1\/2}\Big(K^{-1}\big(v(\cdot,\theta)-w\big),v(\cdot,\theta)-w\Big).
$$
Then
$$
D_w'(\theta)=
\Big(K^{-1}\fr{\pa v(\cdot,\theta)}{\pa\theta},v(\cdot,\theta)-w\Big).
$$
Substituting \er{derv} into this identity, we obtain
\[
\lb{Dw'1}
D_w'(\theta)=-\a\big(\nabla\F[v](\cdot,\theta),v(\cdot,\theta)-w\big).
\]
By convexity of $\F$,
$$
\F[w]-\F[v]\ge (\nabla\F[v],w-v).
$$
The identity \er{Dw'1} implies
$$
D_w'(\theta)\le\a\big(\F[w]-\F[v](\theta)\big)
=\a\big(\F[w]-H(\theta)\big).
$$
Integrating from $0$ to $\theta$, we obtain
$$
\int_0^\theta H(s)ds\le\theta\F[w]-\frac{D_w(\theta)-D_w(0)}{\a}
\le\theta\F[w]+\frac{D_w(0)}{\a}.
$$
Since $H$ is non-increasing,
$$
H(\theta)\le\frac{1}{\theta}\int_0^\theta H(s)ds\le
\F[w]+\frac{D_w(0)}{\a\theta}.
$$
This yields \er{estHtheta}.~\BBox

\subsection{Convergence to the stationary solution}

We first prove convergence in $L^1(\T)$.
The next lemma provides a rigorous convergence statement for
periodic Gaussian hills, related to the asymptotic result
of \cite{DPV14}.

\begin{lemma}
\lb{LmconvL1}
Let $ p_0\in\mP_1$ and
let $ p\in C^1(\R_+,\mP_1)$ satisfy Eq.~\er{eqgpwti}
and the initial condition $ p(\cdot,0)= p_0$. Let $\theta\to \iy$.
Then

i) The KL-divergence $H$, defined by \er{defKL}, satisfies
\[
\lb{Hto0}
H(\theta)=o(1).
\]
Moreover,
\[
\lb{gpto0}
\| p(\cdot,\theta)-1\|_{L^1(\T)}=o(1).
\]

ii) Let, in addition, $ p_0\in C^1(\T)$. Then
$ p\in C^1(\R_+,\mP_1\cap C^1(\T))$ and
\[
\lb{asHtheta}
H(\theta)=o\Big({1\/\log\theta}\Big),
\]
\[
\lb{gpto0q}
\| p(\cdot,\theta)-1\|_{L^1(\T)}=o\Big({1\/\sqrt{\log\theta}}\Big).
\]
Moreover, if $ p_0\in C^2(\T)$, then
\[
\lb{asHthetas}
H(\theta)=o\Big({1\/\log^2\theta}\Big),
\]
\[
\lb{gpto0qs}
\| p(\cdot,\theta)-1\|_{L^1(\T)}=o\Big({1\/\log\theta}\Big).
\]
\end{lemma}

\no {\bf Proof.} i) Let $ p\in C^1(\R_+,\mP_1)$
satisfy Eq.~\er{eqgpwti} and let  $ p(\cdot,0)= p_0$.
Then $v_0$, defined by \er{defv0}, satisfies $v_0=C_0(\T)$.
There exists a sequence $(p_N)_{N\in\N}$
of trigonometric polynomials $p_N\in C_0(\T)$ of degree $N$,
for example, Fej\'er's sums,
such that
the functions $w_N=v_0-p_N$ satisfy
\[
\lb{wNto0}
\|w_N\|_{\cX_1}\to 0,\qq\text{as}\ \ N\to+\iy.
\]
We have
$$
\|p_N\|_{K^{-1}}^2=\sum_{0<|n|\le N}\frac{|\wh p_{N,n}|^2}{\wh G_n}<\iy,
$$
where $\wh p_{N,n}$ are the Fourier coefficients of the function $p_N$
and $\wh G_n>0$ are given by \er{FCG}.
The estimate \er{estHtheta} and the definition \er{defDw} imply
$$
H(\theta)\le\F[w_N]+{\|p_N\|_{K^{-1}}^2\/2\alpha\theta},
$$
for all $N\in\N$ and for all $\theta>0$, where
$v$, $H$, and $\F$ are given by \er{vissol}, \er{defKL}, and
\er{deffuncFi}, respectively.
This estimate gives
\[
\lb{estlimH}
\lim_{\theta\to\iy}H(\theta)\le\F[w_N],
\]
for all fixed $N\in\N$. Moreover, the definition \er{deffuncFi}
and the relation \er{wNto0} imply
$$
\F[w_N]=\log\int_0^1 e^{w_N}dx\to 0,\qq\text{as}\ \ N\to+\iy.
$$
The estimate \er{estlimH} gives \er{Hto0}.

Pinsker's inequality \cite[Th~4.19]{BLM12} gives
\[
\lb{Pineq}
\| p(\theta)-1\|_{L^1(\T)}^2\le 2H(\theta).
\]
The relation  \er{Hto0} gives  \er{gpto0}.

ii) Let $(x,\theta)\in\T\ts\R_+$. Eq.~\er{eqforu} implies
$$
u(x,\theta)=u_0(x)
+\a\Big(\int_0^\theta\langle K p\rangle_{ p}(s)ds
-\int_0^\theta (K p)(x,s)ds\Big),
$$
where $u=\log p$, $K$ is defined by \er{defK}.
Then
$$
\frac{\pa u(x,\theta)}{\pa x}=u_0'(x)
-\a\int_0^\theta  U_1'(x,s)ds,
$$
which yields
$$
\Big\|\frac{\pa u(\cdot,\theta)}{\pa x}\Big\|_{\cX_1}=\|u_0'\|_{\cX_1}
+\a\theta\| G_1'\|_{\cX_1}.
$$
If $ p_0\in C^1(\T)$, then $\|\frac{\pa u(\cdot,\theta)}{\pa x}\|_{\cX_1}<\iy$
for all finite $\theta\ge 0$ and we have
$ p(\cdot,\theta)\in C^1(\T)$.

Let $ p_0\in C^1(\T)$ and let $v_0$ has the form \er{defv0}.
Then $v_0\in C_0(\T)\cap C^1(\T)$
and there exists the sequence $(p_N)_{N\in\N}$ of trigonometric polynomials
$p_N$ (Jackson's polynomials)
such that the sequence $w_N=v_0-p_N$ satisfies
$\d_N=\|w_N\|_{\cX_1}=o(N^{-1})$, as $N\to\iy$, see \cite[Ch~7]{DL93}.

Let $N\in\N$.
Repeating the previous arguments, we obtain
\[
\lb{estHth}
H(\theta)\le\F[w_N]+{\|p_N\|_{K^{-1}}^2\/2\alpha\theta},
\]
where the function $H$ and the functional $\F$ are defined by
\er{defKL} and \er{deffuncFi}, respectively.
The definition \er{deffuncFi} gives
\[
\lb{estFwn}
\F[w_N]=\log\int_0^1 e^{w_N}dx\le C\d_N^2,
\]
for some $C>0$.
Moreover, the estimate
$$
\|p_N\|_{L^2(\T)}^2=\sum_{0<|k|\le N}|\wh p_{N,k}|^2<\iy
$$
yields $|\wh p_{N,k}|<C$ for all $k\in\Z\sm\{0\}$ and for some $C>0$.
Then
\[
\lb{est|pN|}
\|p_N\|_{K^{-1}}^2=\sum_{0<|k|\le N}e^{ak^2}|\wh p_{N,k}|^2\le Ce^{aN^2},
\]
for some $C>0$. Substituting the estimates \er{estFwn} and \er{est|pN|}
into \er{estHth}, we obtain
\[
\lb{estH2}
H(\theta)\le C\Big(\d_N^2+{e^{aN^2}\/\theta}\Big),
\]
for some $C>0$.
Assuming $N=N(\theta)\sim \sqrt{{1-\ve\/a}\log\theta},0<\ve<1$, we obtain
$$
{e^{aN^2}\/\theta}=\theta^{-\ve},\qq
\d_N^2=o(N^{-2})=o\Big({1\/\log\theta}\Big).
$$
The estimate \er{estH2} yields \er{asHtheta}.
The estimate \er{Pineq} then gives \er{gpto0q}.

Let $ p_0\in C^2(\T)$.
Then $v_0\in C_0(\T)\cap C^2(\T)$
and there exists the sequence $(p_N)_{N\in\N}$ of trigonometric polynomials
$p_N$
such that
$\d_N=\|v_0-p_N\|_{\cX_1}=o(N^{-2})$, as $N\to\iy$, see \cite[Ch~7]{DL93}.
Assuming again $N=N(\theta)\sim \sqrt{{1-\ve\/a}\log\theta},0<\ve<1$, we obtain
$$
{e^{aN^2}\/\theta}=\theta^{-\ve},\qq
\d_N^2=o(N^{-4})=o\Big({1\/\log^2\theta}\Big).
$$
The estimate \er{estH2} yields \er{asHthetas}.
Then the estimate \er{Pineq} gives \er{gpto0qs}.~\BBox

\medskip

Introduce the function
\[
\lb{defR}
R=\a V+\b F.
\]

\begin{lemma}
Let  $F\in C^1(\T)$, $ V_0\in C_0^1(\T)$,
where $C_0^1(\T)=C_0(\T)\cap C^1(\T)$.
Let $ V\in C^1(\R_+,C_0(\T))$ be
 the solution to Eq.~\er{ideq2}, for $j=1$,
satisfying the condition
$ V(x,0)= V_0(x),x\in\T$.
Then $ V\in C^1(\R_+,C_0^1(\T))$,
the function $R$, defined by \er{defR}, satisfies:
$R\in C^1(\R_+,C_0^1(\T))$,
$R_0=R(\cdot,0)\in C_0^1(\T)$,
and
\[
\lb{derRtheta}
\frac{\pa R}{\pa\theta}=\a K( p-1),\qq\text{where}\ \
 p={e^{-R}\/\int_0^1e^{-R}dx},
\]
$ p\in C^1(\R_+,\mP_1(\T))$, $ p_0= p(\cdot,0)\in \mP_1\cap C^1(\T)$,
$K$ is defined by \er{defK}.
Moreover,
\[
\lb{asR}
\|R(\cdot,\theta)\|_{L^1(\T)}=o\Big(\frac{1}{\sqrt{\log\theta}}\Big),\qq
\theta\to\iy.
\]
If  $F\in C^2(\T)$, $ V_0\in C_0^2(\T)$,
where $C_0^2(\T)=C_0(\T)\cap C^2(\T)$, then
\[
\lb{asRs}
\|R(\cdot,\theta)\|_{L^1(\T)}=o\Big(\frac{1}{\log\theta}\Big),\qq
\theta\to\iy.
\]
\end{lemma}

\no {\bf Proof.}
Lemma~\ref{LmconvL1}~ii) yields
$ V\in C^1(\R_+,C_0^1(\T))$ and
$R\in C^1(\R_+,C_0^1(\T))$.
Eq.~\er{ideq2} implies \er{derRtheta}.
The identity \er{derRtheta} gives
$$
u=\log p=-R-\log\int_0^1e^{-R}dx.
$$
Integrating this identity and using \er{defKL} and $\int_0^1Rdx=0$, we obtain
$H=\log\int_0^1e^{-R}dx$, which yields
\[
\lb{uRH}
u=-R-H.
\]
Note that $u=u_+-u_-,|u|=u_++u_-$, where
$f_+=\sup\{f,0\}$, $f_-=\sup\{-f,0\}$. Then
$$
H=-\int_0^1udx=-\int_0^1u_+dx+\int_0^1u_-dx,
$$
which yields
$$
\|u\|_{L^1(\T)}=\int_0^1u_+dx+\int_0^1u_-dx=H+2\int_0^1u_+dx.
$$
Using the estimate $\log r\le r-1$, as $r\ge 1$,
we obtain $u_+\le( p-1)_+$, which yields
$$
\|u\|_{L^1(\T)}=H+2\int_0^1( p-1)_+dx.
$$
Moreover, the identity $\int_0^1( p-1)dx=0$ yields
$\| p-1\|_{L^1(\T)}=2\int_0^1( p-1)_+dx$, then
$$
\|u\|_{L^1(\T)}\le H+\| p-1\|_{L^1(\T)}.
$$
The identity \er{uRH} implies
$$
\|R\|_{L^1(\T)}\le\|u\|_{L^1(\T)}+H\le\| p-1\|_{L^1(\T)}+2H.
$$
The asymptotics \er{asHtheta} and \er{gpto0q} provide \er{asR}.

If  $F\in C^2(\T)$, $ V_0\in C_0^2(\T)$, then
the asymptotics \er{asHthetas} and \er{gpto0qs} give \er{asRs}.~\BBox

\medskip

The next lemma strengthens this convergence result.
In particular, we establish, for periodic Gaussian hills,
a convergence statement related to that in \cite{D16}.

\begin{lemma}
\lb{Thuthvpc}
Let  $F\in C^1(\T)$, $ V_0\in C_0^1(\T)$.
Let $ V\in C^1(\R_+,C_0(\T))$ be
 the solution to Eq.~\er{ideq2}, for $j=1$,
satisfying the condition
$ V(x,0)= V_0(x),x\in\T$.
Then
\[
\lb{Rto0}
\|R(\cdot,\theta)\|_{\cX_1}\to 0,
\]
\[
\lb{asgpunif}
\| p(\cdot,\theta)-1\|_{\cX_1}\to 0,
\]
as $\theta\to\iy$.
If $F\in C^2(\T)$, $ V_0\in C_0^2(\T)$, then
\[
\lb{Rto0s}
\|R(\cdot,\theta)\|_{C^1(\T)}\to 0,
\]
\[
\lb{asgpunifs}
\| p(\cdot,\theta)-1\|_{C^1(\T)}\to 0,
\]
as $\theta\to\iy$.
\end{lemma}

\no {\bf Proof.}
Let  $F\in C^1(\T)$, $ V_0\in C_0^1(\T)$. Then
$R_0=\a V_0+\b F\in C_0^1(\T)$. Then there exists
the sequence $(q_N)_{N\in\N}$, of trigonometric polynomials $q_N$
of degree $N$ such that
$$
\|R_0-q_N\|_{\cX_1}=O(N^{-1}),\qq N\to\iy,
$$
see \cite[Ch~7]{DL93}.

Let $\wh R_k$ be the Fourier coefficients of the function
$R$, see the definition \er{FC}.
Eq.~\er{derRtheta} yields
$$
\frac{\pa\wh R_k}{\pa\theta}=\a\wh G_k\wh p_k,\qq k\in\Z\sm\{0\},
$$
where $\wh G_k=e^{-ak^2}$, $a=2\s^2\pi^2>0$, see \er{FCG}.
Using $|\wh p_k(\theta)|=|\int_0^1e^{-i2\pi kx} p(x,\theta)dx|\le 1$,
we obtain
\[
\lb{WhRk-R0}
|\wh R_k(\theta)-\wh R_k(0)|\le \a\theta e^{-ak^2},\qq k\in\Z\sm\{0\}.
\]
This estimate gives
\[
\lb{sumwhRk0}
\Big\|\sum_{|k|>N}(\wh R_k(\theta)-\wh R_k(0))e^{i2\pi kx}\Big\|_{\cX_1}
\le\a\theta \sum_{|k|>N}e^{-ak^2}
\le C_1\theta\int_N^\iy e^{-ax^2}dx
\]
for some $C_1,C>0$.

Let $N=N(\theta)\sim\sqrt{\frac{1+\ve}{ a}\log\theta},\ve>0 $.
Then $\theta e^{-aN^2}\sim \theta^{-\ve}$
and we have
$$
R=R_0+(R-R_0)=P_N+(R-P_N),\qq N\to\iy,
$$
where $P_N$ is the trigonometric polynomials of degree $N$, given by
\[
\lb{PNtheta}
P_N(\theta)=q_N+\sum_{|k|\le N}(\wh R_k-\wh R_k(0))e^{i2\pi kx},
\]
and
\[
\lb{aseta}
\eta(\theta):=\|R(\cdot,\theta)-P_N(\cdot,\theta)\|_{\cX_1}
=o\Big({1\/N(\theta)}\Big).
\]
Nikol'skii's inequality, see \cite[Th~4.2.6]{DL93}, gives
\[
\lb{Nikineq}
\|P_N(\cdot,\theta)\|_{\cX_1}\le CN(\theta)\|P_N(\cdot,\theta)\|_{L^1(\T)},
\]
for some $C>0$. This yields
$$
\begin{aligned}
\|R(\cdot,\theta)\|_{\cX_1}\le\eta(\theta)+\|P_N(\cdot,\theta)\|_{\cX_1}
\le\eta(\theta)+CN(\theta)\|P_N(\cdot,\theta)\|_{L^1(\T)}
\\
\le\eta(\theta)+CN(\theta)\big(\|R(\cdot,\theta)\|_{L^1(\T)}
+\eta(\theta)\big).
\end{aligned}
$$
The asymptotics \er{asR} and \er{aseta} give \er{Rto0}.
The definition \er{derRtheta} implies \er{asgpunif}.

Let  $F\in C^2(\T)$, $ V_0\in C_0^2(\T)$. Then
$R_0\in C_0^2(\T)$. Then there exists
the sequence $(q_N)_{N\in\N}$, of trigonometric polynomials $q_N$
of degree $N$ such that
\[
\lb{estR0-qNs}
\|R_0-q_N\|_{\cX_1}=o(N^{-2}),\qq \|R_0'-q_N'\|_{\cX_1}=o(N^{-1}),\qq N\to\iy,
\]
see \cite[Ch~7]{DL93}.
The estimate \er{WhRk-R0} implies
\[
\lb{estsumwhRk-0s}
\Big\|\Big(\sum_{|k|>N}(\wh R_k(\theta)-\wh R_k(0))e^{i2\pi kx}\Big)'\Big\|_{\cX_1}
\le 2\pi\a\theta \sum_{|k|>N}|k|e^{-ak^2}
\le C\theta e^{-aN^2},
\]
for some $C>0$.
Let $N=N(\theta)\sim\sqrt{\frac{1+\ve}{ a}\log\theta},\ve>0 $.
Then $\theta e^{-aN^2}\sim \theta^{-\ve}$.
The estimates \er{sumwhRk0}, \er{estR0-qNs}, and \er{estsumwhRk-0s} give
\[
\lb{estetaR'}
\eta(\theta)=o(N^{-2}(\theta)),\qq
\|R'(\cdot,\theta)-P_N'(\cdot,\theta)\|_{\iy}=o(N^{-1}(\theta)),
\]
where $\eta$ is given by \er{aseta} and  $P_N$ is defined by \er{PNtheta}.

The estimate \er{Nikineq} and Bernstein's inequality
$$
\|P_N'(\cdot,\theta)\|_{\cX_1}\le 2\pi N(\theta)\|P_N(\cdot,\theta)\|_{\cX_1},
$$
see \cite[Ch~4.1]{DL93}, give
$$
\|P_N'(\cdot,\theta)\|_{\cX_1}\le C N^2(\theta)\|P_N(\cdot,\theta)\|_{L^1(\T)},
$$
for some $C>0$.
Using the estimate
$$
\|P_N(\cdot,\theta)\|_{L^1(\T)}\le \|R(\cdot,\theta)\|_{L^1(\T)}
+\|R(\cdot,\theta)-P_N(\cdot,\theta)\|_{L^1(\T)}
\le \|R(\cdot,\theta)\|_{L^1(\T)}+\eta(\theta),
$$
we obtain
$$
\|P_N'(\cdot,\theta)\|_{\cX_1}\le C N^2(\theta)
\big(\|R(\cdot,\theta)\|_{L^1(\T)}+\eta(\theta)\big),
$$
As $\theta\to\iy$,
The estimates \er{asRs}, \er{estetaR'}, and $N(\theta)\sim\sqrt{\log\theta}$
provide
$\|P_N'(\cdot,\theta)\|_{\cX_1}=o(1)$.
The second estimate in \er{estetaR'} gives
$\|R'(\cdot,\theta)\|_{\cX_1}=o(1)$. Using $R(\cdot,\theta)\in C_0(\T)$,
we obtain $\|R(\cdot,\theta)\|_{\cX_1}=o(1)$, which yields \er{Rto0s}.
The definition \er{derRtheta} implies \er{asgpunifs}.~\BBox

\medskip

{\bf Proof of Theorem~\ref{ThPC}.}
The asymptotics \er{gpto0} gives \er{Rto00}.
The asymptotics \er{Rto02} and \er{Rto0s2}
are determined in Lemma~\ref{Thuthvpc}.~\BBox

\section{Finite interval and the ``INTERVAL'' scheme}
\setcounter{equation}{0}

\subsection{Finite interval}
We consider the replicator equation \er{eqgpwti} for $j=2$,
on the interval $I_2=[0,1]$ with kernel $ G_2(x,s)=g(x-s)$,
see \er{I_j} and \er{kernels}.
Theorem~\ref{ThGS}~ii) ensures global solvability
of Eq.~\er{eqgpwti}.

We show that no stationary solution exists
in the class of strictly positive continuous densities.
Due to Eq.~\er{eqgpwti},
the corresponding stationary equation has the form
\[
\lb{steqfi}
g* p(x):=\int_0^1g(x-s) p(s)ds
=C,\qq x\in[0,1],\qq C=\int_0^1\int_0^1g(u-s) p(s)duds.
\]
We consider this equation in the class $ p\in\mP_2$, where $\mP_2$
is defined by \er{spmP}.

\begin{lemma}
\lb{Lmstsolfi}
Eq.~\er{steqfi} is not solvable in the class $ p\in\mP_2$.

\end{lemma}

\no {\bf Proof.}
Let $ p\in\mP_2$.
The function $g* p$ has an analytic continuation
from the interval $[0,1]$ onto the whole complex plane.
Then Eq.~\er{steqfi} takes the form
$g* p=C$ on $\C$.
The function $g$, and then the function  $g* p$,
belongs to the Schwartz class. Therefore, $(g* p)(x)\to 0$,
as $|x|\to\iy$.
Then $C=0$ and we have the equation $g* p=0$ on $\R$.
It follows that
$\wh g(\x)\wh  p(\x)=0$, where $\wh g$ and $\wh  p$ are Fourier transforms
of $g$ and $ p$, respectively, $\wh p$ is an entire function,
$\wh g(\x)>0$ for all $\x\in\R$. Thus, we have $\wh  p=0$, then $ p=0$,
$ p\not\in\mP_2$.~\BBox

\medskip
We study the quasi-stationary distribution $ p=1$
and prove Proposition~\ref{Prop1qs}.

\medskip

\no {\bf Proof of Proposition~\ref{Prop1qs}.} Eq.~\er{eqgpwti} implies
\[
\lb{dergpatp=1}
{d p(x,\theta)\/d\theta}\Big|_{\theta=0}
=-\a\big((g\star 1)(x)-\int_0^1(g\star 1)(x)dx\big).
\]
The identity
\[
\lb{gF-F}
(g\star 1)(x)=\F\Big({x\/\s}\Big)-\F\Big({1-x\/\s}\Big),
\]
gives
$$
\int_0^1(g\star 1)(x)dx=2\s\int_0^{1\/\s}\F(x)dx.
$$
Using the identity
$$
\int_0^\iy\F(x)dx={1\/\sqrt{2\pi}}\int_0^\iy x e^{-{x^2\/2}}dx
={1\/\sqrt{2\pi}},
$$
we obtain
\[
\lb{asmconj}
\int_0^1(g\star 1)(x)dx
=2\s\Big({1\/\sqrt{2\pi}}-\int_{1\/\s}^\iy\F(x)dx\Big).
\]
If $x\in[M\s,1-M\s]$ for some $M>0$ small enough,
then the identity \er{gF-F} gives
$$
(g\star 1)(x)\le2\F(M).
$$
Substituting this estimate and the identity \er{asmconj} into
\er{dergpatp=1}, we obtain \er{asdergpp=1centre}.

Let $x=\x\s\in[0,M\s]$. Then the identity \er{gF-F} implies
$$
(g\star 1)(x)=\F(\x)-\F\Big({1\/\s}-\x\Big).
$$
Substituting this estimate and the identity \er{asmconj} into
\er{dergpatp=1}, we obtain \er{asdergpp=1edge}.
The proof of \er{asdergpp=1edge+} is similar.~\BBox

\subsection{Quasi-stationary solution in the scheme ``INTERVAL''}
By analogy with Lemma~\ref{Lmstsolfi}, one can prove
the nonexistence of a stationary solution in the scheme ``INTERVAL''.
We next study the quasi-stationary bias $ V_L$, given by \er{defcVL},
and prove Theorem~\ref{ThQSSg}.

Let $j=3$, let $ V\in\cX_3$, and let $p$ be defined by \er{pcV}.
Then $p\in\mP_3$. Assume, in addition,
that $F$ satisfies \er{Fint}
for some $L>0$ large enough. Introduce the density $p_L\in\mP_3$
\[
\lb{defqstd}
p_L(x)={1\/Z_L}\ca
e^{\b(F(0)-F(x))}, x<-L,\\
\qq 1,\qq -L\le x\le L+1,\\
e^{\b(F(1)-F(x))}, x>L+1,
\ac
\]
where
\[
\lb{ZL}
Z_L=1+2L+\int_{-\iy}^{-L}e^{\b(F(0)-F(x))}dx
+\int_{L+1}^\iy e^{\b(F(1)-F(x))}dx.
\]
We show that $p_L$ is a quasi-stationary distribution.
Let $ V_L$ denote the biased potential \er{defcVL}
corresponding to the distribution $p_L$ via formula \er{pcV}.
In the set $\cX_3$, consider a ball \er{bincX} centered at $ V_L$.

\medskip

\no {\bf Proof of Theorem~\ref{ThQSSg} .} Eqs.~\er{ideq2} and \er{defKjxs}
and the definition \er{kernels} give
\[
\lb{V-VL}
{d( V(x,\theta )- V_L(x))\/d\theta }
={d V(x,\theta )\/d\theta }=(\wt G V)(x,\theta)
=(\wt G V)(x,\theta)-(\wt G V_L)(x)+(\wt G V_L)(x),
\]
as $(x,\theta)\in\R\ts\R_+$, where
$$
(\wt G V)(x,\theta)=
\int_\R K_3(x,s) p(s,\theta; V)ds
=\int_\R\Big( G_3(x,s)-\int_0^1 G_3(x,u)du\Big)
 p(s,\theta; V)ds,
$$
$ G_3$ is given by \er{kernels}. This definition implies
\[
\lb{defwtG}
(\wt G V_L)=a_L-\int_0^1a_L(x)dx,\qq
a_L(x)=\int_\R  G_3(x,s)p_L(s)ds,\qq x\in\R.
\]
The definitions \er{kernels} and \er{defqstd} give
$$
\begin{aligned}
a_L(x)={1\/Z_L}
\Big(\int_{-\iy}^{-L}g(-s)e^{\b(F(0)-F(s))}ds
+\int_{-L}^{0}g(-s)ds+\int_{0}^{1}g(x-s)ds
+\int_{1}^{L+1}g(1-s)ds
\\
+\int_{L+1}^\iy g(1-s)e^{\b(F(1)-F(s))}ds\Big),\qq x\in\R.
\end{aligned}
$$
where $Z_L$ is given by \er{ZL}.
Using the estimates
$$
\sup_{x\in[0,1]}\Big|\int_{0}^{1}g'(x-s)ds\Big|
=\sup_{x\in[0,1]}|g(x-1)-g(x)|
\le{1\/\sqrt{2\pi}\s}\sup_{x\in[0,1]}\Big(e^{-{(x-1)^2\/2\s^2}}
+e^{-{x^2\/2\s^2}}\Big)
 \le {\sqrt2\/\sqrt\pi\s},
$$
and $Z_L\ge 2L$,
we obtain
$$
\sup_{x\in[0,1]}|a_L'(x)|\le{1\/\sqrt{2\pi}\s L}.
$$
The identity \er{defwtG} gives
\[
\lb{cA=cA-cA}
\begin{aligned}
\sup_{x\in[0,1]}|(\wt G V_L)(x)|
=\sup_{x\in[0,1]}|a_L(x)-\int_0^1a_L(x)dx|\le
\sup_{x\in[0,1]}a_L(x)-\inf_{x\in[0,1]}a_L(x)
\\
\le\sup_{x\in[0,1]}|a_L'(x)|\le{1\/\sqrt{2\pi}\s L}.
\end{aligned}
\]
Due to the conditions of Theorem,
$$
\sup_{(x,\theta)\in[0,1]\ts[0,\cT]}| V(x,\theta)- V_L(x)|\le R.
$$
Then the estimates
 $\sup_{x\in\R}\int_\R |K_3(x,s)|ds\le 2$ and \er{estgp_j} give
\[
\lb{wtG-GLpr}
\begin{aligned}
\sup_{(x,\theta)\in[0,1]\ts[0,\cT]}|(\wt G V)(x,\theta)-(\wt G V_L)(x)|
=\sup_{(x,\theta)\in[0,1]\ts[0,\cT]}\Big|\int_\R K_3(x,s)
\big( p(s,\theta; V)- p(s,\theta; V_L)\big)ds\Big|
\\
\le 2A_2\sup_{(x,\theta)\in[0,1]\ts[0,\cT]}| V(x,\theta)- V_L(x,\theta)|,
\end{aligned}
\]
where $A_2$ is given by \er{defmA}. Moreover, the definitions \er{defmA},
\er{defmA1}, and \er{defcVL} yield
$$
2A_2\le{2\a B_2^2\/ B_1}e^{-\b\inf_{x\in\R}F(x)}
={2\a e^{4\a(\sup_{x\in\R}| V_L(x)|+R)
-\b\inf_{x\in\R}F(x)}\/\int_{\R}e^{-\b F(s)}ds}
={2\a e^{4\a R-5\b\inf_{x\in\R}F(x)}\/\int_{\R}e^{-\b F(s)}ds}=K_R,
$$
where $K_R$ is given by \er{defkRLc}.
The estimate \er{wtG-GLpr} gives
$$
\sup_{(x,\theta)\in[0,1]\ts[0,\cT]}|(\wt G V)(x,\theta)-(\wt G V_L)(x)|
\le K_R\sup_{(x,\theta)\in[0,1]\ts[0,\cT]}| V(x,\theta)- V_L(x,\theta)|.
$$
Then the identity \er{V-VL} implies
$$
D_\theta^+\sup_{x\in[0,1]}| V(x,\theta )- V_L(x)|
\le\sup_{x\in[0,1]}\Big|{d( V(x,\theta )- V_L(x))\/d\theta }\Big|
\le K_R\sup_{x\in[0,1]}| V(x,\theta)- V_L(x)|+\ve_L,
$$
for all $\theta\in[0,\cT]$
such that $ V(\cdot,\theta )$ remains in the ball $\cB_R$, here
$D^+f(x)=\lim\sup_{h\to+0}{f(x+h)-f(x)\/h}$ is the upper right Dini derivative.
Applying Gronwall's lemma, we obtain \er{estcV-cV0}.

Let $\theta_{R,L}$ denote the time at which the solution $ V$
reaches the boundary of the ball
$\cB_R$. The estimate \er{estcV-cV0} gives
$$
R\le{e^{K_R\theta_{R,L}}-1\/\sqrt{2\pi}\s LK_R},
$$
which yields the estimate
$$
\theta_{R,L}\ge {\log(\sqrt{2\pi}\s LK_R R+1)\/K_R}.
$$
This establishes \er{tstgtVinBR}.~\BBox

\section{Spaces of probability measures}
\setcounter{equation}{0}

\subsection{Compactness of probability-measure spaces}
Recall that {\it
the set $\cP(\cX)$ of probability measures on a metric space $\cX$
is called tight if, for every $\ve>0$, there exists a compact set $K_\ve\ss\cX$
such that $\m(K_\ve)\ge 1-\ve$ for all $\m\in\cP(\cX)$.
}

In the following lemma we analyze the sets $\cP_j,j=1,2,3$.
The proof of part iii) uses Prokhorov's theorem;
see \cite[Th~5.1]{Bi99}.

\no {\it Let $\cX$ be a complete metric space and let $\cP(\cX)$ be a space
of probability measures on $\cX$ equipped with the weak topology.
Let $\cK\ss\cP(\cX)$ be a family of probability measures on $\cX$. Then
$$
\cK\ \ \text{is sequentially compact}\qq\Leftrightarrow\qq
\cK\ \ \text{is tight}.
$$
}

\begin{lemma}
\lb{LmcPjcomp}
 Let $j\in\{1,2,3\}$.
Then

i) The set $\cP_j$ is convex and closed in the weak topology.

ii) The set $\cP_j$ is the closure of $\cP_j^{ac}$ in the weak topology
\[
\lb{MisclP}
\ol{\cP_j^{ac}}^w=\cP_j,
\]
where $\cP_j^{ac}\ss\cP_j$ is the space of absolutely continuous
measures whose densities belong to $\mP_j$,
 defined by \er{spmP}.

iii) The set $\cP_j$ is compact in the weak topology.
\end{lemma}

\no {\bf Proof.}
i) Let $j\in\{1,2,3\}$, let $\m,\n\in\cP_j$, and let $t\in[0,1]$.
Then the measure $\l=(1-t)\m+t\n\in\cP_j$ satisfies:
$\l(B)=(1-t)\m(B)+t\n(B)\ge 0$, for any Borel set $B\ss I_j$,
$\l(I_j)=(1-t)\m(I_j)+t\n(I_j)=1$, and $\l$ is countably additive.
Then $\l\in\cP_j$, hence $\cP_j$ is convex.

Let $j\in\{1,2\}$.
Let $\mu_n\in\cP_j$ for all $n\in\N$ and let $\mu_n\rightharpoonup\mu$,
as $n\to\iy$. Then
$\int_{I_j} f(x)\m(dx)=\lim_{n\to\iy}\int_{I_j} f(x)\m_n(dx)\ge 0$,
for any $f\in C({I_j}),f\ge 0$. Therefore, $\m$ is a non-negative measure.
Moreover, $\m({I_j})=\int_{I_j} \m(dx)=\lim_{n\to\iy}\int_{I_j} \m_n(dx)=1$.
Thus, $\m\in\cP_j$, which proves that $\cP_j$ is closed.

We will show that $\cP_3$ is closed in the weak topology \er{wcoonvt}.
Let $\mu_n\in M_3,n\in\N$, $\m_n\rightharpoonup\m$. The representation
\er{mcMF} implies
$$
\m_n=\eta_n+c_{-,n}\n_-+c_{+,n}\n_+,\qq
\n_-(dx)=w_-(x)\1_{(-\iy,0)}dx,\qq
\n_+(dx)=w_+(x)\1_{(1,+\iy)}dx,
$$
\[
\lb{normmes3}
\m_n(\R)=\eta_n([0,1])+A_-c_{-,n}+A_+c_{+,n}=1,
\]
where $0\le \eta_n([0,1])\le 1$,
$0\le c_{-,n}\le \frac{1}{A_-}$, $0\le c_{+,n}\le \frac{1}{A_+}$,
for all $n\in\N$. Then $c_{n,\pm}\to c_{\pm}$ for some $c_\pm\ge 0$
on some subsequence. Moreover, $\eta_n$ are concentrated on
the compact set $[0,1]$, then the family $\{\eta_n\}$ is sequentially compact.
Then $\eta_n\rightharpoonup\eta$ for some $\eta\in\cP_{\le 1}[0,1]$
on some subsequence. Thus,
$$
\int_\R f(x)\mu_n(dx)=\int_{[0,1]} f(x)\eta_n(dx)
+c_{-,n}\int_\R f(x)\n_-(dx)+c_{+,n}\int_\R f(x)\n_+(dx),
$$
for any $f\in C_b(\R)$, and we have in the limit
$$
\int_\R f(x)\mu(dx)=\int_{[0,1]} f(x)\eta(dx)
+c_-\int_\R f(x)\n_-(dx)+c_+\int_\R f(x)\n_+(dx).
$$
By continuity, $\eta_n([0,1])\to\eta([0,1])$, and the normalization condition
\er{normmes3} passes to \er{nmcMF}, which yields $\mu\in\cP_3$.
Thus, the set $\cP_3$ is closed in the weak topology.

ii) Let $\mu\in\cP_1$. Consider the measure
$d\m_\s(x)=\big(\int_\T G_{1,\s}(x,s)\m(ds)\big)dx$, $x\in\T$, where
$ G_{1,\s}$ is given by \er{kernels} and \er{gaussp}, $\s>0$
is small enough. Then $\m_\s\in\cP^{ac}_1$. Moreover,
$\int_\T G_{1,\s}(\cdot,s)f(s)ds\to f$, as $\s\to 0$,
uniformly on $\T$, for any $f\in C(\T)$. Therefore,
$$
\int_\T f(x)d\mu_\s(x)=
\int_\T f(x)\int_\T G_{1,\s}(x,s)\m(ds)dx\to
\int_\T f(s)\m(ds).
$$
Thus, $\mu_\s\rightharpoonup\mu$, that is, the set $\cP^{ac}_1$ is dense in $\cP_1$.
Since $\cP_1$ is closed, we get \er{MisclP} for $j=1$.

Consider $j=2$. Let $\m\in\cP_2$ and let $\ve\in (0,\frac{1}{4})$.
Introduce the linear mapping $T_\ve(x)=\ve+(1-2\ve)x$ on $[0,1]$.
Then $T_\ve([0,1])=[\ve,1-\ve]$. Let $\rho_\ve$ be a standard
Friedrichs mollifier
with compact support inside $(-\frac{\ve}{2},\frac{\ve}{2})$.
Introduce the measures
$$
d\mu_\ve(x)=\Big(\int_{[0,1]}\rho_\ve\big(x-T_\ve(s)\big)\mu(ds)\Big)dx.
$$
Then $\m_\ve([0,1])=1$ and, therefore, $\mu_\ve\in\cP^{ac}_2$.
Moreover, $T_\ve\to\text{id}$ uniformly on $[0,1]$, which yields
$\mu_\ve\rightharpoonup\mu$. Thus, the set $\cP^{ac}_2$ is dense in $\cP_2$.
Since $\cP_2$ is closed, we obtain \er{MisclP} for $j=2$.

Consider $j=3$. Let $\m\in\cP_3$ have the form \er{mcMF}. Denote
$m=\eta([0,1])$. If $m=0$, then $\m\in\cP^{ac}_3$. Assume $m\ne 0$.
Introduce the measure $\eta_m=\frac{\eta}{m}$. Then $\eta_m([0,1])=1$,
therefore, $\eta_m\in\cP_2$. Then, due to ii), there exists the sequence
$\eta_{m,\ve}\in\cP^{ac}_2$, $\ve>0$ is small enough, such that $\eta_{m,\ve}\rightharpoonup\eta_m$,
as $\ve\to 0$. Then the measures  $\m_\ve$, $\ve>0$ is small enough, given by
$$
\m_\ve(dx)=m\eta_{m,\ve}(dx)+c_-w_-(x)\1_{(-\iy,0)}dx+
c_+w_+(x)\1_{(1,+\iy)}dx,
$$
satisfy: $\m_\ve\in\cP^{ac}_3$ and $\m_\ve\rightharpoonup\mu$,
as $\ve\to 0$. This proves \er{MisclP} for $j=3$.

iii) The sets $\T$ and $[0,1]$ are compact. Then the sets
$\cP_1$ and $\cP_2$ are tight (the compact set $K_\ve$
from the definition of tightness is $\T$ or $[0,1]$ itself).
In both cases, compactness follows immediately
from Prokhorov's theorem.

Consider $\cP_3$ and prove that it is tight.
The identity \er{nmcMF} implies
$0\le c_-\le \frac{1}{A_-}$, $0\le c_+\le \frac{1}{A_+}$,
where $A_\pm$ are given by \er{Apm}.
Let $R>1$. The support of the measure $\eta$ lies in $[0,1]$, then
$$
\mu(\R\sm[-R,R])=c_-\int_{-\iy}^{-R}w_-(x)dx+c_+\int_R^\iy w_+(x)dx
\le \frac{1}{A_-}\int_{-\iy}^{-R}w_-(x)dx
+\frac{1}{A_+}\int_R^\iy w_+(x)dx,
$$
where $w_\pm$ are defined by \er{wpm}.
This yields
$$
\sup_{\m\in\cP_3}\mu(\R\sm[-R,R])\to 0,\qq \text{as}\ \ R\to\iy.
$$
Thus, $\cP_3$ is tight.
Due to Prokhorov's theorem, $\cP_3$ is sequentially compact.
Thus, $\cP_3$ is closed and sequentially compact,
and hence compact.~\BBox

\subsection{Gaussian potentials}
We use the following properties of the Gaussian potentials
$\cU_{\m,j},j=1,2,3$, given by \er{gp}.

\begin{lemma}
\lb{Lmgp1}
i) Each function $\cU_{\m,j},\m\in\cP_j,j=1,2,3$,
and $\cA_\mu,\mu\in\cP_3$ is continuous and positive on $I_j$ and satisfies
\[
\lb{estUmu}
\sup_{x\in I_j} \cU_{\mu,j}(x)\le {\alpha_j\/\s},\qq
\Lip_{I_j}\cU_{\mu,j}\le {\beta_j\/\s^2},
\]
where $\alpha_j$ and $\beta_j$ are given by \er{2estp} and
\er{estgaus'}, respectively.

Moreover, $\cU_{\m,j},j=1,2$, and $\cA_\mu$ have analytic extensions
onto the whole complex plane.

ii) Each mapping $\m\mapsto \cU_{\m,j},j=1,2,3$, is a continuous mapping
from $\cP_j$ to $C(I_j)$, that is,
the weak convergence $\m_n\rightharpoonup\m$ implies
$\sup_{x\in I_j}|\cU_{\m,j}(x)-\cU_{\m_n,j}(x)|\to 0$.

\end{lemma}

\no {\bf Proof.}
i) The functions $ G_j,j=1,2,3$, and $g$
are positive and continuous on $\R$,
then $\cU_{\m,j}$ are also positive and continuous. Moreover, the definition \er{gp} gives
\[
\lb{estUmupr}
\cU_{\m,j}(x)\le\sup_{x,s\in I_j} G_j(x,s),\qq
|\cU_{\m,j}'(x)|\le\sup_{x,s\in I_j}\Big|\frac{\pa G_j(x,s)}{\pa x}\Big|.
\]
The estimates \er{estUmupr} and  \er{estgaus} give \er{estUmu}.
The Gaussian $g$ is an entire function.
The integrals in $\cU_{\m,j},j=1,2$, are taken on a compact set,
then they have entire extensions. The integral in $\cA_{\m}$
is taken on $\R$. However, the Gaussian decays rapidly along the real axis,
which also yields an entire extension of $\cA_{\m}$.

ii) Let $j\in\{1,2,3\}$. Since $ G_j(x,\cdot)\in\mC_j$ for all $x\in I_j$,
the definition \er{gp} gives
$$
|\cU_{\m}(x)-\cU_{\m_n}(x)|
=\Big|\int_{I_j} G_j(x,s)\big(\m(ds)-\m_n(ds)\big)\Big|\to 0,\qq\forall\ \
x\in I_j,
$$
where $\cU_{\m}=\cU_{\m,j},\cU_{\m_n}=\cU_{\m_n,j}$, $\mC_j$ is the space
of test functions, given by \er{sptf}.
Thus, we have pointwise convergence  $\cU_{\m_n}(x)\to \cU_{\m}(x)$
for all $x\in I_j$.
Moreover,
$$
\sup_{x\in I_j}|\cU_{\m_n}(x)|\le\sup_{(x,s)\in\R^2}| G_j(x,s)|.
$$
Then the family $\{\cU_{\m_n}\}$ is uniformly bounded on $I_j$.
Furthermore, the estimate
$$
| G_j(x,s)- G_j(y,s)|\le \o(|x-y|),\qq \forall\ \ x,y,s\in[0,1],
$$
gives
$$
|\cU_{\m_n}(x)-\cU_{\m_n}(y)|=
\Big|\int_{I_j}\big( G_j(x,s)- G_j(y,s)\big)\m_n(ds)\Big|
\le\o(|x-y|),
$$
where $\o$ is the modulus of continuity of the function $ G_j$ on $I_j$.
Thus, the family $\{\cU_{\m_n}\}$ is equicontinuous.
Therefore, the sequence $\cU_{\m_n}$ converges to $\cU_\m$ uniformly on $[0,1]$.

In fact, let $\ve>0$. Let $\d>0$ be such that $|x-y|<\d$ implies
$|\cU_{\m_n}(x)-\cU_{\m_n}(y)|\le{\ve\/3}$ and
$|\cU_{\m}(x)-\cU_{\m}(y)|\le{\ve\/3}$.
Let $x_1,...,x_K$ be such that $|x-x_j|<\d$ for all $x\in[0,1]$
and for some $j=1,...,K$. Then $|\cU_{\m_n}(x_j)-\cU_{\m}(x_j)|\le{\ve\/3}$
for all $n>N$, where $N\in\N$ is large enough.
Let $x\in[0,1]$, let $x_j$ be the nearest to $x$ point, and let $n>N$.
Then
$$
|\cU_{\m_n}(x)-\cU_{\m}(x)|\le
|\cU_{\m_n}(x)-\cU_{\m_n}(x_j)|+|\cU_{\m_n}(x_j)-\cU_{\m}(x_j)|
+|\cU_{\m}(x_j)-\cU_{\m}(x)|\le\ve.
$$
This yields the statement for $j=1,2$.
The identity \er{Uconstout} shows that $\cU_{\m_n,3}$ converges to $\cU_{\m,3}$
uniformly on $[0,1]$; hence $\cU_{\m_n,3}$ converges to $\cU_{\m,3}$
uniformly on $\R$. This proves the case $j=3$.~\BBox

\medskip

Define the average of the Gaussian potential $\cU_{\m_j,j}$
with respect to the measure $\m_j\in\cP_j$ by
\[
\lb{defavU}
\langle \cU_{\m_j,j}\rangle_{\m_j,j}
=\int_{I_j}\cU_{\m,j}(x)\m_j(dx).
\]
We will need the following results.

\begin{lemma}
\lb{Lmaviscont}
 Let $j\in\{1,2,3\}$ and let $\m\in\cP_j$. Then
\[
\lb{estavUmu}
\big|\langle \cU_{\m,j}\rangle_{\m,j}\big|\le{\alpha_j\/\s}.
\]
Moreover,  the functional $\langle \cU_{\m,j}\rangle_{\m,j}$ is continuous
with respect to weak convergence,
that is, the weak convergence $\m_n\rightharpoonup\m$
implies $\langle \cU_{\m_n,j}\rangle_{\m_n,j}\to\langle \cU_{\m,j}\rangle_{\m,j}$,
as $n\to\iy$.

\end{lemma}

\no {\bf Proof.}
The estimates \er{estUmu} and the definition \er{defavU} imply \er{estavUmu}.
Let $\m_n\rightharpoonup\m$ in $\cP_j$. Due to
Lemma~\ref{Lmgp1}~i), $\cU_{\m_n,j}$ and $\cU_{\m,j}$ are continuous on $I_j$
and, due to Lemma~\ref{Lmgp1}~ii),
$\sup_{x\in I_j}|\cU_{\m,j}(x)-\cU_{\m_n,j}(x)|\to 0$, as $n\to\iy$.
The definition \er{defavU} gives
$$
|\langle \cU_{\m,j}\rangle_{\m,j}-\langle \cU_{\m_n,j}\rangle_{\m_n,j}|\le
\int_{I_j}|\cU_{\m,j}(x)-\cU_{\m_n,j}(x)|\m(dx)
+\Big|\int_{I_j}\cU_{\m_n,j}(x)\m(dx)-\int_{I_j}\cU_{\m_n,j}(x)\m_n(dx)\Big|,
$$
which yields
$\langle \cU_{\m_n,j}\rangle_{\m_n,j}\to\langle \cU_{\m,j}\rangle_{\m,j}$,
as $n\to\iy$.~\BBox

\section{Weak solutions}
\setcounter{equation}{0}

\subsection{Existence and uniqueness}
Oechssler and Riedel \cite{OR01}, \cite{OR02} establish existence
and uniqueness of strong solutions to replicator equations in spaces
of finite signed measures equipped with the total variation norm.
They also study stability and long-time behavior
using the weak topology on spaces of probability measures.
For our purposes, it will be more convenient
to work directly with the weak solutions.
This allows us to work directly with non-negative measures,
without introducing a space of signed measures.
The next lemma establishes existence and uniqueness
of weak solutions.
These results may be obtained from the results of \cite[Th~3]{OR01}.
We instead give a direct proof by approximation,
passing from absolutely continuous measures in $\cP^{ac}_j$
to general measures in $\cP_j$. The existence of the solution
in the space $\cP^{ac}_j$ was proved in Lemma~\ref{Lmglclsol}.

Let $j\in\{1,2,3\}$.
Define the bounded-Lipschitz metric $d_{BL}$ on $\cP_j$ by
\[
\lb{defdBL}
d_{BL}(\m,\n)=\sup_{\|f\|_{BL}\le 1}
\Big|\int_{I_j}f(x)\m(dx)-\int_{I_j}f(x)\n(dx)\Big|,
\]
where the bounded-Lipschitz norm $\|\cdot\|_{BL}$
for the functions, defined on $I_j$,
is given by
\[
\lb{defnBL}
\|f\|_{BL}=\sup_{x\in I_j}|f(x)|+\Lip_{I_j}f,
\]
see Dudley \cite[Ch~11]{Du02}.

For $\cT>0$, define $AC_w([0,\cT],\cP_j)$
to consist of paths taking values in $\cP_j$
that are weakly absolutely continuous on $[0,\cT]$;
that is $\m_\theta\in AC_w([0,\cT],\cP_j)$ means that $\m_\theta\in\cP_j$
for all $\theta\in[0,\cT]$ and the function
$\theta\mapsto\int_{I_j}\vp(x)\m_{\theta}(dx)$ is absolutely continuous
 on  $[0,\cT]$ for any $\vp\in\mC_j$:
$$
AC_w([0,\cT],\cP_j)=\{\m:[0,\cT]\to\cP_j:
\theta\mapsto\int_{I_j}\vp(x)\m_{\theta}(dx)\in AC([0,\cT])
\ \forall \ \vp\in\mC_j\}.
$$

\begin{lemma}
\lb{LmWS}

Let $j\in\{1,2,3\}$.

i) Let, in addition, $\m_0\in\cP_j$.
Then there exists a unique solution
$\m_\theta\in AC_w([0,+\iy),\cP_j)$,
to Eq.~\er{eqgpwtfj},
satisfying the initial condition $\m_\theta|_{\theta=0}=\m_0$.
Moreover,
this solution is continuous with respect to the initial conditions in
the BL-metrics,
that is, if $\n_\theta$ is the solution to Eq.~\er{eqgpwtfj}, satisfying
the initial condition $\n_\theta|_{\theta=0}=\n_0$, then
\[
\lb{estdifmn}
d_{BL}(\m_\theta,\n_\theta)\le e^{q\theta}d_{BL}(\m_0,\n_0),
\qq q=\a\Big(\frac{5\alpha_j}{\s}+\frac{3\beta_j}{\s^2}\Big),
\]
where $\a_j$ and $\b_j$ are given by \er{estgaus} and \er{estgaus'},
respectively.
Furthermore, $\m_\theta$ satisfies the identities \er{idmutheta}
and \er{suppmu=0}.

ii) Assume that the initial measure is absolutely continuous with respect to
the Lebesgue measure
$\mu_0(dx)= p_0(x)dx$,
where
$ p_0\in L^1(I_j)$, $ p_0\ge0$, $\int_{I_j} p_0(x)dx=1$.
Then for each $\theta\ge0$ the measure $\mu_\theta$,
satisfying Eq.~\er{eqgpwtfj}
and the initial condition $\mu_\theta|_{\theta=0}=\mu_0$,
is absolutely continuous
$\mu_\theta(dx)= p(x,\theta)dx$.
Moreover,
$ p\in C^1\bigl([0,\infty);L^1(I_j)\bigr)$,
$ p$ satisfies Eq.~\er{eqgpwti} in $L^1(I_j)$ for all $\theta\ge0$
and the initial condition
$ p(\cdot,0)= p_0 $ in $L^1(I_j)$.
Furthermore, the function $ p$ satisfies the estimates \er{2estp}.
\end{lemma}

\no {\bf Proof.}
i) Let $j\in\{1,2,3\}$.
Choose a sequence of densities $ p_0^{(n)}\in\mP_j$
such that $\m_0^{(n)}(dx)= p_0^{(n)}dx\rightharpoonup\m_0(dx)$
weakly in $\cP_j$.
We proved in Lemma~\ref{Lmglclsol}~ii) that for each density
$ p_0^{(n)}\in\mP_j$ for all $\theta>0$ there exists a unique
solution $ p_n(\cdot,\theta)\in\mP_j$
to Eq.~\er{eqgpwti}, satisfying the condition
$ p_n(x,0)= p_0^{(n)}(x),x\in I_j$.
Introduce the sequence of measures $\m_\theta^{(n)}\in\cP_j^{ac}$ by
\[
\lb{appmes}
\m_\theta^{(n)}(dx)= p_n(x,\theta)dx,\qq x\in I_j.
\]
The measures $\m_\theta^{(n)}$ satisfy Eq.~\er{eqgpwtfj}.

Let $\vp\in \mC_j$, $\cU_\mu=\cU_{\mu,j}$. Then \er{eqgpwtfj} gives
\[
\lb{deqfsol}
{df_n(\theta)\/d\theta}
=-\a\int_{I_j}\vp(x)
\big(\cU_{\m_\theta^{(n)}}(x)-\langle \cU_{\m_\theta^{(n)}}\rangle_{\m_\theta^{(n)}}\big)
\m_\theta^{(n)}(dx),\qq\text{where}\  \
f_n(\theta)=\int_{I_j}\vp(x)\m_\theta^{(n)}(dx),
\]
where the measures $\m_\theta^{(n)}$ are given by \er{appmes}.
This yields
\[
\lb{derfnbnd}
\Big|{df_n(\theta)\/d\theta}\Big|
\le\a\sup_{x\in I_j}|\vp(x)|\sup_{x\in I_j}
\big|\cU_{\m_\theta^{(n)}}(x)-\langle \cU_{\m_\theta^{(n)}}\rangle_{\m_\theta^{(n)}}\big|
\le C, \qq C={2\a \alpha_j\/\s}\sup_{x\in I_j}|\vp(x)|,
\]
wwhere we have used the estimates \er{estUmu} and \er{estavUmu}.
Consider  any finite interval $[0,\cT]$.
The mean value theorem gives
$$
|f_n(\theta_1)-f_n(\theta_2)|\le C|\theta_1-\theta_2|,\qq \theta_1,\theta_2\in[0,\cT].
$$
Therefore, the family $\{f_n\}$ is equicontinuous on the interval $[0,\cT]$.
Moreover,
$$
|f_n(\theta)|\le |f_n(0)|+C\theta,\qq \forall\ \ \theta\in[0,\cT].
$$
The set $f_n(0)=\int_{I_j}\vp(x)\m_0^{(n)}(dx)$ is bounded, then the family $\{f_n\}$ is uniformly bounded on the interval $[0,\cT]$.
By the Arzel\`a--Ascoli theorem, there exists a subsequence
$\m_\theta^{(n_k)}\in\cP_j,\theta\in[0,\cT]$, converging weakly and uniformly
in $\theta\in[0,\cT]$ to some family $\m_\theta\in\cP_j$, that is,
\[
\lb{limfnk}
\sup_{\theta\in[0,\cT]}|f_{n_k}(\theta)-f(\theta)|\to 0,\qq\text{where}\  \
f(\theta)=\int_{I_j}\vp(x)\m_\theta(dx).
\]
A diagonal argument gives this convergence on every interval $[0,\cT]$,
$\cT=1,2,3,4,...$ using a diagonal process.
Thus, there exists a sequence of absolutely continuous
measures $\m_\theta^{(n)}\in AC_w([0,+\iy),\cP_j^{ac})$, $n\in\N$, such that each
$\m_\theta^{(n)}$ satisfies Eq.~\er{eqgpwtfj} and
$\m_\theta^{(n)}\rightharpoonup\m_\theta$, as $n\to\iy$,
uniformly on any interval $[0,\cT],\cT>0$.
Moreover, due to \er{derfnbnd}, the limit function $f$ is
Lipschitz continuous with Lipschitz constant $C$, then it is
absolutely continuous.

Lemma~\ref{Lmgp1} gives
$\cU_{\m_n}\to \cU_\m$ uniformly in $x\in I_j$, then
$\langle \cU_{\m_n}\rangle_{\m_n}\to\langle \cU_{\m}\rangle_{\m}$.
Therefore, the sequence
\[
\lb{limUmn}
\sup_{\theta\in[0,\cT]}
\Big|\int_{I_j}\vp(x)\cU_{\m_\theta^{(n)}}(x)\m_\theta^{(n)}(dx)
-\int_{I_j}\vp(x)\cU_{\m_\theta}(x)\m_\theta(dx)\Big|\to 0.
\]

Let $\cT>0$ and let $\theta\in[0,\cT]$.
Eq.~\er{deqfsol} gives
$$
\int_{I_j}\vp(x)\m_\theta^{(n)}(dx)-\int_{I_j}\vp(x)\m_0^{(n)}(dx)
=-\a\int_0^\theta\int_{I_j}\vp(x)
\big(\cU_{\m_t^{(n)}}(x)-\langle \cU_{\m_t^{(n)}}\rangle_{\m_t^{(n)}}\big)
\m_t^{(n)}(dx)dt.
$$
Passing to the limit on both sides and using \er{limfnk} and \er{limUmn},
we obtain
\[
\lb{intfws}
\int_{I_j}\vp(x)\m_\theta(dx)-\int_{I_j}\vp(x)\m_0(dx)
=-\a\int_0^\theta\int_{I_j}\vp(x)
\big(\cU_{\m_t}(x)-\langle \cU_{\m_t}\rangle_{\m_t}\big)
\m_t(dx)dt.
\]
This identity shows that $\m_\theta$ satisfies \er{eqgpwtfj}, then it
is a global solution to Eq.~\er{eqgpwtfj},
satisfying the initial condition $\m_\theta|_{\theta=0}=\m_0$.

Let $\n_\theta$ be the solution to Eq.~\er{eqgpwtfj}, satisfying
the initial condition $\n_\theta|_{\theta=0}=\n_0$. Then Eq.~\er{intfws} yields
\[
\lb{eqwf1}
\int_{I_j}\vp(x)\big(\m_\theta(dx)-\n_\theta(dx)\big)
-\int_{I_j}\vp(x)\big(\m_0(dx)-\n_0(dx)\big)
=\int_0^\theta\big(R_\vp(\mu_t)-R_\vp(\nu_t)\big)dt,
\]
for all $\vp\in\mC_j$,
where
\[
\lb{defRmu}
R_\vp(\mu)
=-\a\int_{I_j}\vp(x)\big(\cU_{\m}(x)-\langle \cU_{\m}\rangle_{\m}\big)\m(dx).
\]
The estimates \er{estgaus} and \er{estgaus'} give
$$
\sup_{x\in I_j}\| G_j(x,\cdot)\|_{BL}
\le \frac{\alpha_j}{\s}+\frac{\beta_j}{\s^2}.
$$
Then the definitions \er{gp} and \er{defdBL} imply
\[
\lb{Um-UnBL}
\sup_{x\in I_j}|\cU_{\m}(x)-\cU_{\n}(x)|\le
\sup_{x\in I_j}\Big|\int_{I_j} G_j(x,s)\m(ds)-\int_{I_j} G_j(x,s)\n_j(ds)\Big|
\le \Big(\frac{\alpha_j}{\s}+\frac{\beta_j}{\s^2}\Big) d_{BL}(\m,\n).
\]
Moreover, the estimates \er{estUmu} yield
$$
\|\cU_{\m}\|_{BL}\le \frac{\alpha_j}{\s}+\frac{\beta_j}{\s^2}.
$$
The definition \er{defdBL} provides
\[
\lb{estUm-n}
\Big|\int_{I_j}\cU_{\n}(x)\big(\m(dx)-\n(dx)\big)\Big|\le
\Big(\frac{\alpha_j}{\s}+\frac{\beta_j}{\s^2}\Big) d_{BL}(\m,\n).
\]
The definition \er{defavU} and the estimates \er{estUmu} and \er{estUm-n}
imply
\[
\lb{avUm-UnBL}
\big|\langle \cU_{\m}\rangle_{\m}-\langle \cU_{\n}\rangle_{\n}\big|
=\big|\int_{I_j}\big(\cU_{\m}(x)-\cU_{\n}(x)\big)\m(dx)
+\int_{I_j}\cU_{\n}(x)\big(\m(dx)-\n(dx)\big)\big|
\le 2\Big(\frac{\alpha_j}{\s}+\frac{\beta_j}{\s^2}\Big) d_{BL}(\m,\n).
\]

Let $\|\vp\|_{BL}\le 1$.
The definition \er{defRmu} yields
\[
\lb{Rm-Rnpr}
\begin{aligned}
|R_\vp(\mu)-R_\vp(\nu)|=
\a\Big|\int_{I_j}\vp(x)\big(\cU_{\m}(x)-\cU_{\n}(x)
-\langle \cU_{\m}\rangle_{\m}+\langle \cU_{\n}\rangle_{\n}\big)\m(dx)
\\
+\int_{I_j}\vp(x)\big(\cU_{\n}(x)-\langle \cU_{\n}\rangle_{\n}\big)
\big(\m(dx)-\n(dx)\big)\Big|.
\end{aligned}
\]
The definition \er{defdBL} and the estimates \er{estUmu}, \er{estavUmu}
give
$$
\Big|\int_{I_j}\vp(x)\big(\cU_{\n}(x)-\langle \cU_{\n}\rangle_{\n}\big)
\big(\m(dx)-\n(dx)\big)\Big|\le{2\alpha_j\/\s}d_{BL}(\m,\n).
$$
The estimates \er{Um-UnBL} and \er{avUm-UnBL} imply
$$
\Big|\int_{I_j}\vp(x)\big(\cU_{\m}(x)-\cU_{\n}(x)
-\langle \cU_{\m}\rangle_{\m}+\langle \cU_{\n}\rangle_{\n}\big)\m(dx)\Big|\le
3\Big(\frac{\alpha_j}{\s}+\frac{\beta_j}{\s^2}\Big) d_{BL}(\m,\n).
$$
Substituting these estimates into \er{Rm-Rnpr}, we obtain
$$
|R_\vp(\mu)-R_\vp(\nu)|\le q d_{BL}(\m,\n).
$$

Eq.~ \er{eqwf1} gives
$$
\Big|\int_{I_j}\vp(x)\big(\m_\theta(dx)-\n_\theta(dx)\big)\Big|
\le\Big|\int_{I_j}\vp(x)\big(\m_0(dx)-\n_0(dx)\big)\Big|
+q\int_0^\theta d_{BL}(\m_t,\n_t)dt,
$$
for all $\vp:\|\vp\|_{BL}\le 1$. Taking the supremum over all
$\|\vp\|_{BL}\le 1$, we obtain
$$
d_{BL}(\m_\theta,\n_\theta)\le d_{BL}(\m_0,\n_0)
+q\int_0^\theta d_{BL}(\m_t,\n_t)dt.
$$
Gr\"onwall's inequality gives \er{estdifmn}. The uniqueness follows.

Let $\mu_\theta$ satisfy Eq.~\er{eqgpwtfj}
and the initial condition $\mu_\theta|_{\theta=0}=\mu_0$, where
$\mu_0\in\cP_j$.
Introduce the function
\[
\lb{deffR}
R(x,\theta)=-\a\bigl(\cU_{\mu_\theta}(x)-
\langle \cU_{\mu_\theta}\rangle_{\mu_\theta}\bigr),\qq
(x,\theta)\in I_j\ts\R_+.
\]
The path $\mu_\theta$ is weakly absolutely continuous on $\R_+$;
hence, by Lemma~\ref{Lmgp1}, the function $\cU_{\mu_\theta}(x)$
is continuous in $\theta\in\R_+$ and $x\in I_j$.
Then the function $R$ is continuous in $\theta$ and $x$, that is,
$R\in C(\R_+,C(I_j))$.
The estimates \er{estUmu} and \er{estavUmu} imply
\[
\lb{estbxgt}
 \sup_{x\in I_j}|R(x,\theta)|\le {2\a \alpha_j\/\s}.
\]
Consider the measure
\[
\lb{defRmes}
\nu_\theta (dx)= B(x,\theta )\mu_0(dx),\qq\text{where}\ \
B(x,\theta )=e^{\int_0^\theta  R(x,s)ds}.
\]
Direct differentiation shows that $\nu_\theta $ satisfies the equation
$$
{d\/d\theta}\int_{I_j}\vp(x)\n_{\theta}(dx)
=\int_{I_j}\vp(x)R(x,\theta)\n_{\theta}(dx),\qq\forall\ \ \vp\in \mC_j,\qq\theta>0,
$$
and the initial condition $\nu|_{\theta=0}=\mu_0$.
On the other hand, the original measure $\mu_\theta $
satisfies the same linear equation.
By part~i),
this solution is unique.
Therefore,
\[
\lb{idmuthetapr}
\mu_\theta =\nu_\theta =B(\cdot,\theta )\mu_0,\qq\theta>0.
\]
The definitions \er{defRmes} and \er{deffR} yield \er{idmutheta},
which implies \er{suppmu=0}.

ii) Let $\mu_0(dx)= p_0(x)dx$, $ p_0\in L^1(I_j)$, $ p_0\ge0$,
$\int_{I_j} p_0(x)dx=1$.
The identity \er{idmuthetapr} implies
\[
\lb{idpp0}
\mu_\theta(dx)= p(x,\theta )dx,\qq\text{where}\ \
 p(x,\theta )= B(x,\theta ) p_0(x),\qq (x,\theta)\in I_j\ts\R_+.
\]
Moreover, $B$ is continuously differentiable and bounded in $\theta$
on any compact time interval, so
$$
{\pa  p(\cdot ,\theta )\/\pa \theta}=
R(\cdot,\theta ) p(\cdot ,\theta )
$$
in $L^1(I_j)$, that is, $ p$ satisfies Eq.~\er{eqgpwti}.
Substituting the estimate \er{estbxgt} into the definitions \er{defRmes},
we obtain
$$
 e^{- 2\a {\alpha_j\/\s}\theta}\le\inf_{x\in I_j}|B(x,\theta )|
 \le\sup_{x\in I_j}|B(x,\theta )|\le e^{ 2\a {\alpha_j\/\s}\theta}.
$$
The identity \er{idpp0} gives the estimate \er{2estp}.~\BBox

\medskip

\no {\bf Proof of Theorem~\ref{ThwS}.} Lemma~\ref{LmWS}
gives the result.~\BBox

\subsection{Growth of the oscillation of the bias potential}
Define the functionals $J_j$, $j\in\{2,3\}$, on $\cP_j$ by
\[
\lb{defJj}
J_j[\m]=\osc_{[0,1]}\cU_{\m,j},\qq \m\in\cP_j.
\]
where $\osc$ is defined in \er{defosc}.

\begin{lemma} Let $j\in\{2,3\}$. Then the functional $J_j$ is weakly
continuous on $\cP_j$, that is
$\m_n \rightharpoonup\m$ in $\cP_j$ implies $J[\m_n]\to J[\m]$, as $n\to\iy$.
Moreover, the functional $J_j$ satisfies
\[
\lb{Jjpos}
\min_{\m\in\cP_j}J_j[\m]>0.
\]
\end{lemma}

\no {\bf Proof.} Let $j\in\{2,3\}$.
Lemma~\ref{Lmgp1}~ii) and the definition \er{defJj}
yield that the functional $J_j$ is weakly continuous on $\cP_j$.
Due to Lemma~\ref{LmcPjcomp}~iii), the set $\cP_j$
is compact in the weak topology, then $\min_{\m\in\cP_j}J_j[\m]$
is attained.

The identity \er{Uconstout} yields
$$
J_3[\m]=\osc_{[0,1]}\cA_{\m}=\max_{x\in [0,1]}\cA_{\m}(x)
-\min_{x\in [0,1]}\cA_{\m}(x),\qq \m\in\cP_3.
$$
Due to Lemma~\ref{Lmgp1}~i), the functions
$\cU_{\m,2}(\cdot)$ and $\cA_\m(\cdot)$ have an analytic extension onto the
whole complex plane. Moreover, the definitions
\er{gp} and \er{Uconstout} show $\cU_{\m,2}(x)\to 0$ and $\cA_\m(x)\to 0$,
as $x\to\pm\iy$, then $\cU_{\m,2}(\cdot)$ and $\cA_\m(\cdot)$ cannot be
constant. This yields \er{Jjpos}.~\BBox

\medskip

\no {\bf Proof of Theorem~\ref{Thosc}.}
Let $j\in\{2,3\}$.
Let $ V\in C^1(\R_+,\cX_j)$ satisfy Eq.~\er{ideq2},
where $ p\in C^1(\R_+,\mP_j)$ is given by \er{defKrho}.
The definitions \er{defKjxs} and \er{djavg} imply
\[
\lb{eqcVgp}
{d V(x,\theta )\/d\theta }
= U_j(x,\theta)
-\int_0^1 U_j(s,\theta)ds,\qq
(x,\theta)\in I_j\ts\R_+.
\]

For each $\theta>0$, we introduce the measures
$\m_\theta\in\cP_j^{ac}\ss\cP_j$, $\theta\ge 0$, by
$\m_\theta(dx)= p(x,\theta)dx$, and
the measure
$$
\n_\theta=\frac{1}{\theta}\int_0^\theta\m_t dt.
$$
Due to Lemma~\ref{LmcPjcomp}~i), the set $\cP_j$ is convex,
then $\n_\theta\in\cP_j$. Moreover, $\n_\theta\in\cP_j^{ac}$, since
each $\m_t\in\cP_j^{ac}$. Integrating Eq.~\er{eqcVgp}, we get
$$
 V(x,\theta)- V(x,0)
=\int_0^\theta\Big( U_j(x,t)
-\int_0^1 U_j(s,t)ds\Big)dt
=\theta\Big(\cU_{\n_\theta,j}(x)
-\int_0^1 \cU_{\n_\theta,j}(s)ds\Big),\qq
x\in I_j,
$$
where we used the definitions \er{gp} and \er{djavg}.
This yields
$$
\osc_{[0,1]}\big( V(x,\theta)- V(x,0)\big)=\theta\osc_{[0,1]}\cU_{\n_\theta,j}
=\theta J_j[\n_\theta].
$$
The estimate \er{Jjpos} implies
$\osc_{[0,1]}\big( V(x,\theta)- V(x,0)\big)\ge\theta \d$,
for some $\d>0$, which yields \er{oscVinc}.~\BBox

\subsection{Weakly stationary measures}
Let $j\in\{1,2,3\}$.
Due to Eq.~\er{eqgpwtfj}, a weak stationary measure $\m_j\in\cP_j$ satisfies
\[
\lb{steqfifj}
\int_{I_j}\vp(x)
\big(\cU_{\m_j,j}(x)-\langle \cU_{\m_j,j}\rangle_{\m_j,j}\big)\m_j(dx)=0,
\]
for all $\vp\in \mC_j$. This is equivalent to
\[
\lb{cstmcMF}
\cU_{\m_j,j}(x)=\langle \cU_{\m_j,j}\rangle_{\m_j,j}=\const,
\]
on $I_j$, $\m_j$-a.e.
Thus, a measure is weakly stationary if and only if its potential
is constant on its support.

\begin{lemma}
\lb{Lmfsmdi}
i) Each weakly stationary measure in $\cP_1$ is either uniform,
or purely atomic with no accumulation points.

ii) Each weakly stationary measure in $\cP_2$
is a finitely atomic measure.

iii) Let $\m\in\cP_3$ be a weakly stationary measure.
Then
\[
\lb{stm3}
\m=\sum_{j=1}^N a_j\d_{x_j}+c_-w_-(x)\1_{(-\iy,0)}+c_+w_+(x)\1_{(1,+\iy)},\qq
0\le x_1<x_2<...<x_N\le 1,
\]
for some $N\ge 0$, $a_j>0$, $c_\pm\ge 0$,  where $\d_{x_j}$ denotes the atomic (Dirac) measure
concentrated at the point $x_j$, $w_\pm$ are given by \er{wpm}.
Moreover,
$c_-\ne 0$ iff $\cA_\m(0)=\langle \cU_{\m,3}\rangle_{\m,3}$,
$c_+\ne 0$ iff $\cA_\m(1)=\langle \cU_{\m,3}\rangle_{\m,3}$,
where $\cA_\m$ is given by  \er{Uconstout}.

\end{lemma}

\no {\bf Proof.}
i)  Let $\m\in\cP_1$ be a weakly stationary measure.
Let $h=\cU_{\m,1}-\langle \cU_{\m,1}\rangle_{\m,1}$.
Due to Lemma~\ref{Lmgp1},
the function $h$ has an analytic continuation to an entire function.
The identity \er{cstmcMF} gives $h=0$ on $\supp\m$.
If $\supp\m$ is an infinite set, then it has a limit point on $\T$,
since $\T$ is compact. This yields  $h=0$, then $\cU_{\m,1}=\const$,
on the whole complex plane. Therefore, $\wh{(\cU_{\m,1})}_k=\wh G_k\wh\m_k=0$,
for all $k\ge 1$,
where $\wh{(\cU_{\m,1})}_k$, $\wh G_k$, and $\wh\m_k$ are Fourier coefficients
of the functions $\cU_{\m,1}$, $ G_1$, and the measure $\m_k$, respectively.
Using $\wh G_k>0$, we obtain $\wh\m_k=0$
for all $k\ge 1$. Therefore, the measure $\m$ is uniform on $\T$, which proves
the statement.

ii) Let $\m\in\cP_2$ be a weakly stationary measure.
Let $h=\cU_{\m,2}-\langle \cU_{\m,2}\rangle_{\m,2}$.
The function $h$ has an analytic continuation to an entire function.
The identity \er{cstmcMF} gives $h=0$ on $\supp\m$.
The set of zeros of $h$ on the interval $[0,1]$
is finite. Thus, $\supp\m$ is a finite set.

iii)
Let $\m\in\cP_3$ be a weakly stationary measure, then it satisfies \er{mcMF}.
Then,
according to \er{cstmcMF}, the identity $\cU_\m(x)=c$
holds $\m$-almost everywhere, where $c=\langle \cU_{\m,3}\rangle_{\m,3}$.
The identity  \er{Uconstout} implies that $\cU_{\m,3}=\cA_\m$ on the interval $[0,1]$.
Since $\cA_\m$ admits an analytic continuation to the entire complex plane,
the equation $\cA_\m=c$ has only a finite number
of solutions on the interval $[0,1]$.
It follows that $\eta$ in \er{mcMF}
is a finitely atomic measure.
Moreover, the identity \er{Uconstout} shows that
$\cU_{\m,3}=\cA_\m(0)=\const$ on $(-\iy,0]$. Therefore, the interval $(-\iy,0]$
belongs to the support of the stationary measure, that is, $c_-\ne 0$, iff
$\cA_\m(0)=\langle \cU_{\m,3}\rangle_{\m,3}$. Similarly,
 the interval $[1,+\iy)$
belongs to the support of the stationary measure, that is, $c_+\ne 0$, iff
$\cA_\m(1)=\langle \cU_{\m,3}\rangle_{\m,3}$.
This yields \er{stm3}.~\BBox

\section{Dynamics of the weak solution}
\setcounter{equation}{0}

\subsection{Energy functional}
For $\m\in\cP_j,j=1,2,3$, we define the energy functional
\[
\lb{defcEfj}
\cE_j[\m]={1\/2}\int_{I_j^2} G_j(x,y)\m(dx)\m(dy)
={1\/2}\int_{I_j}\cU_{\m,j}(x)\m(dx)={1\/2}\langle \cU_{\m,j}\rangle_\m,
\]
and the dissipation functional
\[
\lb{defcDfj}
\cD_j[\m]=\int_{I_j}\big(\cU_{\m,j}(x)-\langle \cU_{\m,j}\rangle_{\m}\big)^2\m(dx).
\]
In the following lemma we show that the functionals $\cE_j$ and $\cD_j$
extend the functionals $E_j$ and $D_j$ from the set $\cP_j^{ac}$
to the set $\cP_j$, where $E_j$ and $D_j$
are given by \er{defcE} and \er{defcD},
respectively.

\begin{lemma}
\lb{Lmgp2}
Let $j\in\{1,2,3\}$. Then

i) The functionals $\cE_j$ and $\cD_j$ are continuous
with respect to weak convergence,
that is, the weak convergence $\m_n\rightharpoonup\m$
implies $\cE_j[\m_n]\to\cE_j[\m]$
and $\cD_j[\m_n]\to\cD_j[\m]$, as $n\to\iy$.

ii) The ranges of the functionals $\cE_j$ and $\cD_j$
on $\cP_j$ are compact.

iii)  The functionals $\cE_j$ and $\cD_j$ satisfy

a) $\cE_j[\m]=E_j[p]$ and $\cD_j[\m]=D_j[p]$
for all $\m\in\cP_j^{ac}$, $\m(dx)=p(x)dx$, where $p\in\mP_j$,

b) if $\m_n\in\cP_j^{ac}$ for all $n\in\N$,
$\m_n(dx)=p_n(x)dx$, $p_n\in\mP_j$,
and $\m_n\rightharpoonup\m$ for some $\m\in\cP_j$, then $E_j[p_n]\to\cE_j[\m]$
and $D_j[p_n]\to\cD_j[\m]$.

\end{lemma}

\no {\bf Proof.}
i) Let $j\in\{1,2,3\}$ and let $\m_n\rightharpoonup\m$ in $\cP_j$, as $n\to\iy$.
Lemma~\ref{Lmgp1}~ii) gives $\sup_{x\in I_j}|\cU_{\m,j}(x)-\cU_{\m_n,j}(x)|\to 0$.
Lemma~\ref{Lmaviscont} yields
$\langle \cU_{\m_n,j}\rangle_{\m_n,j}\to\langle \cU_{\m,j}\rangle_{\m,j}$, then
the definition \er{defcEfj} provides
$\cE_j[\m_n]\to\cE_j[\m]$.
Moreover, $f_n(x)=\cU_{\m_n}(x)-\langle \cU_{\m_n}\rangle_{\m_n}$
converges to $f(x)=\cU_{\m}(x)-\langle \cU_{\m}\rangle_{\m}$,
uniformly on $I_j$.
Then the definition \er{defcDfj} implies
$$
\begin{aligned}
\big|\cD_j[\m]-\cD_j[\m_n]\big|
=\Big|\int_{I_j}f^2(x)\m(dx)-\int_{I_j}f_n^2(x)\m_n(dx)\Big|
\\
\le\int_{I_j}\big|f^2(x)-f_n^2(x)\big|\m(dx)
+\Big|\int_{I_j}f_n^2(x)\m(dx)-\int_{I_j}f_n^2(x)\m_n(dx)\Big|
\end{aligned}
$$
which yields $\cD_j[\m_n]\to\cD_j[\m]$, as $n\to\iy$.

ii) Due to Lemma~\ref{LmcPjcomp}~iii), the set $\cP_j$ is weakly compact, that is,
each sequence $(\m_n)_{n\in\N}$, $\m_n\in\cP_j$, contains a weakly
convergent subsequence.
Consequently, each sequence
$(\cE_j[\mu_n])_{n\in\N}$ and $(\cD_j[\mu_n])_{n\in\N}$ contains a convergent subsequence,
then the ranges of the functionals
$\cE_j[\mu]$ and $\cD_j[\mu],\mu\in\cP_j$, is compact.

iii) a) Let $\m\in\cP_j^{ac}$ and let $\m(dx)=p(x)dx,p\in\mP_j$.
 Then the definitions \er{defcE}, \er{defcD}, \er{defcEfj}, and \er{defcDfj}
show that  $\cE_j[\m]=E_j[p]$ and $\cD_j[\m]=D_j[p]$.

b) Due to Lemma~\ref{LmcPjcomp}~ii),
the set $\cP_j$ is the closure of $\cP_j^{ac}$ in the weak topology.
The functionals $E,D,\cE$, and $\cD$ are continuous with respect
to the weak topology. This yields the statement.~\BBox

\medskip

\no {\bf Remark.} Let $j=2$.
Weak stationary measures (see Lemma~\ref{Lmfsmdi}~ii)
with the same energy $\cE_2$
can form continuous families.
For example, the atomic measures $\d_a$, $a\in[0,1]$,
form a continuous family
of weak stationary measures with the energy $\cE[\d_a]={1\/2}g(0)$.
For any fixed $b\in(0,1]$,
the two-point measures ${1\/2}(\d_a+\d_{a+b})$
with $a\in[0,1-b]$ are weakly stationary and have the same energy
$\cE[{1\/2}(\d_a+\d_{a+b})]={1\/4}(g(0)+g(b))$.

\subsection{The minimizer}

Let $j\in\{1,2\}$. For a compact set $S\ss I_j$, we introduce  the set $\cP_j(S)$
of all measures  $\n\in\cP_j$ with the support in $S$:
\[
\lb{defcPjS}
\cP_j(S)=\{\n\in\cP_j:\supp\n\ss S\}.
\]
Note that $\cP_j(I_j)=\cP_j$.

We shall prove that, for $j\in\{1,2\}$, every trajectory converges
to a minimizer of $\cE_j$. Now
we prove that a minimizer exists, is unique, and
satisfies the Nash condition \er{chrelmm}, see, e.g., \cite[Sect~5.1]{OR01}.
These are standard results in potential theory;
see, e.g., \cite[Ch~2]{L66}. We include complete proofs for the kernels considered here.

\begin{lemma}
\lb{Lmminisunique}
Let $j\in\{1,2\}$ and let $S\ss I_j$ be compact. Then

i) The set
 $\cP_j(S)$ is convex and compact in the weak topology.
 Moreover,  $\cP_j(S)$ is weakly closed in $\cP_j$.

ii)  The functional $\cE_j$, defined by \er{defcEfj},
has a unique minimizer on the set $\cP_j(S)$.

iii) A measure $\m_*$ is a minimizer of the functional $\cE_j$
on the set $\cP_j(S)$ if and only if
\[
\lb{chrelmm}
\cU_{\m_*,j}(x)\ge 2\cE_j[\m_*]=\langle \cU_{\m_*,j}\rangle_{\m_*}\ \
\forall\  x\in S,\qq
\cU_{\m_*,j}=\langle \cU_{\m_*,j}\rangle_{\m_*}=\const\ \ \text{on}\  \supp\m_*.
\]
\end{lemma}

\no {\bf Proof.}
i) Let $j\in\{1,2\}$, let $\m,\n\in\cP_j(S)$, and let $t\in[0,1]$.
Then the measure $\l=(1-t)\m+t\n\in\cP_j(S)$ satisfies:
$\l(B)=(1-t)\m(B)+t\n(B)\ge 0$, for any Borel set $B\ss S$,
$\l(S)=(1-t)\m(S)+t\n(S)=1$, and $\l$ is countably additive.
Then $\l\in\cP_j(S)$, therefore, the set $\cP_j(S)$ is convex.
The set $S$ is compact. Then the set
$\cP_j(S)$ is tight (the compact set $K_\ve$
from the definition of tightness is $S$).
Due to Prokhorov's theorem, $\cP_j(S)$ is compact.
Its canonical embedding into $\cP_j$ is continuous.
Hence $\cP_j(S)$ is weakly closed in $\cP_j$.

ii) By Lemmas~\ref{Lmgp2}~i) and \ref{LmcPjcomp}~i), the functional
$\cE_j,j\in\{1,2\}$,
is continuous and the set $\cP_j$ is compact in the weak topology.
Consequently, a minimum exists.

We next prove uniqueness. The set $\cP_j(S)$ is convex. Assume that the functional
$\cE_j$ is strictly convex on $\cP_j(S)$, that is,
\[
\lb{cEisstrconv}
\cE_j[(1-t)\m+t\n]<(1-t)\cE_j[\m]+t\cE_j[\n],\qq \forall \ \ \m,\n\in\cP_j(S),\ \
0<t<1.
\]
If there are two distinct minimizers $\m_1$ and $\m_2$, then
$\cE_j[\m_1]=\cE_j[\m_2]=m$, where $m=\min_{\m\in\cP_j(S)}\cE_j[\m]$, and
the estimate \er{cEisstrconv} gives
$\cE_j[\l(t)]<m$, for all $t\in(0,1)$,
where $\l(t)=(1-t)\m_1+t\m_2\in\cP_j(S)$.
Therefore, the minimizer is unique.

It remains to establish \er{cEisstrconv}. The definition \er{defcEfj} gives
\[
\lb{cE1-t}
\cE_j[(1-t)\m+t\n]=(1-t^2)\cE_j[\m]+t^2\cE_j[\n]+t(1-t) Q_j(\m,\n),
\]
\[
\lb{cEcEQ}
 Q_j(\m,\n)=\int_{S^2} G_j(x,y)\m(dx)\n(dy).
\]
The identity \er{cE1-t} yields
\[
\lb{t1-tcE}
(1-t)\cE_j[\m]+t\cE_j[\n]-\cE_j[(1-t)\m+t\n]
=(1-t)t\big(\cE_j[\m]+\cE_j[\n]- Q_j(\m,\n)\big).
\]
The definitions \er{defcEfj} and \er{cEcEQ} and the symmetry
$ G_j(x,y)= G_j(y,x)$ imply
$$
\cE_j[\m]+\cE_j[\n]- Q_j(\m,\n)={1\/2} Q_j(\m-\n,\m-\n).
$$
Substituting this identity into \er{t1-tcE}, we obtain
\[
\lb{cEstrpospr}
(1-t)\cE_j[\m]+t\cE_j[\n]-\cE_j[(1-t)\m+t\n]
={(1-t)t\/2}\  Q_j(\m-\n,\m-\n).
\]

Let $j=1$. Then the Parseval identity gives
$$
 Q_j[\l,\l]=\sum_{n\in\Z}\wh G_n|\wh\l_n|^2>0,
$$
where $\l=\m-\n$, $\wh\l_n$ and $\wh G_n=e^{-2\s^2\pi^2n^2}$
are Fourier coefficients
of the signed measure $\l$ and the periodic Gaussian $ G_1$,
respectively.
The identity \er{cEstrpospr} gives the estimate \er{cEisstrconv} for $j=1$.

Let $j=2$.
Extend the signed measure $\l=\m-\n$ by zero beyond
the interval $[0,1]$ and define the Fourier transform
by
$$
\wh\l(\xi)=\int_{[0,1]}e^{ix\xi}\l(dx),\qq \xi\in\R.
$$
The Parseval identity gives
\[
\lb{cQispos}
 Q_j[\l,\l]={1\/2\pi}\int_{\R}\wh g(\xi)|\wh\l(\xi)|^2d\xi>0,
\]
where $\wh g$ is the Fourier transform of the function $g$,
\[
\lb{Ftrgauss}
\wh g(\xi)=2\sqrt\pi \s e^{-{1\/2}\s^2\xi^2}>0.
\]
Then the identity \er{cEstrpospr} gives the estimate
\er{cEisstrconv} for $j=2$.

iii) Let $j\in\{1,2\}$.
Let $\m_*$ be a minimizer of the functional $\cE_j$.
Consider the variations $\m(\ve)=(1-\ve)\m_*+\ve\n$,
$\ve>0$ is small enough, for an arbitrary probability measure
$\n\in\cP_j(S)$. By part~i), $\m(\ve)\in\cP_j(S)$. The definition \er{defcEfj} implies
$$
\begin{aligned}
\cE_j[\m(\ve)]={1\/2}\int_{S}\cU_{(1-\ve)\m_*+\ve\n,j}(x)
\big((1-\ve)\m_*(dx)+\ve\n(dx)\big)
\\
=\cE_j[\m_*]+\frac{\ve}{2}\int_{S}\cU_{\m_*,j}(x)\big(\n(dx)-\m_*(dx)\big)
+O(\ve^2)
\\
=\cE_j[\m_*]+\frac{\ve}{2}\Big(\int_{S}\cU_{\m_*,j}(x)\n(dx)-2\cE_j[\m_*]\Big)
+O(\ve^2),\qq
\ve\to 0.
\end{aligned}
$$
The condition $\cE_j[\m(\ve)]-\cE_j[\m_*]\ge 0$ gives
\[
\lb{qeneq1}
\int_{S}\cU_{\m_*,j}(x)\n(dx)\ge 2\cE_j[\m_*]
\]
for the arbitrary probability measure $\n\in\cP_j(S)$.
If $\cU_{\m_*,j}(x)<2\cE_j[\m_*]$ on some non-empty interval $(a,b)\in S$,
then, taking $\n=0$ on $S\sm(a,b)$ and $\n>0$ on $(a,b)$, we obtain
$$
\int_{S}\cU_{\m_*,j}(x)\n(dx)=\int_a^b \cU_{\m_*,j}(x)\n(dx)
< 2\cE_j[\m_*]\int_a^b \n(dx)=2\cE_j[\m_*]\int_{S} \n(dx)=2\cE_j[\m_*],
$$
which violates \er{qeneq1}. Therefore,
$\cU_{\m_*,j}(x)\ge 2\cE_j[\m_*]$ for all $x\in S$.
Moreover, the identity
$\int_{S}\cU_{\m_*,j}(x)\m_*(dx)=2\cE_j[\m_*]$ gives
$\cU_{\m_*,j}(x)= 2\cE_j[\m_*]$ for all $x\in\supp\m_*$.
The relations \er{chrelmm} follow.

Conversely, let $\m\in\cP_j(S)$ be such that
$$
\cU_{\m,j}(x)\ge c\ \ \forall\ x\in S,\qq
\cU_{\m,j}(x)=c\ \ \forall\ x\in\supp\m,
$$
for some $c>0$. Then $\cE_j[\m]=\frac{1}{2}\int_{S}\cU_{\m,j}(x)\m(dx)=c$.
Let $\n\in\cP_j(S)$ be another probability measure.
Then
$$
\begin{aligned}
\cE_j[\n]=\frac{1}{2}\int_{S}\cU_{\m+(\n-\m),j}(x)\big(\m(dx)+(\n(dx)-\m(dx))\big)
=\cE_j[\m]+\int_{S}\cU_{\m,j}(x)\big(\n(dx)-\m(dx)\big)
\\
+\frac{1}{2} Q_j(\n-\m,\n-\m),
\end{aligned}
$$
where $ Q_j$ is given by \er{cEcEQ}.
This gives
$$
\cE_j[\n]-\cE_j[\m]=\int_{S}\cU_{\m,j}(x)\n(dx)-2\cE_j[\m]
+\frac{1}{2} Q_j(\n-\m,\n-\m).
$$
Using \er{cQispos} and the estimate
$$
\int_{S}\cU_{\m,j}(x)\n(dx)\ge c\int_{S}\n(dx)= c=2\cE_j[\m],
$$
we obtain
$\cE_j[\n]\ge\cE_j[\m]$. Therefore, $\m$ is the minimizer.~\BBox

\subsection{Properties of the minimizer}

The following lemma identifies the minimizer: it is uniform or
finitely atomic in the periodic case, and finitely atomic
on a finite interval.

\begin{lemma}
\lb{Lmstsol}
Let $j\in\{1,2\}$, let $S\ss I_j$ be a compact set, and let $\m_*$ be the minimizer of the functional
$\cE_j$, given by \er{defcEfj}, on $\cP_j(S)$.

i) Let $j=1$. Then

a) if $S=\T$, then $\m_*$ is the uniform measure,

b) if $S\ss\T$ and $S\ne\T$, then $\m_*$ is a finitely atomic measure.

ii) Let $j=2$ and let $S=[0,1]$.
Then $\m_*$ is a finitely atomic measure.  Moreover,

a) the points $0$ and $1$ belong to the support of the measure  $\m_*$,

b) the measure $\m_*$ is symmetric about the point ${1\/2}$,

c) if $\s>1$, then $\m_*={1\/2}(\d_0+\d_1)$,
where $\d_a$ denotes the atomic (Dirac) measure
concentrated at the point $a\in\R$.

\end{lemma}

\no {\bf Proof.}
i) Let $j=1$. The definition \er{defcEfj} and the Parseval identity give
$$
\cE_1[\m]={1\/2}\int_{I_j^2} G_1(x,y)\m(dx)\m(dy)
={1\/2}\sum_{n\in\Z}\wh G_n|\wh\m_n|^2,
$$
where $\wh\m_n$ and $\wh G_n=e^{-2\s^2\pi^2n^2}$
are Fourier coefficients
of the measure $\m$ and the periodic Gaussian $ G_1$,
respectively.
This yields that the uniform measure is the unique minimizer
of $\cE_1$ on $\cP_1=\cP_1(\T)$.

Let $S\ss\T$ and $S\ne\T$ and let $\m_*$ be the minimizer.
The function $\cU_{\m_*,1}$, given by \er{gp},
has an analytic continuation onto the whole complex plane.
The Nash conditions \er{chrelmm} imply
$\cU_{\m_*,1}(z)= \cE_1[\m_*]=\const$ on the whole $\supp\m_*$.
If $\supp\m_*$ has accumulation points, then $\cU_{\m_*}=\const$
on the whole $\C$. The identity
$$
\cU_{\m_*,1}(x)=\sum_{m\in\Z}e^{i2\pi mx}\wh G_m\wh\m_n=\const,\qq x\in\T,
$$
gives $\m_*=\const$. Then $\m_*$ is the uniform measure. This contradicts
the condition  $\supp\m_*=S\ne\T$. Therefore, $\supp\m_*$
has no accumulation points on $\T$, which proves the statement.

ii) Let $j=2$ and let $S=[0,1]$. Let $\m_*\in\cP_2$ be a minimizer.
The function $\cU_{\m_*,2}$, given by  \er{gp},
has an analytic continuation onto the whole
complex plane. Moreover,  $\cU_{\m_*,2}(x)\to 0$ at $x\to\pm\iy$.
The identities \er{chrelmm} show that
$\cU_{\m_*,2}(z)= \cE_2[\m_*]=\const$ on the whole $\supp\m_*$. If
$\supp\m_*$ has accumulation points, then $\cU_{\m_*,2}=\const=0$ on the whole $\C$.
Then $\m_*=0$, contradicting the condition
$\m_*\in\cP_2$. Thus, the measure $\m_*$ cannot have accumulation points,
which gives
$$
\m_*=\sum_{n=1}^Na_n\d_{x_n},\qq a_n>0,\qq 0\le x_1<x_2<...<x_N\le 1,
\qq n=1,...,N,\qq N\in\N.
$$
The definitions \er{gp} and \er{kernels}
give $\cU_{\m_*,2}(x)=\sum_{n=1}^Na_ng(x-x_n)$, then
\[
\lb{Am*'}
\cU_{\m_*,2}'(x)=\sum_{n=1}^Na_ng'(x-x_n)
=-{1\/\s^2}\sum_{n=1}^Na_n(x-x_n)g(x-x_n),
\]
for all $x\in\R$, here we used $g'(x)=-{x\/\s^2}g(x)$, see \er{gaussp}.

a) Let us show that $x_1=0$. Assume $x_1>0$. Then \er{Am*'} gives
$\cU_{\m_*,2}'(x)>0$ for all $x\in[0,x_1)$.
This yields $\cU_{\m_*,2}(0)<\cU_{\m_*,2}(x_1)$, which contradicts \er{chrelmm}.
Therefore, $x_1=0$.

Similarly, assume $x_N<1$. Then \er{Am*'} implies
$\cU_{\m_*}'(x)<0$ for all $x\in(x_N,1]$.
This yields $\cU_{\m_*}(x_N)>\cU_{\m_*}(1)$, contradicting \er{chrelmm}.
Therefore, $x_N=1$.

b) Let $\m_*$ be the minimizer of $\cE_2$, and
let $\m_*^-$ be the measure symmetric to $\m_*$ with respect
to the point ${1\/2}$. Then
$$
\cE_2[\m_*]={1\/2}\int_{[0,1]^2}g(x-y)\m_*(dx)\m_*(dy)
={1\/2}\int_{[0,1]^2}g\big((1-x)-(1-y)\big)\m_*^-(dx)\m_*^-(dy)=\cE[\m_*^-].
$$
Consequently, $\m_*^-$ also minimizes $\cE_2$.
By the uniqueness of the minimizer,
we have $\m_*^- = \m_*$.

c) The definition \er{gaussp} gives
$g'(x)=-{x\/\s^2}g(x)$,
$$
g''(x)=-{xg'(x)+g(x)\/\s^2}={g(x)\/\s^2}\Big({x^2\/\s^2}-1\Big).
$$
If $\s\ge 1$, then $g''(x)\le 0$ for all $x\in[0,1]$. This yields
$g(x)\ge (1-x)g(0)+xg(1)$ for all $x\in[0,1]$.
Let $0\le y\le x\le 1$.
Then
$y-x\ge 2xy-x-y$ and, due to $g(0)>g(1)$, we have
$$
g(x-y)\ge (1-x+y)g(0)+(x-y)g(1)= g(0)+(y-x)\big(g(0)-g(1)\big)
\ge g(0)+(2xy-x-y)\big(g(0)-g(1)\big).
$$
By symmetry, this holds for all $x,y\in[0,1]$.
Integrating with respect to $\mu(dx)\mu(dy)$, we obtain
$$
\cE[\m]\ge
g(0)+\big(g(0)-g(1)\big)\int_{[0,1]^2}(2xy-x-y)\m(dx)\m(dy)
=g(0)-2\big(g(0)-g(1)\big)(m-m^2),
$$
where $m=\int_{[0,1]}x\m(dx)$. It follows from $m\in[0,1]$ that
$2(m-m^2)\le{1\/2}$,
then $\cE[\m]\ge {g(0)+g(1)\/2}$.
The equality is attained at the measure
$\frac{\delta_0+\delta_1}{2}$, which proves the required assertion.~\BBox

\medskip

\no {\bf Remark.}  The relations \er{chrelmm} suggest that as $\s \to 0$
(narrow Gaussians), the number $N$ of points of the support of the measure
increases, while the weights $a_n$ of the atoms decrease.
Both the spacing of the support points and their weights
are of order $\s$. The weights are approximately
uniform across the interval, except for boundary layers
of width $\sim \s$ near the edges,
where the weights increase rapidly as the edges are approached.

\subsection{Kullback-Leibler divergence}

Let $j\in\{1,2\}$ and let the measure $\n\in\cP_j$
be absolutely continuous with respect to the measure $\m\in\cP_j$.
Define the Kullback--Leibler divergence $\cD_{j,KL}(\n\|\m)$ by the identity
\[
\lb{cDKL}
\cD_{j,KL}(\n\|\m)=\int_{I_j}\log\Big(\frac{d\n}{d\m}\Big)(x)\n(dx),
\]
where $\frac{d\n}{d\m}$ is the Radon--Nikodym derivative.
We will need the following standard result.

\begin{lemma}
\lb{LmCdP}
Let $j\in\{1,2\}$, let $S\ss I_j$ be compact, and let $\m\in\cP_j(S)$.
Let $\cC_j(S,\m)$ be the set
of all measures $\n\in\cP_j(S)$, absolutely continuous with respect to
the measure $\m$ and such that $\cD_{j,KL}(\n\|\m)<\iy$:
$$
\cC_j(S,\m)=\{\n\in\cP_j(S):\n\ll\m,\cD_{j,KL}(\n\|\m)<\iy\}.
$$
Then the set $\cC_j(S,\m)$ is weakly dense in the set $\cP_j(S)$:
$$
\ol{\cC_j(S,\m)}^w=\cP_j(S).
$$

\end{lemma}

\no {\bf Proof.} Let $x\in S$. Consider the measure
$$
\n_x^{(\ve)}=\frac{\m|_{B_\ve(x)}}{\m(B_\ve(x))},
$$
where $ \ve>0$ is small enough,
$B_\ve(x)=(x-\ve,x+\ve)\cap S$ is the $\ve$-neighborhood
of the point $x$ in $S$, $\m|_{B_\ve(x)}(\O)=\m(B_\ve(x)\cap\O)$,
for any Borel set $\O\ss S$, $\m(B_\ve(x))>0$, since $x\in\supp\m$.
Then
$$
\cD_{j,KL}(\n_x^{(\ve)}\|\m)=\log\frac{1}{\m(B_\ve(x))}<\iy.
$$
Moreover, the estimate
$$
\Big|\int_{I_j}\vp(y)\n_x^{(\ve)}(dy)-\vp(x)\Big|
\le \sup_{y\in B_\ve(x)}|\vp(y)-\vp(x)|\to 0,\qq\text{as}
\ \ \ve\to 0,\qq \forall\ \ \vp\in\mC_j,
$$
gives
$\n_x^{(\ve)}\rightharpoonup\d_x$.
Therefore, any atomic measure can be weakly approximated by the measure
from $\cC_j(S,\m)$. Finite convex combination of such measures also belong
to $\cC_j(S,\m)$. Finite atomic measures are weakly dense in $\cP_j(S)$.
This proves the result.~\BBox

\subsection{Lyapunov function}
The next lemma shows that the energy functional $\cE_j$
is a Lyapunov functional for Eq.~\er{eqgpwtfj}.

\begin{lemma}
\lb{LmEDtheta}
Let $j\in\{1,2\}$.  Let the sequence $\m_0^{(n)}\in \cP_j^{ac}$, $n\in\N$,
satisfy $\m_0^{(n)}\rightharpoonup\m_0$ for some $\m_0\in\cP_j$.
Let $\m_\theta\in AC_w([0,+\iy),\cP_j)$,
be the solution to Eq.~\er{eqgpwtfj},
satisfying the initial condition $\m_\theta|_{\theta=0}=\m_0$.
Let $\m_\theta^{(n)}\in AC_w([0,+\iy),\cP_j^{ac})$, $n\in\N$,
be the solutions to Eq.~\er{eqgpwtfj},
satisfying the initial condition $\m_\theta^{(n)}|_{\theta=0}=\m_0^{(n)}$.
Then

i) $\m_\theta^{(n)}\rightharpoonup\m_\theta$, as $n\to\iy$, in $\cP_j$,
uniformly on any interval $[0,\cT],\cT>0$.

ii) The functionals $\cE_j[\m_\theta^{(n)}]$ and
$\cD_j[\m_\theta^{(n)}]$
converge to $\cE_j[\m_\theta]$ and $\cD_j[\m_\theta]$, respectively,
uniformly on $[0,\cT]$.
 The functions $\theta\mapsto\cE_j[\m_{\theta}]$
and $\theta\mapsto\cD_j[\m_{\theta}]$ are continuous on $[0,+\iy)$.

iii) The function $\theta\mapsto\cE_j[\m_{\theta}]$
satisfies the identity
\[
\lb{derLFf}
\cE_j[\m_t]+\a\int_s^t\cD_j[\m_u]du
=\cE_j[\m_s],\qq 0\le s\le t<\iy,
\]
where $\cD_j$ is given by \er{defcDfj}. In particular,
the function $\theta\mapsto\cE_j[\m_{\theta}]$ is non-increasing
and has a finite limit.

iv) The function $\theta\mapsto\cD_j[\m_{\theta}]$ satisfies the estimate
\[
\lb{varcDf}
\int_0^\iy\cD_j[\m_{\theta}]d\theta<\iy.
\]
In particular, $\theta\mapsto\cD_j[\m_{\theta}]$
tends to zero along some sequence $\theta_n\to\iy$.

\end{lemma}

\no {\bf Proof.}
i) Let $\m_0^{(n)}\in \cP_j^{ac}$, $n\in\N$,
$\m_0^{(n)}\rightharpoonup\m_0\in\cP_j$. Let $\m_\theta^{(n)}$
be the solutions to Eq.~\er{eqgpwtfj},  satisfying the condition
$\m_\theta^{(n)}|_{\theta=0}=\m_0^{(n)}$. Let $\m_\theta$
be the solutions to Eq.~\er{eqgpwtfj},  satisfying the condition
$\m_\theta|_{\theta=0}=\m_0$.
Then the estimate \er{estdifmn} implies
$$
\sup_{\theta\in[0,\cT]}d_{BL}(\m_\theta^{(n)},\m_\theta)\le
\sup_{\theta\in[0,\cT]}e^{q\theta}d_{BL}(\m_0^{(n)},\m_0)
\le e^{q\cT}d_{BL}(\m_0^{(n)},\m_0),\qq\forall\ \ \cT>0.
$$
For $j\in\{1,2\}$, the definitions \er{defdBL} and \er{defnBL} yield
$$
\sup_{\theta\in[0,\cT]}
\Big|\int_{I_j}\vp(x)\m_\theta(dx)-\int_{I_j}\vp(x)\m_\theta^{(n)}(dx)\Big|\to 0,
\qq\forall\ \ \vp\in\mC_j.
$$
Thus, $\m_\theta^{(n)}\rightharpoonup\m_\theta$,
as $n\to\iy$, in $\cP_j$,
uniformly on any interval $[0,\cT],\cT>0$.

ii) Let $j\in\{1,2,3\}$. If the family $\m_\theta^{(n)}$
converges in $\cP_j$ to the family $\m_\theta$ uniformly
on $[0,\cT]$, then
$$
\sup_{\theta\in[0,\cT]}
\Big|\int_{I_j}\vp(x)\m_\theta(dx)-\int_{I_j}\vp(x)\m_\theta^{(n)}(dx)\Big|\to 0,
\qq\forall\ \ \vp\in\mC_j.
$$
Due to
Lemma~\ref{Lmgp1}~i), $\cU_{\m_\theta^{(n)},j}$ and $\cU_{\m_\theta,j}$
are continuous on $I_j$
and, due to Lemma~\ref{Lmgp1}~ii),
$\sup_{x\in I_j}|\cU_{\m_\theta,j}(x)-\cU_{\m_\theta^{(n)},j}(x)|\to 0$, as $n\to\iy$.
The definition \er{defavU} gives
$$
\begin{aligned}
\sup_{\theta\in[0,\cT]}\Big|\langle \cU_{\m_\theta,j}\rangle_{\m_\theta,j}
-\langle \cU_{\m_\theta^{(n)},j}\rangle_{\m_\theta^{(n)},j}\Big|\le
\sup_{\theta\in[0,\cT]}
\Big(\int_{I_j}|\cU_{\m_\theta,j}(x)-\cU_{\m_\theta^{(n)},j}(x)|\m_\theta(dx)
\\
+\Big|\int_{I_j}\cU_{\m_\theta^{(n)},j}(x)\m_\theta(dx)
-\int_{I_j}\cU_{\m_\theta^{(n)},j}(x)\m_\theta^{(n)}(dx)\Big|\Big)\to 0,\qq
n\to\iy.
\end{aligned}
$$
Then the definition \er{defcEfj} implies
$\sup_{\theta\in[0,\cT]}|\cE_j[\m_\theta]-\cE_j[\m_\theta^{(n)}]|\to 0$.
Therefore, $\cE[\m_\theta^{(n)}]$
converges to $\cE[\m_\theta]$ uniformly on $[0,\cT]$.

Moreover, each
$f_{\theta,n}(x)=\cU_{\m_\theta^{(n)}}(x)-
\langle \cU_{\m_\theta^{(n)}}\rangle_{\m_\theta^{(n)}},\theta\in[0,\cT]$,
converges to
$f_\theta(x)=\cU_{\m_\theta}(x)-\langle \cU_{\m_\theta}\rangle_{\m_\theta}$,
uniformly on $I_j$, as $n\to\iy$.
Then the definition \er{defcDfj} implies
$$
\begin{aligned}
\sup_{\theta\in[0,\cT]}\big|\cD_j[\m_\theta]-\cD_j[\m_\theta^{(n)}]\big|
=\sup_{\theta\in[0,\cT]}\Big|\int_{I_j}f_\theta^2(x)\m_\theta(dx)-\int_{I_j}f_{\theta,n}^2(x)\m_\theta^{(n)}(dx)\Big|
\\
\le\sup_{\theta\in[0,\cT]}\Big(\int_{I_j}\big|f_\theta^2(x)-f_{\theta,n}^2(x)\big|\m(dx)
+\Big|\int_{I_j}f_{\theta,n}^2(x)\m(dx)-\int_{I_j}f_{\theta,n}^2(x)
\m_\theta^{(n)}(dx)\Big|\Big)\to 0,
\end{aligned}
$$
as $n\to\iy$.
Therefore, $\cD[\m_\theta^{(n)}]$ converge to $\cD[\m_\theta]$,
uniformly on $[0,\cT]$.

The function $\m_\theta$ is absolutely continuous in
the weak $\cP_j$-metrics, and, due to Lemma~\ref{Lmgp2}~ii),
$\cE_j[\mu]$ and $\cD_j[\mu]$ are continuous with respect to this metric.
Then the functions
$\theta\mapsto\cE_j[\m_{\theta}]$ and $\theta\mapsto\cD_j[\m_{\theta}]$
are therefore continuous.

iii), iv) Let $j\in\{1,2\}$.
 The identity \er{derLF} gives
$$
{d\cE_j[\m_\theta^{(n)}]\/d\theta}
=-\a\cD_j[\m_\theta^{(n)}],\qq \theta\ge 0,
$$
which yields
$$
\cE_j[\m_t^{(n)}]+\a\int_s^t\cD_j[\m_u^{(n)}]du
=\cE_j[\m_s^{(n)}],\qq 0\le s\le t<\iy.
$$
It follows from the statement i), that one can pass to the limit,
yielding \er{derLFf}.
Eq.~\er{derLFf} implies energy decay and the integrability of the dissipation functional:
$$
\cE_j[\m_t]\le\cE_j[\m_s],\qq \a\int_s^t\cD_j[\m_u]du=\cE_j[\m_s]-\cE_j[\m_t].
$$
Passing to the limit $s\to 0,t\to+\iy$, we obtain \er{varcDf}.~\BBox

\subsection{The trajectory goes to the minimizer}
 Due to Lemma~\ref{Lmminisunique}~ii),
the functional $\cE_j,j\in\{1,2\}$, has the unique minimizer on $\cP_j(S)$,
where $S\ss I_j$ is a compact set.
The next lemma shows that the energy $\cE_j$ approaches
its minimum along each trajectory and that the trajectory
converges weakly to the corresponding minimizer.

\begin{lemma}
\lb{Lmegtm}
Let $j\in\{1,2\}$.  Let $\m_0\in\cP_j$ and let
$\m_\theta\in AC_w([0,+\iy),\cP_j)$
be the solution to Eq.~\er{eqgpwtfj},
satisfying the initial condition $\m_\theta|_{\theta=0}=\m_0$.
Let, in addition, $\n\in\cP_j$ be absolutely continuous with respect
to $\m_0$, and let $\cD_{j,KL}(\n\|\m_0)<\iy$.
Then
\[
\lb{KLineqws}
\cE_j[\m_\theta]\le \cE_j[\n]+\frac{\cD_{j,KL}(\n\|\m_0)}{\a\theta },\qq
\theta\ge 0.
\]
Moreover, if $S=\supp\m_0$, then
\[
\lb{limcEtheta}
\lim_{\theta\to\iy}\cE_j[\m_\theta]=\min_{\n\in\cP_j(S)}\cE_j[\n],
\]
and
\[
\lb{trgomin}
\m_\theta\rightharpoonup\m_\iy,\qq\text{as}\ \ \theta\to\iy,
\]
where $\cP_j(S)$ is given by \er{defcPjS},
$\m_\iy$ is the minimizer of the functional $\cE_j$ on $\cP_j(S)$.
Furthermore,
\[
\lb{smiysssm0}
\supp\m_\iy\ss\supp\mu_0.
\]
\end{lemma}

\no {\bf Proof.}
Let $j\in\{1,2\}$ and let $\n\in\cP_j.$
Introduce the densities
$p_0^{(\s)}\in\mP_j$ and $q^{(\s)}\in\mP_j$ by
\[
\lb{denss}
p_0^{(\s)}(x)=\int_{I_j}\wt G_j(x,y,\s)\m_0(dy),\qq
q^{(\s)}(x)=\int_{I_j}\wt G_j(x,y,\s)\n(dy),
\]
where
\[
\lb{defmGjxy}
\wt G_j(x,y,\s)=\frac{ G_j(x,y,\s)}{\int_{I_j} G_j(z,y,\s)dz},
\]
$ G_j(x,y,\s)$ are given by \er{kernels} and \er{gaussp}.
Let $\m_0^{(\s)}\in\cP_j^{ac}$ and $\n^{(\s)}\in\cP_j^{ac}$ be the corresponding
measures, $\m_0^{(\s)}(dx)=p_0^{(\s)}(x)dx$
and $\n^{(\s)}(dx)=q^{(\s)}(x)dx$.
Then $\m_0^{(\s)}\rightharpoonup\m_0$ and $\n^{(\s)}\rightharpoonup\n$,
as $\s\to 0$.
Moreover, we have the {\it data processing inequality}:
\[
\lb{estDPI}
D_{j,KL}(q^{(\s)}\|p_0^{(\s)})\le \cD_{j,KL}(\n\|\m_0),\qq
\forall\ \ \s>0,
\]
see the proof below.
Let $\m_\theta^{(\s)}\in AC_w([0,+\iy),\cP_j^{ac})$
be the solutions to Eq.~\er{eqgpwtfj},
satisfying the initial condition $\m_\theta^{(\s)}|_{\theta=0}=\m_0^{(\s)}$,
and let $\m_\theta^{(\s)}(dx)=p_\theta^{(\s)}(x)dx$.
Due to Lemma~\ref{LmEDtheta}~i),
$\m_\theta^{(\s)}\rightharpoonup\m_\theta$, as $\s\to 0$,
uniformly on any interval $[0,\cT],\cT>0$.
The estimates \er{KLineq} and \er{estDPI} give
$$
\theta E_j[p_\theta^{(\s)}]\le \theta E_j[q^{(\s)}]
+\frac{D_{j,KL}(q^{(\s)}\|p_0^{(\s)})}{\a }
\le \theta E_j[q^{(\s)}]
+\frac{\cD_{j,KL}(\n\|\m_0)}{\a },\qq \s>0,\qq\theta\in[0,\cT].
$$
Lemma~\ref{Lmgp2}~i) shows that the functional $\cE_j$ is continuous
with respect to $\s$, then $\s\to 0$ gives
$$
\theta\cE_j[\m_\theta]\le \theta\cE_j[\n]
+\frac{\cD_{j,KL}(\n\|\m_0)}{\a },
$$
on the interval $[0,\cT]$, for all $\cT>0$, which yields \er{KLineqws}.

We establish \er{estDPI}; see also \cite[Ch~2.8]{CT06}. The definitions \er{denss} imply
$$
q^{(\s)}(x)=\int_{I_j}\wt G_j(x,y,\s)h(y)\m_0(dy),\qq\text{where}\ \
h(y)=\frac{d\m_0}{d\n}
$$
is the Radon--Nikodym derivative, and then
\[
\lb{qsdiivpos}
\frac{q^{(\s)}(x)}{p_0^{(\s)}(x)}=
\int_{I_j}h(y)\rho_x^{(\s)}(dy),\qq
\rho_x^{(\s)}(dy)=\frac{\wt G_j(x,y,\s)}{p_0^{(\s)}(x)}\m_0(dy),
\]
for each $x\in I_j$, the measure $\rho_x^{(\s)}$ belongs to $\cP_j$,
$\rho_x^{(\s)}\in\cP_j$. Applying Jensen's inequality to the convex
function $F(u)=u\log u$, we obtain
$$
F\Big(\frac{q^{(\s)}(x)}{p_0^{(\s)}(x)}\Big)\le
\int_{I_j}F\big(h(y)\big)\rho_x^{(\s)}(dy).
$$
The definitions \er{defKLqp}, \er{qsdiivpos}, \er{defmGjxy}, and \er{cDKL} give
$$
\begin{aligned}
D_{j,KL}(q^{(\s)}\|p_0^{(\s)})
=\int_{I_j}p_0^{(\s)}(x)F\Big(\frac{q^{(\s)}(x)}{p_0^{(\s)}(x)}\Big)dx\le
\int_{I_j}p_0^{(\s)}(x)\int_{I_j}F\big(h(y)\big)\rho_x^{(\s)}(dy)dx
\\
=\int_{I_j}\int_{I_j}F\big(h(y)\big)\wt G_j(x,y,\s)\m_0(dy)dx
=\int_{I_j}F\big(h(y)\big)\m_0(dy)
=\cD_{j,KL}(\n\|\m_0),\qq
\forall\ \ \s>0,
\end{aligned}
$$
which yields \er{estDPI}.

The identity \er{suppmu=0} gives $S:=\supp\mu_\theta=\supp\mu_0$,
for all $\theta\ge 0$. Due to Lemma~\ref{Lmminisunique}~i),
the set $\cP_j(S)$ is weakly closed in $\cP_j$, then
$\m_\theta\rightharpoonup\m_\iy$ yields $\m_\iy\in\cP_j(S)$.
Then $\supp\m_\iy=\lim_{\theta\to\iy}\supp\mu_\theta\ss\supp\mu_0$,
which yields \er{smiysssm0}.
The estimate \er{estDPI} gives
$$
\limsup_{\theta\to\iy}\cE_j[\m_\theta]\le\cE_j[\n],\qq \forall\ \
\n\in\cC_j(S,\m_0).
$$
Due to Lemma~\ref{LmCdP}
the set $\cC_j(S,\m_0)$ is weakly dense in the set $\cP_j(S)$, then,
due to Lemma~\ref{Lmgp2}~i), the functional $\cE_j$ is continuous
on this set. Therefore,
$$
\limsup_{\theta\to\iy}\cE_j[\m_\theta]\le e_S=\min_{\n\in\cP_j(S)}\cE_j[\n],
$$
the minimum exists, since the set $\cP_j(S)$ is compact.
However, $\m_\theta\in\cP_j(S)$ for all finite $\theta\ge 0$.
Then $\cE_j[\m_\theta]\ge e_S$, which yields \er{limcEtheta}.

We prove \er{trgomin}. Assume that
$\lim_{\theta\to\iy}\m_\theta=\m_\iy\ne\m_*$, where $\m_*$ is a minimizer
 of $\cE_j$ on $\cP_j(S)$.
By Lemma~\ref{Lmminisunique}~i), the set $\cP_j(S)$ is compact,
then there exists the sequence
$(\m_{\theta_k})_{k\in\N}$ such that $\m_{\theta_k}\rightharpoonup\m_\iy$,
as $k\to\iy$. Lemma~\ref{Lmgp2}~i) yields
$\cE_j[\m_{\theta_k}]\to \cE_j[\m_\iy]$. The identity \er{limcEtheta}
implies that $\m_\iy$ is also the minimizer of $\cE_j$ on $\cP_j(S)$.
This contradicts the uniqueness of the minimizer, which proves
\er{trgomin}.~\BBox

\medskip

\no {\bf Remark.}
Oechssler and Riedel proved in \cite[Th~2]{OR02}
for a broad class of replicator equations that the minimizer $\m_*$
of the functional $\cE$ is weakly asymptotically stable.
However, this is a local result: it does not imply that every trajectory eventually enters
the corresponding neighborhood.
For our Eq.~\er{eqgpwtfj} with $j\in\{1,2\}$ that every initial measure generates a trajectory
converging to the minimizer on its initial support.

\subsection{Dynamics of the weak solution}

We recall that, by Lemma~\ref{Lmfsmdi}, any weakly
stationary measure is a finitely atomic measure.

\medskip

\no {\bf Proof of Theorem~\ref{Thfs}.}
i)  Let $j\in\{1,2\}$. Lemma~\ref{LmEDtheta}~iii) shows that
the function $\theta\mapsto\cE_j[\m_{\theta}]$ is non-increasing.
Lemma~\ref{Lmegtm} implies that
$\m_\theta\rightharpoonup\m_\iy$, as $\theta\to\iy$,
where $\m_\iy$ is the unique minimizer of the functional $\cE_j$
on $\cP_j(S)$. The relation \er{smiysssm0} gives \er{smiysssm02}.
Lemma~\ref{Lmminisunique}~iii) gives \er{NC12}.

 For $j=2$, Lemma~\ref{Lmstsol}~ii) yields $\m_\iy$
is a finitely atomic measure.
For $j=1$, Lemma~\ref{Lmstsol}~i) implies $\m_\iy$ is a uniform measure
in the case $\supp\m_0=\T$ and $\m_\iy$
is a purely atomic measure with no accumulation points in the case
$\supp\m_0\ss\T$ and $\supp\m_0\ne\T$.

ii) Let $j=3$. Due to Lemma~\ref{LmcPjcomp}~iii), the set $\cP_3$ is compact
with respect to weak convergence. Hence every sequence
$\theta_n\to\iy$ admits a subsequence $\theta_{n_k}$ such that
$\mu_{\theta_{n_k}}\rightharpoonup\mu_\iy$
for some $\mu_\iy\in\cP_3$. Consequently,
$\o(\m_0)$ is non-empty. Moreover,
$\o(\m_0)$ is an intersection of closed sets,
then it is closed. It is therefore a closed subset of a compact set,
then it is compact. The trajectory $\m_\theta$ is weakly continuous on $[0,+\iy)$; hence
$\ol{\{\m_\theta:\theta>t\}}^w$
is connected, since it is the closure of the continuous image
of the connected set $[t,+\iy)$. The definition \er{omlj} yields that the set $\o(\m_0)$
is connected.

Let $\m_\theta\rightharpoonup\m_\iy\in\cP_3$, as $\theta\to\iy$.
The function $\m_\theta$ satisfies Eq.~\er{eqgpwtfj}, that is,
$$
{d\/d\theta}\int_{\R}\vp(x)\m_{\theta}(dx)
=-\a\int_{\R}\vp(x)
\big(\cU_{\m_{\theta},3}(x)
-\langle \cU_{\m_{\theta},3}\rangle_{\m_{\theta},3}\big)
\m_{\theta}(dx),\qq\forall\ \ \vp\in \mC_3,
$$
holds a.e. $\theta\ge 0$. Due to Lemma~\ref{Lmgp1}~ii), the function
$\m\to\int_{\R}\vp(x)(\cU_{\m,3}(x)-\langle \cU_{\m,3}\rangle_{\m,3})\m(dx)$
is continuous in $\cP_3$. Then
$$
\int_{\R}\vp(x)
\big(\cU_{\m_{\theta},3}(x)
-\langle \cU_{\m_{\theta},3}\rangle_{\m_{\theta},3}\big)
\m_{\theta}(dx)\to \int_{\R}\vp(x)
\big(\cU_{\m_{\iy},3}(x)
-\langle \cU_{\m_{\iy},3}\rangle_{\m_{\iy},3}\big)
\m_{\iy}(dx),
$$
as $\theta\to\iy$. Thus, the functions $\int_{\R}\vp(x)\m_{\theta}(dx)$ and
 ${d\/d\theta}\int_{\R}\vp(x)\m_{\theta}(dx)$ have the limit, as $\theta\to\iy$, therefore,
 ${d\/d\theta}\int_{\R}\vp(x)\m_{\theta}(dx)\to 0$. Thus, $\m_\iy$ is a weakly stationary measure.
Due to Lemma~\ref{Lmfsmdi}~iii), $\m_\iy$ satisfies  \er{stm3}.~\BBox

\end{document}